\documentclass[12pt]{article} 
\usepackage[a4paper,textwidth=490.0pt,textheight=703.1pt]{geometry}
\usepackage{slashed}
\usepackage{graphicx}
\usepackage{epsfig}
\usepackage{amsmath}
\usepackage{amssymb}
\usepackage{ulem}
\usepackage{mdwlist}
\usepackage{color}
\usepackage{float}   
\usepackage[rflt]{floatflt}
\usepackage{slashed}
\usepackage{cite}
\usepackage{mathtools}
\usepackage{subfig}

\newcommand{\HA}{{\rm H}}

\newcommand{\lnMa}[1]{L_1^{#1}}   
\newcommand{\lnMb}[1]{L_2^{#1}}
\newcommand{\Li}{\text{Li}}

\usepackage{array}

\newcommand{\Ahathat}{\hat{\hspace*{0mm}\hat{A}}}
\newcommand{\Athathat}{\hat{\hat{\tilde{A}}}}

\usepackage[english]{babel}
\usepackage[latin1]{inputenc}
\usepackage{ae}

\usepackage{url}

\usepackage{amsmath, amsthm, amssymb}

\newcommand{\GeV}{\rm GeV}
\newcommand{\MS}{\overline{{\rm MS}}}

\newcommand\MOM{{\rm MOM}}
\newcommand\NN{\nonumber}

\newcommand{\Mvec}{{\rm \bf M}}

\newcommand{\ep}{\varepsilon}

\usepackage{rotating}

\usepackage{graphicx}

\newcounter{mmacnt}
\def\restartmma{\setcounter{mmacnt}{0}}
\restartmma \catcode`|=\active
\def|#1|{\mathrm{#1}}
\catcode`|=12

\allowdisplaybreaks[4]

\usepackage{color}

\usepackage{tikz}
\usetikzlibrary{matrix}

\allowdisplaybreaks[4]

\begin{document}
\setlength{\baselineskip}{0.515cm}
\sloppy
\thispagestyle{empty}
\begin{flushleft}
DO--TH 23/16  \hfill  
\\
DESY 25--125 \hfill October 2025
\\
RISC Report series 25--07 
\\
ZU-TH 66/25 \\ 
\end{flushleft}

\setcounter{table}{0}

\begin{center}

\vspace*{1cm}
{\Large\bf \boldmath The two-mass contributions to the three-loop massive
operator matrix elements $\tilde{A}_{Qg}^{(3)}$
and $\Delta \tilde{A}_{Qg}^{(3)}$} 

\vspace{\fill}
\large
J.~Ablinger$^{a}$, 
J.~Bl\"umlein$^{b,c}$,
A.~De Freitas$^{a}$, 
A.~von Manteuffel$^{d}$,\\
C.~Schneider$^a$,
and
K.~Sch\"onwald$^{e}$\footnote{
Present address: CERN, Theoretical Physics Department, CH-1211 Geneva 23, Switzerland.}

\vspace*{1cm}
{\small
{\it $^a$~Johannes Kepler University Linz,
Research Institute for Symbolic Computation (RISC), \newline
Altenberger Stra\ss{}e 69, A-4040, Linz, Austria}

\vspace*{2mm}
{\it $^b$~Deutsches Elektronen-Synchrotron DESY, Platanenallee 6, 
15738 Zeuthen, Germany}

\vspace*{2mm}
{\it $^c$ Institut f\"ur Theoretische Physik III, IV, TU Dortmund, 
Otto-Hahn Stra\ss{}e 4, \newline 44227 Dortmund, Germany}

\vspace*{2mm}
{\it $^d$~Institut f\"ur Theoretische Physik, Universit\"at 
Regensburg,
93040 Regensburg, Germany}

\vspace*{2mm}
{\it $^e$~Physik-Institut, Universit\"at Z\"urich,
Winterthurerstrasse 190, CH-8057 Z\"urich, Switzerland}}
\\

\end{center}
\normalsize
\vspace{\fill}
\begin{abstract}
\noindent 
We calculate the two-mass three-loop contributions to the unpolarized and polarized massive 
operator matrix elements $\tilde{A}_{Qg}^{(3)}$ and $\Delta \tilde{A}_{Qg}^{(3)}$ in 
$x$-space for a general mass ratio by using a semi-analytic approach. We also compute 
Mellin moments up to $N = 2000 (3000)$ by an independent method, to which we compare the 
results in $x$-space. In the polarized case, we work in the Larin scheme. We present numerical 
results. The two-mass contributions amount to about $50 \%$ of the full  
\textcolor{blue}{$O(T_F^2)$} and \textcolor{blue}{$O(T_F^3)$} terms contributing to the 
operator matrix elements. The present result completes the calculation of all unpolarized 
and polarized massive three-loop operator matrix elements.
\end{abstract}

\vspace*{\fill}
\noindent
\numberwithin{equation}{section}

\newpage
\section{Introduction}
\label{sec:1}

\vspace*{1mm}
\noindent
The massive operator matrix elements (OMEs) of local composite operators in Quantum Chromodynamics (QCD) 
are of central importance to describe the heavy-flavor contributions to the massive 
Wilson coefficients 
in deep-inelastic scattering in the single-mass case in the asymptotic region $Q^2 \gg 
m_Q^2$
\cite{Buza:1995ie,Buza:1996xr,Bierenbaum:2007qe,
Bierenbaum:2022biv,Ablinger:2010ty,Ablinger:2014vwa,Ablinger:2014nga,Behring:2014eya,Ablinger:2019etw,
Blumlein:2021xlc,Ablinger:2023ahe,Ablinger:2024xtt}, where $m_Q$ denotes a heavy quark 
mass. From two-loop order 
onward, two-mass effects contribute 
\cite{Blumlein:2018jfm,Bierenbaum:2022biv}. The three-loop corrections have been calculated analytically in 
Refs.~\cite{Ablinger:2017err,Ablinger:2017xml,Ablinger:2018brx,Ablinger:2019gpu,Ablinger:2020snj}, except
for the OMEs $\tilde{A}_{Qg}^{(3)}$ and $\Delta \tilde{A}_{Qg}^{(3)}$. The further transition functions 
in the single-mass variable flavor number scheme (VFNS) were calculated in 
Refs.~\cite{Buza:1996wv,Bierenbaum:2009zt,
Blumlein:2012vq,Ablinger:2014lka,Behring:2021asx,Ablinger:2022wbb}.
In the polarized case, we work in the Larin scheme \cite{Larin:1993tq}.

The two-mass effects require to extend the single heavy mass VFNS \cite{Buza:1996wv,Ablinger:2025joi} 
to the two-mass one~\cite{Ablinger:2017err}. 
This is phenomenologically motivated since the mass ratio of the charm and the bottom 
quarks
\begin{eqnarray}
\eta = \frac{m_c^2}{m_b^2} \sim \frac{1}{9}
\end{eqnarray}
is not a very small quantity.

The moments $N = 2, 4, 6$ for the constant part of the unrenormalized OMEs  
$\tilde{A}_{Qg}^{(3)}$, $\tilde{a}_{Qg}^{(3)}$, have been calculated in 
\cite{Ablinger:2011pb,
Ablinger:2011gdv} expanding in the parameter $\eta$ with the  code {\tt Q2E/Exp} 
\cite{Harlander,Seidensticker:1999bb}. In the expanded form, much higher Mellin  moments 
have been calculated in Ref.~\cite{Schonwald:2019gmn}. An analytic conversion of these results to $x$-space cannot be 
performed 
straightforwardly since with higher powers in $\eta$ the number of $N$-dependent terms in the expansion 
grows. Therefore, we will compute the two-mass corrections in 
$x$-space 
directly.
Both, in $N$- and $x$-space, elliptic contributions are present, as it is already the 
case for the  
single-mass amplitudes, 
cf.~Refs.~\cite{Ablinger:2023ahe,Ablinger:2024xtt,Ablinger:2017bjx,Behring:2023rlq}. 
Because of this we will calculate 
the quantities $\tilde{A}_{Qg}^{(3)}$ and $\Delta \tilde{A}_{Qg}^{(3)}$ by using a semi-analytic 
method, see also Refs.~\cite{Maier:2017ypu,Fael:2022miw,
Fael:2021kyg,Ablinger:2023ahe,Ablinger:2024xtt}. 

The paper is organized as follows. In Section~\ref{sec:2}, we present 
a series of specific aspects 
for the renormalization of the OME $(\Delta) \tilde{A}_{Qg}^{(3)}$ in the two-mass case. 
We describe the main steps of the calculation in Section~\ref{sec:3}.
The calculation of the two-mass OME $(\Delta) \tilde{A}_{Qg}^{(3)}$ 
in Mellin-$N$ space is described in 
Section~\ref{sec:4}.
We also discuss the limitations of this approach for the present OME.\footnote{Also for the pure-singlet
two-mass OMEs $\tilde{A}_{Qq}^{(3),\rm PS}$ and  $\Delta \tilde{A}_{Qq}^{(3),\rm PS}$ 
\cite{Ablinger:2017xml,
Ablinger:2019gpu} it turned out that the $N$-space solution was not possible, since the corresponding 
difference equations did not factorize at first order, unlike the differential equations in $x$-space.}  
In Section~\ref{sec:5}, we compute the constant part 
of the unrenormalized OMEs $\tilde{A}_{Qg}^{(3)}$ and $\Delta \tilde{A}_{Qg}^{(3)}$
in $x$-space, based on the coupled system of first order differential equations obtained
for the master integrals. In  the generating function representation 
\cite{Ablinger:2012qm,Ablinger:2014yaa} they depend on the variable $t$ emerging in the
linear propagators introduced to perform the integration-by-parts (IBP) reduction 
\cite{Studerus:2009ye,vonManteuffel:2012np}. Section~\ref{sec:6} contains the conclusions. In the Appendix,
we list a series of moments for 
$\tilde{a}_{Qg}^{(3)}$ and $\Delta \tilde{a}_{Qg}^{(3)}$ both expanded and unexpanded in 
the mass ratio $\eta$.
\section{Renormalization}
\label{sec:2}

\vspace*{1mm}
\noindent
\subsection{General Formalism}
\label{sec:2.1}

\vspace*{1mm}
\noindent
The renormalization of the operator matrix elements for deep-inelastic scattering up to ${O}(a_s^3)$ 
has been carried out in the single-mass case in Ref.~\cite{Bierenbaum:2009mv} and in the 
two-mass case in Ref.~\cite{Ablinger:2017err}. Here $a_s = \alpha_s/(4\pi) = g_s^2/(16 \pi^2)$ denotes the 
strong coupling constant. In the following, we collect the essential 
steps for the renormalization of the 
OME $A_{Qg}$ to three--loop order. 
In general, the renormalization in the two-mass case 
has to include that of the single-mass cases. 
Later on, the pure two-mass part of the renormalized 
OME is separated. Some steps of the renormalization in \cite{Ablinger:2017err} 
were given for the power-expanded form in the parameter $\eta$ since the full analytic 
results had not been available yet.

The Feynman integrals contributing to the various operator matrix elements contain mass, coupling, 
ultraviolet operator singularities, as well as collinear divergences, due to massless sub-graphs. They 
are 
regularized in $D=4+\ep$ dimensions. The singularities appear as poles in the Laurent series in 
$\ep$, with the highest pole corresponding to the loop order. At one and two-loop order, 
there are no irreducible contributions to the OMEs $(\Delta) \tilde{A}_{ij}$, 
cf. Refs.~\cite{Buza:1995ie,Buza:1996wv,Bierenbaum:2007qe,Bierenbaum:2007dm,
Bierenbaum:2008yu,Bierenbaum:2009zt,Blumlein:2006mh,Behring:2014eya}.
The first single-particle irreducible diagrams with two masses emerge at ${O}(a_s^3)$. In the 
following, we consider the renormalization of the two-mass contributions in individual 
terms together 
with the genuine two-mass contributions. The latter terms will then be obtained subtracting the 
former ones, cf.~Ref.~\cite{Bierenbaum:2009mv}. The unrenormalized OMEs are given by
\begin{eqnarray}
    \Ahathat_{ij}^{(l)}\Bigl(\frac{m_1^2}{\mu^2},\frac{m_2^2}{\mu^2}\Bigr) =
   \Ahathat_{ij}^{(l)} \Bigl(\frac{m_1^2}{\mu^2}\Bigr)
 + \Ahathat_{ij}^{(l)} \Bigl(\frac{m_2^2}{\mu^2}\Bigr)
 + \Athathat_{ij}^{(l)} \Bigl(\frac{m_1^2}{\mu^2},\frac{m_2^2}{\mu^2}\Bigr),
\label{eq:AhathatDecomp}
\end{eqnarray}
where $\Ahathat_{ij}^{(l)} \Bigl( \frac{m_i^2}{\mu^2} \Bigr)$ are the single-mass OMEs 
\cite{Bierenbaum:2009mv} and $\Athathat_{ij}^{(l)}$ are the two-mass contributions.
A change in the renormalization scheme like in Eqs.~\eqref{eq:asmoma} and \eqref{eq:asmsa} 
generally introduces a mixing between the different components of Eq.~\eqref{eq:AhathatDecomp}. 
Here we introduced the two masses $m_1$ and $m_2$. We set $m_2 < m_1$ so that $\eta < 1$.
In the following we will also refer to the logarithms
\begin{align}
	L_1 &= \ln\left( \frac{m_1^2}{\mu^2} \right), 
&
	L_2 &= \ln\left( \frac{m_2^2}{\mu^2} \right),
&
	L_\eta &= \ln\left( \eta \right) = L_2 - L_1,
\label{eq:masslogs}
\end{align}
where $\mu$ is the renormalization scale.

The renormalization procedure follows the one outlined in Ref.~\cite{Bierenbaum:2009mv}, 
incorporating the necessary modifications for the two-mass case. Here the case of $N_F$ massless and two massive 
quark flavors is considered as this covers the physical case of contributions due to the charm and bottom quarks.
The large mass gap to the top quark in general allows to decouple it after the charm and the bottom quark and thus
does not have to be included into a scheme with $N_F$ massless and three massive quarks.
Since in this chapter again two massive quarks are considered, we will use the notation
\begin{eqnarray}
	\tilde{f}(\xi)&=&\frac{f(\xi)}{\xi}~, 
	\\
	\hat{f}(\xi)&=&f(\xi+2)-f(\xi)~
\label{eq:HAT2}
\end{eqnarray}
to abbreviate certain expressions. Note that these notations  are not meant to apply to OMEs, but 
only to the anomalous dimensions $\gamma_{ij}$ in this section. 
The differences with respect to \cite{Ablinger:2017err} mainly lie in the use of Eq.~\eqref{eq:HAT2} instead of the convention with one heavy quark.
Furthermore, the notation $\hat{\tilde{f}}(x)$ was used to represent the coefficient of the 
$N_F^2$ dependent term of certain anomalous dimensions.
However, this is not in accordance with the conventions for anomalous dimensions which start 
with $N_F^0$.
These terms will be therefore denoted by $f^{N_F^2}$ in the following.
Also the inclusion of reducible contributions introduces some subtleties for OMEs with external
gluons. The renormalization concerns the heavy quark masses, the strong coupling
\cite{Tarasov:1980au,Larin:1993tp,vanRitbergen:1997va,Czakon:2004bu,
Chetyrkin:2004mf,Baikov:2016tgj,Herzog:2017ohr,Chetyrkin:2017bjc,Luthe:2017ttg,Luthe:2017ttc,
Chetyrkin:1999ys,
Chetyrkin:1999qi,Melnikov:2000qh,Broadhurst:1991fy,Marquard:2018rwx,Marquard:2016dcn,
Marquard:2015qpa}, 
the local operators 
\cite{Moch:2004pa,Vogt:2004mw,Vermaseren:2005qc,
Ablinger:2010ty,Blumlein:2012vq,Ablinger:2014lka,Ablinger:2014vwa,
Ablinger:2014nga,Moch:2014sna,Anastasiou:2015vya,Ablinger:2017tan,
Mistlberger:2018etf,Behring:2019tus,Luo:2019szz, Duhr:2020seh, Ebert:2020yqt,Ebert:2020unb,
Luo:2020epw,  
Blumlein:2021enk,
Blumlein:2021ryt,
Blumlein:2022gpp,
Baranowski:2022vcn,
Gehrmann:2023ksf}, 
and the subtraction of the collinear singularities due to massless 
sub-graphs \cite{Bierenbaum:2009mv}, which occurs at one-loop order 
in the present case, unlike the single-mass OMEs.

The schemes most frequently used for the mass renormalization are the $\MS$- and the on-mass shell scheme (OMS). 
In the following, the mass is renormalized in the OMS and the finite renormalization to switch to the $\MS$-mass is provided at 
a later stage. The mass renormalization is applied first, i.e., the respective expressions
are still containing the bare coupling $\hat{a}_s = \hat{g}_s^2/(4 \pi)^2$.\footnote{Note that this notation 
therefore agrees with Ref.~\cite{Gray:1990yh}, but, e.g., differs form the notation 
in~\cite{Melnikov:2000zc,Bekavac:2007tk,Marquard:2016dcn}, 
where also the charge renormalization has been carried out.}

The bare masses $\hat{m_i},~ i \in \{1,2\}$ are expressed by the renormalized on-shell masses $m_i$ via
\begin{align}
\hat{m_i}=Z_{m,i}(m_1,m_2)~ m_i 
           =& m_i \Bigl[ 1 
                       + \hat{a}_s
                       \Bigl(\frac{m_i^2}{\mu^2}\Bigr)^{\ep/2}
                                   \delta m_1 
                       + \hat{a}_s^2
                       \Bigl(\frac{m_i^2}{\mu^2}\Bigr)^{\ep} \delta m_{2,i}\left(m_1,m_2\right)                       
                 \Bigr] 
                 + {O}(\hat{a}_s^3)~,
            \label{eq:mren1}
\end{align}
and  
\begin{eqnarray}
\delta m_{2,i}\left(m_1,m_2\right)=
                                     \delta m_2^0 +\tilde{\delta}{m_2}^{i}(m_1,m_2)~.
\label{eq:dm2twomass}
\end{eqnarray}                                    
Here $\delta m_2^0$ is the single mass-contribution, whereas $\tilde{\delta}{m_2}^i$ denotes the additional contribution 
emerging in the case of two massive flavors.  Note that from order ${O}(\hat{a}_s^2)$ onward the $Z$-factor 
renormalizing 
$\hat{m}_1$ depends on $m_2$ and vice versa. For the massive operator matrix elements this can be observed at $3$-loop 
order for the first time. The coefficients $\delta m_1$ and $\delta m_2$ have been derived in
Refs.~\cite{Tarrach:1980up,Nachtmann:1981zg} up to ${O}(\ep^0)$ and ${O}(\ep^{-1})$, respectively. The constant part of 
$\delta m_2$ 
was given in Refs.~\cite{Gray:1990yh,Broadhurst:1991fy,Fleischer:1998dw} and the ${O}(\ep)$-term of $\delta m_1$ in
Ref.~\cite{Bierenbaum:2009mv}. One obtains
\begin{eqnarray}
    \delta m_1 &=&C_F
	\left[\frac{6}{\ep}-4+\left(4+\frac{3}{4}\zeta_2\right)\ep\right] 
	\label{eq:delm1}  \\
               &\equiv&  \frac{\delta m_1^{(-1)}}{\ep}
                        +\delta m_1^{(0)}
                        +\delta m_1^{(1)}\ep~, 
	\label{eq:delm1exp} \\
    \delta m_2^0 &=& C_F
                   \Biggl[\frac{1}{\ep^2}\left(18 C_F-22 C_A+8T_F(N_F+1)
                    \right)
                  +\frac{1}{\ep}\Biggl(-\frac{45}{2}C_F+\frac{91}{2}C_A
\NN\\&&
                  -14T_F
                   (N_F+1)\Biggr)
                  +C_F\left(\frac{199}{8}-\frac{51}{2}\zeta_2+48\ln(2)\zeta_2
                   -12\zeta_3\right)
                   +C_A\Biggl(-\frac{605}{8}
                  \NN\\&&
                  +\frac{5}{2}\zeta_2-24\ln(2)\zeta_2+6\zeta_3
                  \Biggr)
                  +T_F\left[N_F\left(\frac{45}{2}+10\zeta_2\right)+
                  \frac{69}{2}-14\zeta_2\right]\Biggr]
                  \label{eq:delm2}  \\
               &\equiv&  \frac{\delta m_2^{0,(-2)}}{\ep^2}
                        +\frac{\delta m_2^{0,(-1)}}{\ep}
                        +\delta m_2^{0,(0)}~, \label{eq:delm2exp}
\\
\tilde{\delta}{m_2}^{i}(m_1,m_2) &=& C_F T_F \Biggl\{
  \frac{8}{\ep^2}
 -\frac{14}{\ep}
 +8 r_i^4 \HA_{0}^2(r_i)
 -8 (r_i+1)^2 \left(r_i^2-r_i+1\right) \HA_{-1,0}(r_i)
 \NN\\&&
 +8 (r_i-1)^2 \left(r_i^2+r_i+1\right) \HA_{1,0}(r_i)
 +8 r_i^2 \HA_0(r_i)
 +\frac{3}{2} \left(8 r_i^2+15\right)
 \NN\\&&
+2 \Bigl[4 r_i^4-12 r_i^3-12 r_i+5\Bigr] \zeta_2 \Biggr\}
\label{eq:dm2mixfull}
 \\
               &\equiv&  \frac{\tilde{\delta}{m_2}^{(-2)}}{\ep^2}
                        +\frac{\tilde{\delta}{m_2}^{(-1)}}{\ep}
                        +\tilde{\delta}{m_2}^{i,(0)}~, 
\label{eq:delm2mixexp}
\end{eqnarray}
cf. Ref.~\cite{Gray:1990yh}, $i\in \{1,2\}$ and  
\begin{eqnarray}
r_1 = \sqrt{\eta}~~~~\text{and}~~~~r_2=\frac{1}{\sqrt{\eta}}.
\end{eqnarray}
The superscript $i$ for the coefficients 
$\tilde{\delta} m_2^{(-2)}$ and $\tilde{\delta} m_2^{(-1)}$ has been dropped as they are independent of 
the renormalized 
mass $m_i$. The  color factors of QCD are given by 
$\textcolor{blue}{C_A} = N_c = 3$,
$\textcolor{blue}{C_F} = (N_c^2-1)/(2 N_c) = 4/3$, 
$\textcolor{blue}{T_F} = 1/2$. 
The constants 
$\zeta_n,~~n \geq 2$, denote the Riemann $\zeta$ function  \cite{RIEMANN,TITCH1} evaluated at integer argument 
$n$,
\begin{eqnarray}
\zeta_n =  \sum_{k=1}^\infty \frac{1}{k^n}~~~n \geq 2,~~~n \in \mathbb{N}.
\end{eqnarray}
The harmonic polylogarithms \cite{Remiddi:1999ew}
used to express the result in Eq.~\eqref{eq:dm2mixfull} are iterated integrals of the kind
\begin{eqnarray}
\HA_{b,\vec{a}}(x) = \int_0^x dy f_b(y) \HA_{\vec{a}}(y),~~~f_b(x) \in \Biggl\{\frac{1}{x}, 
\frac{1}{1-x}, \frac{1}{1+x}\Biggr\}, a_i,b \in \{-1,0,1\}
\end{eqnarray}
and if all $k$ entries of the indices in $\HA$ are zero, one defines the harmonic
polylogarithm by $\ln^k(x)/k!$.

Applying Eq.~\eqref{eq:mren1} we obtain the mass renormalized operator matrix elements by
\begin{eqnarray}
    \Ahathat_{ij}\Bigl(\frac{m_1^2}{\mu^2},\frac{m_2^2}{\mu^2},\ep,N\Bigr) 
                 &=& \delta_{ij}
                 +\hat{a}_s~ 
                   \Ahathat_{ij}^{(1)}\Bigl(\frac{m_1^2}{\mu^2},\frac{m_2^2}{\mu^2},\ep,N\Bigr) 
                         + \hat{a}_s^2 \Biggl\{
                                        \Ahathat^{(2)}_{ij}
                                        \Bigl(\frac{m_1^2}{\mu^2},\frac{m_2^2}{\mu^2},\ep,N\Bigr) 
\NN\\&&
                                      + {\delta m_1} 
                                        \Bigl[\Bigl(\frac{m_1^2}{\mu^2}\Bigr)^{\ep/2}
                                        m_1 \frac{d}{d m_1}
                                      +\Bigl(\frac{m_2^2}{\mu^2}\Bigr)^{\ep/2}
                                        m_2 \frac{d}{d m_2}
                                      \Bigr]
                                                   \Ahathat_{ij}^{(1)}
                                           \Bigl(\frac{m_1^2}{\mu^2},\frac{m_2^2}{\mu^2},\ep,N\Bigr)                        
                                \Biggr\}
\NN\\ &&
                         + \hat{a}_s^3 \Biggl\{ 
                                         \Ahathat^{(3)}_{ij}
                                           \Bigl(\frac{m_1^2}{\mu^2},\frac{m_2^2}{\mu^2},\ep,N\Bigr) 
\NN\\ &&
                                        +{\delta m_1} 
                                         \left[
                                           \Bigl(\frac{m_1^2}{\mu^2}\Bigr)^{\ep/2}
                                         m_1\frac{d}{d m_1} 
                                         +
                                           \Bigl(\frac{m_2^2}{\mu^2}\Bigr)^{\ep/2}
                                         m_2 \frac{d}{d m_2} 
                                         \right]
                                                    \Ahathat_{ij}^{(2)}
                                           \Bigl(\frac{m_1^2}{\mu^2},\frac{m_2^2}{\mu^2},\ep,N\Bigr)
\NN\\ &&
                                        +
                                        \delta m_{2,1}(m_1,m_2) 
                                    \Bigl(\frac{m_1^2}{\mu^2}\Bigr)^{\ep} m_1    \frac{d}{d m_1}
                                                    \Ahathat_{ij}^{(1)}\Bigl(\frac{m_1^2}{\mu^2},\frac{m_2^2}{\mu^2},\ep,N\Bigr)
\NN\\ &&
                                          +\delta m_{2,2}(m_1,m_2)
                                    \Bigl(\frac{m_2^2}{\mu^2}\Bigr)^{\ep} m_2    \frac{d}{d m_2} 
                                                    \Ahathat_{ij}^{(1)}\Bigl(\frac{m_1^2}{\mu^2},\frac{m_2^2}{\mu^2},\ep,N\Bigr) 
\NN\\&&
                                        +
                                        \frac{(\delta m_1)^2}{2}
                                        \left[
                                          \Bigl(\frac{m_1^2}{\mu^2}\Bigr)^{\ep}         
                                              m_1^2     \frac{d^2}{{d m_1}^2}
                                                   +
                                            \Bigl(\frac{m_2^2}{\mu^2}\Bigr)^{\ep}
                                               m_2^2    \frac{d^2}{{d m_2}^2}
                                               \right]
                                                     \Ahathat_{ij}^{(1)}
                                           \Bigl(\frac{m_1^2}{\mu^2},\frac{m_2^2}{\mu^2},\ep,N\Bigr) 
\NN\\&&
                                         + (\delta m_1)^2 
                                        \Bigl(\frac{m_1^2}{\mu^2}\Bigr)^{\ep/2}
                                          \Bigl(\frac{m_2^2}{\mu^2}\Bigr)^{\ep/2}
                                       m_1  \frac{d}{d m_1}
                                       m_2  \frac{d}{d m_2}
                                                   \Ahathat_{ij}^{(1)}
                                           \Bigl(\frac{m_1^2}{\mu^2},\frac{m_2^2}{\mu^2},\ep,N\Bigr)
    \Biggr\}~,
\NN\\&&
\label{eq:maren}
\end{eqnarray}
which generalizes Eq.~(3.10) of Ref.~\cite{Bierenbaum:2009mv}.
The OMEs are symmetric under the interchange of the masses $m_1$ and $m_2$.

We now turn to the renormalization of the strong coupling constant. It is important to note that the 
factorization relation for the OMEs strictly requires the external massless partonic legs of the 
operator matrix elements to be on-shell, i.e. 
\begin{eqnarray}
	p^2=0~{\label{eq:ONSHELL}}, 
\end{eqnarray} 
with $p$ the external momentum. This condition would be violated by naively applying
massive loop corrections to the gluon propagator. Following \cite{Bierenbaum:2009mv} it is 
possible to absorb these corrections uniquely into the coupling constant by using the background 
field method \cite{Abbott:1980hw,Rebhan:1985yf,Jegerlehner:1998zg} to maintain the Slavnov-Taylor 
identities of QCD.  In this way, one first obtains the coupling constant in a {\text{MOM}}-scheme. 
A finite renormalization to transform to the {$\MS$}-scheme is applied subsequently. 

The light flavor contributions to the unrenormalized coupling constant in terms of the renormalized coupling constant
in the {$\MS$}-scheme read 
\begin{eqnarray}
   \hat{a}_s             &=& {Z_g^{\MS}}^2(\ep,N_F) 
                             a^{\MS}_s(\mu^2) \NN\\
                         &=& a^{\MS}_s(\mu^2)\left[
                                   1 
                                 + \delta a^{\MS}_{s, 1}(N_F) 
                                      a^{\MS}_s(\mu^2)
                                 + \delta a^{\MS}_{s, 2}(N_F) 
                                      {a^{\MS}_s}^2(\mu)    
                                     \right] + {O}({a^{\MS}_s}^3)~. 
                            \label{eq:asrenMSb}
\end{eqnarray}
Here the coefficients $\delta a^{\MS}_{s, i}(N_F)$ are given by
\begin{eqnarray}
    \delta a^{\MS}_{s, 1}(N_F) &=& \frac{2}{\ep} \beta_0(N_F)~,
                             \label{eq:deltasMSb1} \\
    \delta a^{\MS}_{s, 2}(N_F) &=& \frac{4}{\ep^2} \beta_0^2(N_F)
                           + \frac{1}{\ep} \beta_1(N_F),
                             \label{eq:deltasMSb2}
\end{eqnarray}
with $\beta_k(N_F)$ the expansion coefficients of the QCD $\beta$-function
\cite{Gross:1973id,Politzer:1973fx,Khriplovich:1969aa,Caswell:1974gg,Jones:1974mm}
\begin{eqnarray}
   \beta_0(N_F)
                 &=& \frac{11}{3} C_A - \frac{4}{3} T_F N_F \label{eq:beta0}~, \\
   \beta_1(N_F)
                 &=& \frac{34}{3} C_A^2 
               - 4 \left(\frac{5}{3} C_A + C_F\right) T_F N_F \label{eq:beta1}~.
\end{eqnarray}

The renormalized gluon self-energy $\Pi$ can be split into the purely light and the 
heavy-flavor contributions, $\Pi_L$ and 
$\Pi_H$,
\begin{equation}
 {\Pi}\left(p^2,m_1^2,m_2^2\right)={\Pi}_{L}\left(p^2\right)+{\Pi}_{H}\left(p^2,m_1^2,m_2^2\right)~.
\end{equation}
The heavy quarks are required to decouple from the running coupling constant and the renormalized OMEs for 
$\mu^2<m_1^2,m_2^2$, which 
implies \cite{Buza:1995ie} 
\begin{equation}
{\Pi}_{H}(0,m_1^2,m_2^2)=0~ \label{eq:DecouplingCond}. 
\end{equation}
Applying the background field method has the advantage of producing gauge-invariant results also for off-shell 
Green's functions, \cite{DeWitt:1967ub,Abbott:1980hw}.
We obtain for the heavy-flavor contributions to the unrenormalized 
gluon polarization function 

  \begin{eqnarray}
   \hat{\Pi}^{\mu\nu}_{H,ab,\mbox{\tiny{BF}}}(p^2,m_1^2,m_2^2,\mu^2,\ep,\hat{a}_s)&=&
                                i (-p^2g^{\mu\nu}+p^{\mu}p^{\nu})\delta_{ab}
\hat{\Pi}_{H,\mbox{\tiny{BF}}}(p^2,m_1^2,m_2^2,\mu^2,\ep,\hat{a}_s)~, 
\\
   \hat{\Pi}_{H,\mbox{\tiny{BF}}}(0,m_1^2,m_2^2,\mu^2,\ep,\hat{a}_s)&=&
                    \hat{a}_s   \frac{2\beta_{0,Q}}{\ep}
                         \left[\Bigl(\frac{m_1^2}{\mu^2}\Bigr)^{\ep/2}
                           +\Bigl(\frac{m_2^2}{\mu^2}\Bigr)^{\ep/2} \right]
                          \exp \Bigl(\sum_{i=2}^{\infty}\frac{\zeta_i}{i}
                          \Bigl(\frac{\ep}{2}\Bigr)^{i}\Bigr)
\NN\\ &&
                   +\hat{a}_s^2 \left[\Bigl(\frac{m_1^2}{\mu^2}\Bigr)^{\ep}
                   +\Bigl(\frac{m_2^2}{\mu^2}\Bigr)^{\ep}\right]
                        \Biggl[
                       \frac{1}{\ep}\Bigl(
                                          -\frac{20}{3}T_FC_A
                                          -4T_FC_F
                                    \Bigr)
\NN\\&&
                      -\frac{32}{9}T_FC_A
                      +15T_FC_F
\NN\\&&
                     +\ep            \Bigl(
                                          -\frac{86}{27}T_FC_A
                                          -\frac{31}{4}T_FC_F
                                          -\frac{5}{3}\zeta_2T_FC_A
                                          -\zeta_2T_FC_F
                                   \Bigr)
\NN\\ &&
                        +2 \left(\frac{2\beta_{0,Q}}{\ep}\right)^2
                         \Bigl(\frac{m_1^2}{\mu^2}\Bigr)^{\ep/2}
                           \Bigl(\frac{m_2^2}{\mu^2}\Bigr)^{\ep/2} 
                          \exp \Bigl(2 \sum_{i=2}^{\infty}\frac{\zeta_i}{i}
                          \Bigl(\frac{\ep}{2}\Bigr)^{i}\Bigr)
                         \Biggl]
\NN\\&&                         
                         + {O}(\hat{a}_s^3)~, \label{eq:GluSelfBack}
\end{eqnarray}
where the masses $m_1$ and $m_2$ have been renormalized in the on-shell scheme, cf. Eq.~ \eqref{eq:mren1}. In order to write the relation in Eq.~\eqref{eq:GluSelfBack} 
in a more compact form, the following notation
\begin{eqnarray}
   f(\ep)&\equiv&
   \left[
                 \Bigl(\frac{m_1^2}{\mu^2}\Bigr)^{\ep/2}
                 +\Bigl(\frac{m_2^2}{\mu^2}\Bigr)^{\ep/2}\right]
    \exp \left[\sum_{i=2}^{\infty}\frac{\zeta_i}{i}
                       \Bigl(\frac{\ep}{2}\Bigr)^{i}\right]~, \label{eq:fep}
\end{eqnarray}
is used. The expression $f(\ep)$ is kept unexpanded in the dimensional regularization parameter
$\ep$ for the moment.
Furthermore, the contributions to the 
QCD $\beta$-function coefficients are denoted by $\beta_{i,Q}^{(j)}$ 
\cite{Gross:1973id,Politzer:1973fx,Khriplovich:1969aa,Caswell:1974gg,Jones:1974mm,Buza:1995ie,Bierenbaum:2009mv}
  \begin{eqnarray}
   \beta_{0,Q} &=&-\frac{4}{3}T_F~, \label{eq:b0Q} \\
   \beta_{1,Q} &=&- 4 \left(\frac{5}{3} C_A + C_F \right) T_F~, \label{eq:b1Q} \\
   \beta_{1,Q}^{(1)}&=&
                           -\frac{32}{9}T_FC_A
                           +15T_FC_F~, \label{eq:b1Q1} \\
   \beta_{1,Q}^{(2)}&=&
                               -\frac{86}{27}T_FC_A
                               -\frac{31}{4}T_FC_F
                               -\zeta_2\left(\frac{5}{3}T_FC_A
                                        +T_FC_F\right)~. \label{eq:b1Q2}
  \end{eqnarray}
Eq.~\eqref{eq:GluSelfBack} differs from the sum of the two individual single-mass 
contributions \cite{Bierenbaum:2009mv} by the last term 
only, which is due to additional reducible Feynman diagrams in the cases of two heavy 
quark flavors of different mass.

The background field is renormalized using the $Z$-factor $Z_A$ which is split into light and 
heavy 
quark contributions, $Z_{A,L}$ 
and $Z_{A,H}$. It is related to the $Z$-factor renormalizing the coupling constant $g$ via
  \begin{eqnarray} 
   Z_g=Z_A^{-\frac 1 2}=\frac{1}{\left(Z_{A,L}+Z_{A,H}\right)^{1/2}}~. \label{eq:ZAZg}
  \end{eqnarray}
Concerning the light flavors, we require the renormalization to correspond to the $\MS$-scheme with $N_F$ light flavors
\begin{eqnarray}
Z_{A,l}(N_F)&=&{Z_g^{\MS}}^{1/2}~ \label{eq:ZAl}.
\end{eqnarray}
The heavy-flavor contributions are fixed by condition \eqref{eq:DecouplingCond}, which 
implies
\begin{eqnarray}
   \Pi_{H,\mbox{\tiny{BF}}}(0,\mu^2, a_s, m_1^2,m_2^2)+Z_{A,H}\equiv 0~. 
	\label{eq:ZAcond}
\end{eqnarray}
The $Z$-factor in the MOM-scheme  is read off by combining Eqs.~\eqref{eq:ZAZg},\eqref{eq:DecouplingCond},\eqref{eq:GluSelfBack} and \eqref{eq:ZAcond}
\begin{eqnarray}
   Z^{\MOM}_g(\ep,N_F+2,\mu,m_1^2,m_2^2)
         \equiv \frac{1}{(Z_{A,l}+Z_{A,H})^{1/2}}~. \label{eq:Zgnfp1}
\end{eqnarray}
Up to ${O}({a^{\MOM}_s}^3)$, one obtains the renormalization constant
\begin{eqnarray}
   {Z_g^{\MOM}}^2(\ep,N_F+2,\mu,m_1^2,m_2^2)&=&
                  1+a^{\MOM}_s(\mu^2) \Bigl[
                              \frac{2}{\ep} (\beta_0(N_F)+\beta_{0,Q}f(\ep))
                        \Bigr]
\NN\\ &&
                  +{a^{\MOM}_s}^2(\mu^2) \Bigl[
                                \frac{\beta_1(N_F)}{\ep}
                         +\frac{4}{\ep^2} (\beta_0(N_F)+\beta_{0,Q}f(\ep))^2
\NN\\ &&
                          +\frac{1}{\ep}\left(\Bigl(\frac{m_1^2}{\mu^2}\Bigr)^{\ep}
                            +\Bigl(\frac{m_2^2}{\mu^2}\Bigr)^{\ep}\right)
                           \Bigl(\beta_{1,Q}+\ep\beta_{1,Q}^{(1)}
                                            +\ep^2\beta_{1,Q}^{(2)}
                           \Bigr)
                          \Bigr]
\NN\\ &&
+ {O}({a^{\MOM}_s}^3)~. \label{eq:Zgheavy2}
\end{eqnarray}.
  
  The coefficients of the $\MOM$-scheme $Z$-factor, $\delta
  a_{s,1}^{\MOM}$ and $\delta a_{s,2}^{\MOM}$, are defined analogously to those of the
  $\MS$-coefficients in Eq.~\eqref{eq:asrenMSb}
  \begin{eqnarray}
   \delta a_{s,1}^{\MOM}&=&\frac{2\beta_0(N_F)}{\ep}
                           +\frac{2\beta_{0,Q}}{\ep}f(\ep)
                            ~,\label{eq:dela1} \\
   \delta a_{s,2}^{\MOM}&=&\frac{\beta_1(N_F)}{\ep}+
                            \left(\frac{2\beta_0(N_F)}{\ep}
                              +\frac{2\beta_{0,Q}}{\ep}f(\ep)\right)^2
\NN\\&&
                          +\frac{1}{\ep}\left(\Bigl(\frac{m_1^2}{\mu^2}\Bigr)^{\ep}+
                            \Bigl(\frac{m_2^2}{\mu^2}\Bigr)^{\ep}\right)
                           \Bigl(\beta_{1,Q}+\ep\beta_{1,Q}^{(1)}
                                            +\ep^2\beta_{1,Q}^{(2)}
                           \Bigr) + {O}(\ep^2)~.\label{eq:dela2}
  \end{eqnarray}
Finally, we express our results in the $\MS$-scheme. For this transition 
the decoupling of the heavy quark flavors is assumed. 
The transformation to the ${\MS}$ scheme is then implied by
  \begin{eqnarray}
      {Z_g^{\MS}}^2(\ep,N_F+2) a^{\MS}_s(\mu^2) = 
      {Z_g^{\MOM}}^2(\ep,N_F+2,\mu,m_1^2,m_2^2) a^{\MOM}_s(\mu^2) \label{eq:condas1}~.
  \end{eqnarray}
Solving Eq.~\eqref{eq:condas1} perturbatively one obtains
  \begin{eqnarray}
   a_s^{\MOM}&=& a_s^{\MS}
                -\beta_{0,Q} \left(\ln \Bigl(\frac{m_1^2}{\mu^2}\Bigr)+\ln \Bigl(\frac{m_2^2}{\mu^2}\Bigr)\right) {a_s^{\MS}}^2
                +\Biggl[ \beta^2_{0,Q} \left(\ln
                  \Bigl(\frac{m_1^2}{\mu^2}\Bigr)+\ln
                  \Bigl(\frac{m_2^2}{\mu^2}\Bigr)\right)^2 
    \NN\\&&
                        -\beta_{1,Q} \left(\ln \Bigl(\frac{m_1^2}{\mu^2}\Bigr) +\ln \Bigl(\frac{m_2^2}{\mu^2}\Bigr)\right)
                        -2 \beta_{1,Q}^{(1)}
                 \Biggr] {a_s^{\MS}}^3+O\left({a_s^{\MS}}^4\right)~, 
	\label{eq:asmoma}
  \end{eqnarray}
  or, 
  \begin{eqnarray}
   a_s^{\MS}&=&
               a_s^{\MOM}
              +{a_s^{\MOM}}^2\Biggl(
                          \delta a^{\MOM}_{s, 1}
                         -\delta a^{\MS}_{s, 1}(N_F+2)
                             \Biggr)
              +{a_s^{\MOM}}^{3}\Biggl(
                          \delta a^{\MOM}_{s, 2}
                         -\delta a^{\MS}_{s, 2}(N_F+2)
     \NN\\ &&
                        -2\delta a^{\MS}_{s, 1}(N_F+2)\Bigl[
                             \delta a^{\MOM}_{s, 1}
                            -\delta a^{\MS}_{s, 1}(N_F+2)
                                                      \Bigr]
                             \Biggr)+{O}({a_s^{\MOM}}^4)~. 
	\label{eq:asmsa}
  \end{eqnarray}
Note that, unlike in Eq.~\eqref{eq:asrenMSb}, in Eqs.~\eqref{eq:asmoma} and \eqref{eq:asmsa} $a_s^{\MS} 
\equiv a_s^{\MS}\left(N_F+2\right)$. Applying the coupling renormalization, cf. 
Eq.~\eqref{eq:Zgheavy2}, to Eq.~\eqref{eq:maren}, the OME after mass and coupling renormalization is obtained by
\begin{eqnarray}
   {\hat{A}}_{ij}  &=&  
\delta_{ij}
                 +{a}_s^{\MOM}~ 
                   \Ahathat_{ij}^{(1)} 
                         + {{a}_s^{\MOM}}^2 \Biggl[
                                        \Ahathat^{(2)}_{ij}                                        
                                        +\delta a^{\MOM}_{s, 1}   \Ahathat_{ij}^{(1)}                                           
\NN\\&&
                                      + {\delta m_1} 
                                        \left(\Bigl(\frac{m_1^2}{\mu^2}\Bigr)^{\ep/2}
                                        m_1 \frac{d}{dm_1}
                                      +\Bigl(\frac{m_2^2}{\mu^2}\Bigr)^{\ep/2}
                                        m_2 \frac{d}{dm_2}
                                      \right)
                                                   \Ahathat_{ij}^{(1)}                                                                   
                                \Biggr]
\NN\\ &&
                         + {{a}_s^{\MOM}}^3 \Biggl[ 
                                         \Ahathat^{(3)}_{ij}                                           
                                    +\delta a^{\MOM}_{s, 2}
                                    \Ahathat_{ij}^{(1)}
                                    + 2 \delta a^{\MOM}_{s,1} 
                                    \Biggl[
                                        \Ahathat^{(2)}_{ij}                                        
\NN\\&&
                                      + {\delta m_1} 
                                        \left(\Bigl(\frac{m_1^2}{\mu^2}\Bigr)^{\ep/2}
                                        m_1 \frac{d}{dm_1}
                                      +\Bigl(\frac{m_2^2}{\mu^2}\Bigr)^{\ep/2}
                                        m_2 \frac{d}{dm_2}
                                      \right)
                                                   \Ahathat_{ij}^{(1)}                                                                   
\Biggr]
\NN\\ &&
                                        +{\delta m_1} 
                                         \left(
                                           \Bigl(\frac{m_1^2}{\mu^2}\Bigr)^{\ep/2}
                                         m_1 \frac{d}{dm_1} 
                                         +
                                           \Bigl(\frac{m_2^2}{\mu^2}\Bigr)^{\ep/2}
                                         m_2 \frac{d}{dm_2} 
                                         \right)
                                                    \Ahathat_{ij}^{(2)}                                           
\NN\\ &&
                                        + \left(\delta m_{2,1}(m_1,m_2) \Bigl(\frac{m_1^2}{\mu^2}\Bigr)^{\ep}
                                          m_1 \frac{d}{dm_1}
                                               +\delta m_{2,2}(m_1,m_2) \Bigl(\frac{m_2^2}{\mu^2}\Bigr)^{\ep}
                                          m_2 \frac{d}{dm_2} \right) 
                                                    \Ahathat_{ij}^{(1)}                                            
\NN\\&&
                                        +
                                        \frac{(\delta m_1)^2}{2}
                                        \left(
                                          \Bigl(\frac{m_1^2}{\mu^2}\Bigr)^{\ep}
                                                   m_1^2
                                                   \frac{d^2}{{dm_1}^2}
                                                   +
                                            \Bigl(\frac{m_2^2}{\mu^2}\Bigr)^{\ep}
                                                   m_2^2  \frac{d^2}{{dm_2}^2}
                                               \right)
                                                     \Ahathat_{ij}^{(1)}                                            
\NN\\&&
                                         + (\delta m_1)^2 
                                        \Bigl(\frac{m_1^2}{\mu^2}\Bigr)^{\ep/2}
                                          \Bigl(\frac{m_2^2}{\mu^2}\Bigr)^{\ep/2}
                                        m_1 \frac{d}{dm_1}
                                        m_2 \frac{d}{dm_2}
                                                   \Ahathat_{ij}^{(1)}                                           
    \Biggr]~,
    \label{eq:macoren}
  \end{eqnarray}
  where the dependence on the masses, $\ep$ and $N$ in the arguments of the OMEs has been suppressed for brevity.

The renormalization steps for the local operators and removing the collinear singularities proceed
very closely to Ref.~\cite{Bierenbaum:2009mv}.

Finally, one has to account for the one-particle reducible contributions. 
In the present case, these 
are gluon self-energy contributions to the external legs of lower order one-particle irreducible 
diagrams. From 3-loop order onward, the reducible contributions to $A_{Qg}$ may contain 
three
different heavy flavors, while this is not the case for the irreducible contributions. Note that 
the inclusion of the top quark in a loop of the irreducible terms for $A_{ij}^{(3)}$ would demand 
to consider the energy range $Q^2 \gg m_t^2$. At a scale $\mu^2 \simeq m_t^2$, both charm and bottom 
can be dealt with as effectively massless. The emergence of massive top loops in the reducible 
contributions is accounted for by renormalization. In the following, we will 
strictly consider the case of two heavy flavors only.

The scalar self-energies are obtained by projecting out the Lorentz-structure
\begin{eqnarray}
       \hat{\Pi}_{\mu\nu}^{ab}(p^2,\hat{m}_1^2,\hat{m}_2^2,\mu^2,\hat{a}_s) &=& i\delta^{ab}
                            \left[-g_{\mu\nu}p^2 +p_\mu p_\nu\right] 
                            \hat{\Pi}(p^2,\hat{m}_1^2,\hat{m}_2^2,\mu^2,\hat{a}_s)~,  \\
                            \hat{\Pi}(p^2,\hat{m}_1^2,\hat{m}_2^2,\mu^2,\hat{a}_s)&=&
                            \sum_{k=1}^{\infty}\hat{a}_s^k
                            \hat{\Pi}^{(k)}(p^2,\hat{m}_1^2,\hat{m}_2^2,\mu^2)
                            ~.
\end{eqnarray}
In the same way as the OMEs themselves, the irreducible two-mass self-energies can be 
divided into contributions which depend on one mass 
only and an additional part stemming from diagrams containing both heavy quark flavors
\begin{eqnarray}
\hat{\Pi}^{(k)}\left(p^2,\hat{m}_1^2,\hat{m}_2^2,\mu^2\right)&=&
\hat{\Pi}^{(k)}\Bigl(p^2,\frac{\hat{m}_1^2}{\mu^2}\Bigr)+
\hat{\Pi}^{(k)}\Bigl(p^2,\frac{\hat{m}_2^2}{\mu^2}\Bigr)
+\hat{\tilde{\Pi}}^{(k)}\left(p^2,\hat{m}_1^2,\hat{m}_2^2,\mu^2\right)~.
\label{eq:gSelf2m}
\end{eqnarray}
Up to two-loop order no diagrams with two heavy flavors contribute

\begin{eqnarray}
\hat{\tilde{\Pi}}^{(k)}(p^2,\hat{m}_1^2,\hat{m}_2^2,\mu^2)&=&0~~~~\text{for}~k\in\{1,2\}~.
\end{eqnarray}
The single-mass contributions for the gluon are known from
\cite{Chetyrkin:1999qi,Chetyrkin:1999ys,Chetyrkin:2008jk,Bierenbaum:2009mv}
  \begin{eqnarray}
  \label{eqPI1}
   \hat{\Pi}^{(1)}\Bigl(0,\frac{\hat{m}^2}{\mu^2}\Bigr)&=&
            T_F\Bigl(\frac{\hat{m}^2}{\mu^2}\Bigr)^{\ep/2}
                        \left[
             \frac{8}{3\ep}
              \exp \Bigl(\sum_{i=2}^{\infty}\frac{\zeta_i}{i}
                       \Bigl(\frac{\ep}{2}\Bigr)^{i}\Bigr)
             \right]~,
               ~\label{GluSelf1} 
               \\
   \hat{\Pi}^{(2)}\Bigl(0,\frac{\hat{m}^2}{\mu^2}\Bigr)&=&
      T_F\Bigl(\frac{\hat{m}^2}{\mu^2}\Bigr)^{\ep}\Biggl\{
      -\frac{4}{\ep^2} C_A + \frac{1}{\ep} \left(5 C_A-12 C_F\right) 
      + C_A \Bigl(\frac{13}{12} -\zeta_2\Bigr)
      - \frac{13}{3} C_F   
\NN\\ &&
      + \ep \left[C_A \Bigl(\frac{169}{144} + \frac{5}{4} \zeta_2 - 
      \frac{\zeta_3}{3} \Bigr) 
     - C_F \Bigl(\frac{35}{12}+3 \zeta_2\Bigr) 
     \right]\Biggr\}
       + 
            {O}(\ep^2)~,  \label{eq:GluSelf2}
\\
             \hat{\Pi}^{(3)}\Bigl(0,\frac{\hat{m}^2}{\mu^2}\Bigr)&=&
       T_F\Bigl(\frac{\hat{m}^2}{\mu^2}\Bigr)^{3\ep/2}\Biggl\{
                        \frac{1}{\ep^3}\left[
                                 -\frac{32}{9}T_F C_A \left(2 N_F+1\right)
                                 + \frac{164}{9} C_A^2      
                                       \right]
\NN\\ &&
                       +\frac{1}{\ep^2}\left[
                               \frac{80}{27} (C_A-6 C_F) N_FT_F
                              +\frac{8}{27} (35 C_A-48 C_F) T_F
                              -\frac{781}{27} C_A^2 \right.                                                 
\NN\\ &&
                       \left.       +\frac{712}{9}C_AC_F
                                       \right]
                       +\frac{1}{\ep}\biggl[ 
                                \frac{4}{27}\big(
                                                       C_A(-101-18\zeta_2)
                                                      -62C_F
                                                \big)N_FT_F 
\NN\\ &&
                              -\frac{2}{27}   \big(
                                                       C_A(37+18 \zeta_2)
                                                       +80 C_F
                                                \big) T_F
                              +C_A^2            \Bigl(
                                                  -12\zeta_3
                                                  +\frac{41}{6}\zeta_2
                                                  +\frac{3181}{108}
                                                \Bigr)
\NN\\ &&
                              +C_A C_F           \Bigl(
                                                   16\zeta_3
                                                  -\frac{1570}{27}
                                                \Bigr)
                              +\frac{272}{3}C_F^2
                                       \biggr]
\NN\\ &&
                       +N_FT_F    \biggl[
                                       C_A\Bigl(
                                             \frac{56}{9}\zeta_3
                                            +\frac{10}{9}\zeta_2
                                            -\frac{3203}{243}
                                          \Bigr)
                                      -C_F\Bigl(
                                            \frac{20}{3}\zeta_2
                                            +\frac{1942}{81}
                                          \Bigr)
                                       \biggr]
\NN\\ &&
                       +T_F      \biggl[
                                       C_A\Bigl(
                                            -\frac{295}{18}\zeta_3
                                            +\frac{35}{9}\zeta_2
                                            +\frac{6361}{486}
                                          \Bigr)
                                      -C_F\Bigl(
                                            7\zeta_3
                                            +\frac{16}{3}\zeta_2
                                            +\frac{218}{81}
                                          \Bigr)
                                   \biggr]
\NN\\ &&
                       +C_A^2      \biggl(
                                       4\text{B}_4
                                      -27\zeta_4
                                      +\frac{1969}{72}\zeta_3
                                      -\frac{781}{72}\zeta_2
                                      +\frac{42799}{3888}
                                   \biggr)
\NN\\ &&
                       +C_A C_F      \biggl(
                                      -8\text{B}_4
                                      +36\zeta_4
                                      -\frac{1957}{12}\zeta_3
                                                                            +\frac{89}{3}\zeta_2
                                      +\frac{10633}{81}
                                   \biggr)
\NN\\ &&
                       +C_F^2      \biggl(
                                      \frac{95}{3}\zeta_3
                                      +\frac{274}{9}
                                   \biggr)
                                                   \Biggr\} + {O}(\ep)~.
                                          \label{eq:GluSelf3}
\end{eqnarray}
Similarly to other massive processes 
   \cite{Broadhurst:1991fi,Avdeev:1994db,Laporta:1996mq,
    Broadhurst:1998rz,Boughezal:2004ef,Bierenbaum:2009mv},
   the constant
   \begin{eqnarray}
          \text{B}_4&=&-4\zeta_2\ln^2(2) +\frac{2}{3}\ln^4(2) 
           -\frac{13}{2}\zeta_4
                  +16 \text{Li}_4\Bigl(\frac{1}{2}\Bigr)
                 ~\approx~  -1.762800093...~  \label{eq:B4}
   \end{eqnarray}
emerges in Eq.~\eqref{eq:GluSelf3}.
At ${O}(a_s^3)$, irreducible diagrams with two different masses contribute
for the first time. 
In \cite{Ablinger:2017err} the gluonic case was calculated to ${O}(\eta^3)$ using the codes 
\texttt{Q2E}/\texttt{Exp}\cite{Harlander,Seidensticker:1999bb}. 
However, the full $\eta$ dependence is needed in the following.
All the diagrams can be expressed through a one-dimensional Mellin-Barnes
integral, and the residue sums are easily evaluated using the \texttt{Mathematica}
package \texttt{EvaluateMultiSums} \cite{Schneider:2013zna} which is built 
on \texttt{Sigma} \cite{SIG1,SIG2} and \texttt{HarmonicSums} \cite{Vermaseren:1998uu,
Blumlein:1998if,
Remiddi:1999ew,
Blumlein:2003gb,
Ablinger:2011te,
Ablinger:2013cf,
Ablinger:2014bra,
Blumlein:2009ta,
Blumlein:2009cf,
Blumlein:2009fz,
Ablinger:2010kw,
Ablinger:2013hcp,
Ablinger:2014rba,
Ablinger:2015gdg,
ALL2016,
ALL2018,
Ablinger:2018cja,
Ablinger:2019mkx,
Ablinger:2021fnc}. 
The result is given by \cite{Blumlein:2017mtk} 
\begin{eqnarray}
\hat{\tilde{\Pi}}^{(3)}(0,m_1^2,m_2^2,\mu^2) &=& 
-C_F T_F^2 \Biggl\{
\frac{256}{9 \varepsilon^2}
+\frac{64}{3 \varepsilon} \left[\lnMa{}+\lnMb{}+\frac{5}{9}\right]
-5 \eta -\frac{5}{\eta}
\nonumber \\ &&
+\left(-\frac{5 \eta }{8}-\frac{5}{8 \eta }+\frac{51}{4}\right) \ln ^2(\eta )+\left(\frac{5}{2 \eta }-\frac{5 \eta }{2}\right) \ln (\eta)
+\frac{32 \zeta_2}{3}
\nonumber \\ &&
+32 \lnMa{} \lnMb{}
+\frac{80}{9} \lnMa{}+\frac{80}{9} \lnMb{}
+\frac{1246}{81}
\nonumber \\ &&
+\left(\frac{5 \eta^{3/2}}{2}
+\frac{5}{2 \eta^{3/2}}
+\frac{3 \sqrt{\eta}}{2}
+\frac{3}{2 \sqrt{\eta}}\right) \Biggl[\frac{1}{8} \ln \left(\frac{1+\sqrt{\eta}}{1-\sqrt{\eta}}\right) \ln^2(\eta)
\nonumber \\ &&
-\Li_3\left(-\sqrt{\eta}\right)
+\Li_3\left(\sqrt{\eta}\right)
-\frac{1}{2} \ln (\eta) \left(\Li_2\left(\sqrt{\eta}\right)
-\Li_2\left(-\sqrt{\eta}\right)\right)\Biggr]
\Biggr\}
\nonumber \\ &&
-C_A T_F^2 \Biggl\{
\frac{64}{9 \varepsilon^3} 
+\frac{16}{3 \varepsilon^2} \Biggl[ \left(\lnMa{}+\lnMb{}\right)
-\frac{35}{9}\Biggr]
+\frac{4}{\varepsilon} \Biggl[
\lnMa{2}+\lnMb{2}
-\frac{35}{9} \lnMa{}
\nonumber \\ &&
-\frac{35}{9} \lnMb{}
+\frac{2}{3} \zeta_2
+\frac{37}{27}
\Biggr]
+2 \left(\lnMa{3}+\lnMb{3}\right)
-\frac{70}{3} \lnMa{} \lnMb{}
-\frac{4}{9} \ln^3(\eta)
\nonumber \\ &&
+\left(2  \zeta_2+\frac{37}{9}\right) \left(\lnMa{}+\lnMb{}\right)
+\left[\frac{8}{3} \ln(1-\eta)- \frac{2}{3} \left(\eta+\frac{1}{\eta}\right) -\frac{179}{18}\right] 
\nonumber \\ && \times \ln^2(\eta)
-\frac{16}{3} \left(\eta+\frac{1}{\eta}\right)
-\frac{70}{9} \zeta_2
-\frac{56}{9} \zeta_3 
-\frac{3769}{243}
\nonumber \\ &&
+ \frac{8}{3} \left(\frac{1}{\eta}-\eta\right) \ln(\eta)
+\frac{16}{3} \big( \Li_2(\eta) \ln(\eta) - \Li_3(\eta) \big)
\nonumber \\ &&
+\left[ 8 \frac{1+\eta^3}{3 \eta^{3/2}}+ 10  \frac{1+\eta}{\sqrt{\eta}}\right]
\Biggl[\frac{1}{8} \ln \left(\frac{1+\sqrt{\eta}}{1-\sqrt{\eta }}\right) \ln^2(\eta)
- \Li_3\left(-\sqrt{\eta}\right)
\nonumber \\ &&
+ \Li_3\left(\sqrt{\eta}\right)
-\frac{1}{2} \ln(\eta) \left( \Li_2\left(\sqrt{\eta}\right) - \Li_2\left(-\sqrt{\eta}\right)\right)\Biggr] + 
{O}(\ep)
\Biggr\}.
\label{eq:pi2m}
\end{eqnarray}
Here the classical polylogarithm \cite{Devoto:1983tc,LEWIN1,LEWIN2} is defined by
\begin{eqnarray}
\Li_n(x) = \sum_{k=1}^\infty \frac{x^k}{k^n},~~n \geq 0,~~x \in [-1, 1].
\end{eqnarray}
After adjusting notations complete agreement is found with the earlier calculation in Ref.~\cite{Grozin:2011nk}. For the reducible parts in the case of 
external massless fermion lines, see e.g.~\cite{Schonwald:2019gmn} and references 
therein.

The two-mass OMEs at
one-loop order and the irreducible OMEs at ${O}(a_s^2)$ are defined by
    \begin{eqnarray}
        \Ahathat^{(1)}_{ij}\Bigl(\frac{\hat{m}_1^2}{\mu^2},\frac{\hat{m}_2^2}{\mu^2}\Bigl)&=&
        \Ahathat^{(1)}_{ij}\left(\frac{\hat{m}_1^2}{\mu^2}\right)
        +\Ahathat^{(1)}_{ij}\left(\frac{\hat{m}_2^2}{\mu^2}\right)~,
\label{eq:Aij12m} \\
        \Ahathat^{(\prime),(2),\text{irr}}_{ij}\Bigl(\frac{\hat{m}_1^2}{\mu^2},\frac{\hat{m}_2^2}{\mu^2}\Bigl)&=&
        \Ahathat^{(\prime),(2),\text{irr}}_{ij}\left(\frac{\hat{m}_1^2}{\mu^2}\right)
        +\Ahathat^{(\prime),(2),\text{irr}}_{ij}\left(\frac{\hat{m}_2^2}{\mu^2}\right)~,
        \label{eq:Aij2irr2m}
    \end{eqnarray}
    where the $A_{ij}$'s with one argument denote the usual single-mass OMEs.
The irreducible contributions from OMEs with external gluons need a further discussion.
These have to be included using a consistent projection, either physical
or unphysical,
since the ghost contributions restore gauge invariance globally.
Therefore the superscript ${}^\prime$ is included for irreducible OMEs with external 
gluons.
    Using previous definitions the reducible massive operator matrix elements at 
${O}(a_s^2)$ are composed by
    \begin{eqnarray}
     \Ahathat_{qq}^{(2),\text{NS}}\Bigl(\frac{\hat{m}_1^2}{\mu^2},\frac{\hat{m}_2^2}{\mu^2}\Bigl)&=&
     \Ahathat_{qq}^{(2),\text{NS},\text{irr}}\Bigl(\frac{\hat{m}_1^2}{\mu^2},\frac{\hat{m}_2^2}{\mu^2}\Bigl)
     -\hat{\Sigma}^{(2)}\left(0,\hat{m}_1^2,\hat{m}_2^2,\mu^2\right)~, 
{\label{eq:Aqq2NSred}} \\
\Ahathat_{Qg}^{(2)}\Bigl(\frac{\hat{m}_1^2}{\mu^2},\frac{\hat{m}_2^2}{\mu^2}\Bigl)&=&
\Ahathat_{Qg}^{\prime,(2),\text{irr}}\Bigl(\frac{\hat{m}_1^2}{\mu^2},\frac{\hat{m}_2^2}{\mu^2}\Bigl)
-\Ahathat_{Qg}^{(1)}\Bigl(\frac{\hat{m}_1^2}{\mu^2},\frac{\hat{m}_2^2}{\mu^2}\Bigl)
\hat{\Pi}^{(1)}\left(0,\hat{m}_1^2,\hat{m}_2^2,\mu^2\right)~, 
{\label{eq:AQg2red}} \\
 \Ahathat_{gg}^{(2)}\Bigl(\frac{\hat{m}_1^2}{\mu^2},\frac{\hat{m}_2^2}{\mu^2}\Bigl)&=&
 \Ahathat_{gg}^{\prime,(2),\text{irr}}\Bigl(\frac{\hat{m}_1^2}{\mu^2},\frac{\hat{m}_2^2}{\mu^2}\Bigl)
 -\hat{\Pi}^{(2)}\left(0,\hat{m}_1^2,\hat{m}_2^2,\mu^2\right)
\NN\\&&
 -\Ahathat_{gg}^{(1)}\Bigl(\frac{\hat{m}_1^2}{\mu^2},\frac{\hat{m}_2^2}{\mu^2}\Bigl)
  \hat{\Pi}^{(1)}\left(0,\hat{m}_1^2,\hat{m}_2^2,\mu^2\right)~, 
{\label{eq:Agg2red}}
     \end{eqnarray}
and at ${O}(a_s^3)$ by
     \begin{eqnarray}
     \Ahathat_{qq}^{(3),\text{NS}}\Bigl(\frac{\hat{m}_1^2}{\mu^2},\frac{\hat{m}_2^2}{\mu^2}\Bigl)&=&
     \Ahathat_{qq}^{(3),\text{NS},\text{irr}}\Bigl(\frac{\hat{m}_1^2}{\mu^2},\frac{\hat{m}_2^2}{\mu^2}\Bigl)
     -\hat{\Sigma}^{(3)}\left(0,\hat{m}_1^2,\hat{m}_2^2,\mu^2\right)
\\
 \Ahathat_{Qg}^{(3)}\Bigl(\frac{\hat{m}_1^2}{\mu^2},\frac{\hat{m}_2^2}{\mu^2}\Bigl)&=&
 \Ahathat_{Qg}^{\prime,(3),\text{irr}}\Bigl(\frac{\hat{m}_1^2}{\mu^2},\frac{\hat{m}_2^2}{\mu^2}\Bigl)
 -
 \Ahathat_{Qg}^{(2)}\Bigl(\frac{\hat{m}_1^2}{\mu^2},\frac{\hat{m}_2^2}{\mu^2}\Bigl)
 \hat{\Pi}^{(1)}\left(0,\hat{m}_1^2,\hat{m}_2^2,\mu^2\right)
 \NN\\&&
  -
 \Ahathat_{Qg}^{(1)}\Bigl(\frac{\hat{m}_1^2}{\mu^2},\frac{\hat{m}_2^2}{\mu^2}\Bigl)
 \hat{\Pi}^{(2)}\left(0,\hat{m}_1^2,\hat{m}_2^2,\mu^2\right)
 \\
 \Ahathat_{gg}^{(3)}\Bigl(\frac{\hat{m}_1^2}{\mu^2},\frac{\hat{m}_2^2}{\mu^2}\Bigl)&=&
 \Ahathat_{gg}^{\prime,(3),\text{irr}}\Bigl(\frac{\hat{m}_1^2}{\mu^2},\frac{\hat{m}_2^2}{\mu^2}\Bigl)
 -\Ahathat_{gg}^{(2)}\Bigl(\frac{\hat{m}_1^2}{\mu^2},\frac{\hat{m}_2^2}{\mu^2}\Bigl) 
  \hat{\Pi}^{(1)}\left(0,\hat{m}_1^2,\hat{m}_2^2,\mu^2\right)
 \NN\\&&
 -\Ahathat_{gg}^{(1)}\Bigl(\frac{\hat{m}_1^2}{\mu^2},\frac{\hat{m}_2^2}{\mu^2}\Bigl) 
  \hat{\Pi}^{(2)}\left(0,\hat{m}_1^2,\hat{m}_2^2,\mu^2\right)
 -\hat{\Pi}^{(3)}\left(0,\hat{m}_1^2,\hat{m}_2^2,\mu^2\right)
\,.
\end{eqnarray}
One can subtract the single-mass contributions to these equations using Eq.~\eqref{eq:AhathatDecomp}, keeping
only the genuine two-mass contributions. At three loops one obtains

\begin{eqnarray}
\Athathat_{qq}^{(3),\text{NS}}\Bigl(\frac{\hat{m}_1^2}{\mu^2},\frac{\hat{m}_2^2}{\mu^2}\Bigl)&=&
\Athathat_{qq}^{(3),\text{NS},\text{irr}}\Bigl(\frac{\hat{m}_1^2}{\mu^2},\frac{\hat{m}_2^2}{\mu^2}\Bigl)
-\hat{\tilde{\Sigma}}^{(3)}\left(0,\hat{m}_1^2,\hat{m}_2^2,\mu^2\right)
\\
\Athathat_{Qg}^{(3)}\Bigl(\frac{\hat{m}_1^2}{\mu^2},\frac{\hat{m}_2^2}{\mu^2}\Bigl)&=&
\Athathat_{Qg}^{\prime,(3),\text{irr}}\Bigl(\frac{\hat{m}_1^2}{\mu^2},\frac{\hat{m}_2^2}{\mu^2}\Bigl)
 \NN\\&&
+\Ahathat_{Qg}^{(1)}\left(\frac{\hat{m}_1^2}{\mu^2}\right)
 \left[2 \hat{\Pi}^{(1)}\Bigl(0,\frac{\hat{m}_1^2}{\mu^2}\Bigr)+\hat{\Pi}^{(1)}\Bigl(0,\frac{\hat{m}_2^2}{\mu^2}\Bigr)\right]
 \hat{\Pi}^{(1)}\Bigl(0,\frac{\hat{m}_2^2}{\mu^2}\Bigr)
 \NN\\&&
+\Ahathat_{Qg}^{(1)}\left(\frac{\hat{m}_2^2}{\mu^2}\right)
 \left[2 \hat{\Pi}^{(1)}\Bigl(0,\frac{\hat{m}_2^2}{\mu^2}\Bigr)+\hat{\Pi}^{(1)}\Bigl(0,\frac{\hat{m}_1^2}{\mu^2}\Bigr)\right]
 \hat{\Pi}^{(1)}\Bigl(0,\frac{\hat{m}_1^2}{\mu^2}\Bigr)
 \NN\\&&
-\Ahathat_{Qg}^{(2)}\left(\frac{\hat{m}_1^2}{\mu^2}\right)
 \hat{\Pi}^{(1)}\Bigl(0,\frac{\hat{m}_2^2}{\mu^2}\Bigr)
-\Ahathat_{Qg}^{(2)}\left(\frac{\hat{m}_2^2}{\mu^2}\right)
 \hat{\Pi}^{(1)}\Bigl(0,\frac{\hat{m}_1^2}{\mu^2}\Bigr)
 \NN\\&&
-\Ahathat_{Qg}^{(1)}\left(\frac{\hat{m}_1^2}{\mu^2}\right)
 \hat{\Pi}^{(2)}\Bigl(0,\frac{\hat{m}_2^2}{\mu^2}\Bigr)
-\Ahathat_{Qg}^{(1)}\left(\frac{\hat{m}_2^2}{\mu^2}\right)
 \hat{\Pi}^{(2)}\Bigl(0,\frac{\hat{m}_1^2}{\mu^2}\Bigr)
 \\
\Athathat_{gg}^{(3)}\Bigl(\frac{\hat{m}_1^2}{\mu^2},\frac{\hat{m}_2^2}{\mu^2}\Bigl)&=&
\Athathat_{gg}^{\prime,(3),\text{irr}}\Bigl(\frac{\hat{m}_1^2}{\mu^2},\frac{\hat{m}_2^2}{\mu^2}\Bigl)
-\hat{\tilde{\Pi}}^{(3)}\left(0,\hat{m}_1^2,\hat{m}_2^2,\mu^2\right)
 \NN\\&&
 -\Ahathat_{gg}^{\prime,(2),\text{irr}}\left(\frac{\hat{m}_1^2}{\mu^2}\right) 
  \hat{\Pi}^{(1)}\Bigl(0,\frac{\hat{m}_2^2}{\mu^2}\Bigr)
 -\Ahathat_{gg}^{\prime,(2),\text{irr}}\left(\frac{\hat{m}_2^2}{\mu^2}\right) 
  \hat{\Pi}^{(1)}\Bigl(0,\frac{\hat{m}_1^2}{\mu^2}\Bigr)
 \NN\\&&
 -2 \Ahathat_{gg}^{(1)}\left(\frac{\hat{m}_1^2}{\mu^2}\right) 
  \hat{\Pi}^{(2)}\Bigl(0,\frac{\hat{m}_2^2}{\mu^2}\Bigr)
 -2 \Ahathat_{gg}^{(1)}\left(\frac{\hat{m}_2^2}{\mu^2}\right) 
  \hat{\Pi}^{(2)}\Bigl(0,\frac{\hat{m}_1^2}{\mu^2}\Bigr)
 \NN\\&&
 +\Ahathat_{gg}^{(1)}\left(\frac{\hat{m}_1^2}{\mu^2}\right) 
  \left[2 \hat{\Pi}^{(1)}\Bigl(0,\frac{\hat{m}_1^2}{\mu^2}\Bigr)+ \hat{\Pi}^{(1)}\Bigl(0,\frac{\hat{m}_2^2}{\mu^2}\Bigr)\right]
  \hat{\Pi}^{(1)}\Bigl(0,\frac{\hat{m}_2^2}{\mu^2}\Bigr)
 \NN\\&&
 +\Ahathat_{gg}^{(1)}\left(\frac{\hat{m}_2^2}{\mu^2}\right) 
  \left[2 \hat{\Pi}^{(1)}\Bigl(0,\frac{\hat{m}_2^2}{\mu^2}\Bigr)+\hat{\Pi}^{(1)}\Bigl(0,\frac{\hat{m}_1^2}{\mu^2}\Bigr)\right]
  \hat{\Pi}^{(1)}\Bigl(0,\frac{\hat{m}_1^2}{\mu^2}\Bigr)
\,.
\end{eqnarray}

\subsection{The structure of \boldmath{$(\Delta)A_{Qg}$}}
\label{SubSec-AQqg}
The single-mass and two-mass contributions to the OME $A_{Qg}$ are given by
\begin{eqnarray}
   A_{Qg}&=&
                        a_s  A_{Qg}^{(1)}
                       +a_s^2A_{Qg}^{(2)}
                       +a_s^3A_{Qg}^{(3)}
                       +{O}(a_s^4), \label{eq:AQgpert}\\
   A_{qg,Q}&=&
                        a_s^3A_{qg,Q}^{(3)}
                       +{O}(a_s^4), \label{eq:AqgQpert}
\end{eqnarray}
depending on whether the operator couples to a light or heavy quark. Of these OMEs, only 
$A_{Qg}$ contains two-flavor 
contributions starting from ${O}(a_s^2)$
\begin{eqnarray}
   \tilde{A}_{Qg}&=&                      
                       a_s^2 \tilde{A}_{Qg}^{(2)}
                       +a_s^3 \tilde{A}_{Qg}^{(3)}
                       +{O}(a_s^4)~. \label{eq:AQgpert2m}
\end{eqnarray}
In Eq.~\eqref{eq:AQgpert2m} the ${O}(a_s^2)$ contribution consists of one-particle reducible diagrams only, 
see Eq.~\eqref{eq:AQg2red}. As a consequence, the flavor dependence factorizes in the 
${O}(a_s^2)$ terms.
  
In the renormalized $\text{MOM}$-scheme, cf.~Ref.~\cite{Bierenbaum:2009mv}, the two-loop 
contribution is obtained by
\begin{eqnarray} 
    A_{Qg}^{(2), \MOM}&=&
                    \hat{A}_{Qg}^{(2), \MOM}
                   +Z^{-1,(2)}_{qg}(N_F+2)
                   -Z^{-1,(2)}_{qg}(N_F)
                   +Z^{-1,(1)}_{qg}(N_F+2)\hat{A}_{gg,Q}^{(1), \MOM}
 \NN\\ &&
                   +Z^{-1,(1)}_{qq}(N_F+2)\hat{A}_{Qg}^{(1), \MOM}
                   +\Bigl[ \hat{A}_{Qg}^{(1), \MOM}
                          +Z_{qg}^{-1,(1)}(N_F+2)
\NN\\ &&
                          -Z_{qg}^{-1,(1)}(N_F)
                    \Bigr]\Gamma^{-1,(1)}_{gg}(N_F)~. \label{eq:RenAQg2MOM}
\end{eqnarray}
The unrenormalized terms are given by 
\begin{eqnarray}
   \Athathat_{Qg}^{(2)}&=& - \frac{2}{\ep^2} \beta_{0,Q} \hat{\gamma}_{qg}^{(0)}
			    -\frac{1} {\ep} \beta_{0,Q} \hat{\gamma}_{qg}^{(0)}
			     \left( L_1+L_2 \right)
			     +\tilde{a}_{Qg}^{(2)}
			     \NN\\&&
			     +\ep \overline{\tilde{a}}_{Qg}^{(2)}
                         ~.\label{eq:AhhhQg2}
\end{eqnarray}
The coefficients $\tilde{a}_{Qg}^{(2)}$ and $\overline{\tilde{a}}_{Qg}^{(2)}$ 
are read off from Eq.~\eqref{eq:AQg2red} 
\begin{eqnarray}
\label{eq:aqg2a}
\tilde{a}_{Qg}^{(2)} &=&
- \frac{1}{2} \beta_{0,Q} \hat{\gamma}_{qg}^{(0)} 
\Biggl\{
\frac{1}{2} \left(L_1+L_2\right)^2
+\zeta_2
\Biggr\}~,
\\
\label{eq:aqg2b}
\overline{\tilde{a}}_{Qg}^{(2)} &=&
\frac{1}{2} \beta_{0,Q} \hat{\gamma}_{qg}^{(0)}
\Biggl\{
-\frac{1}{12} \left(L_1+L_2\right)^3
-\frac{1}{2}  \zeta_2 \left(L_1+L_2\right)
-\frac{1}{3} \zeta_3 
\Biggr\}~.
\end{eqnarray}
The renormalized expression at 2 loops then reads
\begin{eqnarray}
    \tilde{A}_{Qg}^{(2), \MS}&=& \frac{1}{4} \beta_{0,Q} \hat{\gamma}_{qg}^{(0)} 
    \left(L_1^2+L_2^2\right)
    + \frac{1}{2} \zeta_2 \beta_{0,Q} \hat{\gamma}_{qg}^{(0)} +\tilde{a}_{Qg}^{(2)}
\nonumber \\
	&=& - \frac{1}{2} \beta_{0,Q} \hat{\gamma}_{qg}^{(0)} L_1 \, L_2~.
\end{eqnarray}
The renormalized 3-loop OMEs in the $\MOM$-scheme are obtained from the charge- and mass-renormalized OMEs by
\begin{eqnarray}
    && A_{Qg}^{(3), \MOM}+A_{qg,Q}^{(3), \MOM}
            =
                     \hat{A}_{Qg}^{(3), \MOM}
                    +\hat{A}_{qg,Q}^{(3), \MOM}
                    +Z^{-1,(3)}_{qg}(N_F+2)
                    -Z^{-1,(3)}_{qg}(N_F)
\NN\\ && \phantom{abc} 
                    +Z^{-1,(2)}_{qg}(N_F+2)\hat{A}_{gg,Q}^{(1), \MOM}
                    +Z^{-1,(1)}_{qg}(N_F+2)\hat{A}_{gg,Q}^{(2), \MOM}
                    +Z^{-1,(2)}_{qq}(N_F+2)\hat{A}_{Qg}^{(1), \MOM}
\NN\\ && \phantom{abc} 
                    +Z^{-1,(1)}_{qq}(N_F+2)\hat{A}_{Qg}^{(2), \MOM}
                    +\Bigl[
                            \hat{A}_{Qg}^{(1), \MOM}
                           +Z^{-1,(1)}_{qg}(N_F+2)
\NN\\ && \phantom{abc} 
                           -Z^{-1,(1)}_{qg}(N_F)
                     \Bigr]\Gamma^{-1,(2)}_{gg}(N_F)
                    +\Bigl[ \hat{A}_{Qg}^{(2), \MOM}
                           +Z^{-1,(2)}_{qg}(N_F+2) 
                           -Z^{-1,(2)}_{qg}(N_F)
\NN\\ && \phantom{abc} 
                           +Z^{-1,(1)}_{qq}(N_F+2)A_{Qg}^{(1), \MOM}
                           +Z^{-1,(1)}_{qg}(N_F+2)A_{gg,Q}^{(1), \MOM}
                     \Bigr]\Gamma^{-1,(1)}_{gg}(N_F) 
\NN\\ && \phantom{abc} 
                    +\Bigl[ \hat{A}_{Qq}^{(2), \text{PS}, \MOM}
                           +Z^{-1,(2), \text{PS}}_{qq}(N_F+2)
                           -Z^{-1,(2), \text{PS}}_{qq}(N_F)
                     \Bigr]\Gamma^{-1,(1)}_{qg}(N_F)
\NN\\ && \phantom{abc} 
                    +\Bigl[ \hat{A}_{qq,Q}^{(2), \text{NS}, \MOM}
                           +Z^{-1,(2), \text{NS}}_{qq}(N_F+2)
                           -Z^{-1,(2), \text{NS}}_{qq}(N_F)
                     \Bigr]\Gamma^{-1,(1)}_{qg}(N_F)~.
\end{eqnarray}
It is explicitly given by\footnote{The different symbols used here are defined in Ref.~\cite{Bierenbaum:2009mv} and 
above.}
\begin{eqnarray}
   \Athathat_{Qg}^{(3)}&=&
\frac{1}{\ep^3} \Biggl[
 \frac{14}{3} \beta_{0} \beta_{0,Q} \hat{\gamma}_{qg}^{(0)}
-\frac{4}{3} \hat{\gamma}_{qg}^{(0)} \gamma_{qq}^{(0)} \beta_{0,Q}
+\frac{7}{3} \beta_{0,Q} \hat{\gamma}_{qg}^{(0)} \gamma_{gg}^{(0)}
+12 \beta_{0,Q}^2 \hat{\gamma}_{qg}^{(0)}
+\frac{1}{12} \gamma_{gq}^{(0)} \left(\hat{\gamma}_{qg}^{(0)}\right)^2\Biggr]
\NN\\&&
+\frac{1}{\ep^2} \Biggl[
\Biggl\{
\frac{1}{16} \gamma_{gq}^{(0)} \left(\hat{\gamma}_{qg}^{(0)}\right)^2
+9 \beta_{0,Q}^2 \hat{\gamma}_{qg}^{(0)}
+\frac{7}{2} \beta_{0} \beta_{0,Q} \hat{\gamma}_{qg}^{(0)}
- \hat{\gamma}_{qg}^{(0)} \gamma_{qq}^{(0)} \beta_{0,Q}
+\frac{7}{4} \beta_{0,Q} \hat{\gamma}_{qg}^{(0)} \gamma_{gg}^{(0)}
\Biggr\} 
\NN\\&&
\times \left(L_1+L_2\right)
+\frac{1}{12} \hat{\gamma}_{qg}^{(0)} \hat{\gamma}_{qq}^{\text{PS},(1)}
+\frac{1}{12} \hat{\gamma}_{qg}^{(0)} \hat{\gamma}_{qq}^{\text{NS},(1)}
-\frac{5}{3} \beta_{0,Q} \hat{\gamma}_{qg}^{(1)}
+\frac{1}{6} \hat{\gamma}_{qg}^{(0)} \hat{\gamma}_{gg}^{(1)}
-\frac{1}{3} \hat{\gamma}_{qg}^{(0)} \beta_{1,Q}
\NN\\&&
+5 \hat{\gamma}_{qg}^{(0)} \beta_{0,Q} \delta m_1^{(-1)}\Biggr] 
+\frac{1}{\ep} 
\Biggl[
\Biggl\{
 \frac{1}{16} \hat{\gamma}_{qg}^{(0)} \hat{\gamma}_{qq}^{\text{NS},(1)}
+\frac{15}{4} \hat{\gamma}_{qg}^{(0)} \beta_{0,Q} \delta m_1^{(-1)}
+\frac{1}{16} \hat{\gamma}_{qg}^{(0)} \hat{\gamma}_{qq}^{\text{PS},(1)}
-\frac{5}{4} \beta_{0,Q} \hat{\gamma}_{qg}^{(1)}
\NN\\&&
+\frac{1}{8} \hat{\gamma}_{qg}^{(0)} \hat{\gamma}_{gg}^{(1)}
-\frac{1}{4} \hat{\gamma}_{qg}^{(0)} \beta_{1,Q}
\Biggr\} \left(L_1+L_2\right)
+\Biggl\{
\frac{13}{8} \beta_{0} \beta_{0,Q} \hat{\gamma}_{qg}^{(0)}
+\frac{13}{16} \beta_{0,Q} \hat{\gamma}_{qg}^{(0)} \gamma_{gg}^{(0)}
+\frac{15}{4} \beta_{0,Q}^2 \hat{\gamma}_{qg}^{(0)}
\NN\\&&
+\frac{3}{64} \gamma_{gq}^{(0)} \left(\hat{\gamma}_{qg}^{(0)}\right)^2
-\frac{1}{2} \hat{\gamma}_{qg}^{(0)} \gamma_{qq}^{(0)} \beta_{0,Q}
\Biggr\} \left(L_1^2+L_2^2\right)
+ \Biggl\{
-\frac{1}{2} \hat{\gamma}_{qg}^{(0)} \gamma_{qq}^{(0)} \beta_{0,Q}
+2 \beta_{0} \beta_{0,Q} \hat{\gamma}_{qg}^{(0)}
+6 \beta_{0,Q}^2 \hat{\gamma}_{qg}^{(0)}
\NN\\&&
+ \beta_{0,Q} \hat{\gamma}_{qg}^{(0)} \gamma_{gg}^{(0)}
\Biggr\} L_1 L_2
+\frac{2}{3} \gamma_{qg}^{(2),N_F^2}
-8 \beta_{0,Q} a_{Qg}^{(2)}
-\frac{1}{32} \left(\hat{\gamma}_{qg}^{(0)}\right)^2 \zeta_2 \gamma_{gq}^{(0)}
+ \hat{\gamma}_{qg}^{(0)} a_{gg,Q}^{(2)}
- \hat{\gamma}_{qg}^{(0)} \tilde{\delta}m_2^{(-1)}
\NN\\&&
+\frac{9}{2} \hat{\gamma}_{qg}^{(0)} \zeta_2 \beta_{0,Q}^2
+\frac{1}{8} \beta_{0,Q} \zeta_2 \hat{\gamma}_{qg}^{(0)} \gamma_{gg}^{(0)}
+\frac{1}{4} \hat{\gamma}_{qg}^{(0)} \zeta_2 \beta_{0,Q} \beta_{0}
+4 \delta m_1^{(0)} \beta_{0,Q} \hat{\gamma}_{qg}^{(0)}
\Biggr]
\NN\\&&
+\tilde{a}_{Qg}^{(3)}\left(m_1^2,m_2^2,\mu^2\right)~.
\label{AhhhQg3} 
\end{eqnarray}
For the renormalized operator matrix element in the $\MS$ scheme one finally obtains,
\begin{eqnarray}
   \tilde{A}_{Qg}^{(3), \MS}&=&
   \Biggl\{
-\frac{9}{8} \beta_{0,Q}^2 \hat{\gamma}_{qg}^{(0)}
-\frac{7}{384} \gamma_{gq}^{(0)} \left(\hat{\gamma}_{qg}^{(0)}\right)^2
+\frac{1}{6} \hat{\gamma}_{qg}^{(0)} \gamma_{qq}^{(0)} \beta_{0,Q}
-\frac{25}{96} \beta_{0,Q} \hat{\gamma}_{qg}^{(0)} \gamma_{gg}^{(0)}
-\frac{25}{48} \beta_{0} \beta_{0,Q} \hat{\gamma}_{qg}^{(0)}
\Biggr\}
\NN\\&&
\times
\left(L_1^3+L_2^3\right)
+\Biggl\{
\frac{1}{8} \hat{\gamma}_{qg}^{(0)} \gamma_{qq}^{(0)} \beta_{0,Q}
-\frac{1}{2} \beta_{0} \beta_{0,Q} \hat{\gamma}_{qg}^{(0)}
-\frac{1}{4} \beta_{0,Q} \hat{\gamma}_{qg}^{(0)} \gamma_{gg}^{(0)}
-\frac{3}{2} \beta_{0,Q}^2 \hat{\gamma}_{qg}^{(0)}
\Biggr\} 
\NN\\&&
\times
\left(L_1^2 L_2+L_2^2 L_1\right)
+\Biggl\{
-\frac{1}{64} \hat{\gamma}_{qg}^{(0)} \hat{\gamma}_{qq}^{\text{PS},(1)}
-\frac{1}{64} \hat{\gamma}_{qg}^{(0)} \hat{\gamma}_{qq}^{\text{NS},(1)}
+\frac{9}{16} \beta_{0,Q} \hat{\gamma}_{qg}^{(1)}
-\frac{1}{16} \hat{\gamma}_{qg}^{(0)} \hat{\gamma}_{gg}^{(1)}
\NN\\&&
-\frac{29}{16} \hat{\gamma}_{qg}^{(0)} \beta_{0,Q} \delta m_1^{(-1)}
+\frac{1}{16} \hat{\gamma}_{qg}^{(0)} \beta_{1,Q}
\Biggr\} \left(L_1^2+L_2^2\right)
-2 L_1 L_2 \hat{\gamma}_{qg}^{(0)} \beta_{0,Q} \delta m_1^{(-1)}
  +\Biggl\{
  \frac{3}{4} \hat{\gamma}_{qg}^{(0)} \tilde{\delta}m_2^{(-1)}
\NN\\&&
+\frac{1}{128} \left(\hat{\gamma}_{qg}^{(0)}\right)^2 \zeta_2 \gamma_{gq}^{(0)}
+\frac{1}{8} \hat{\gamma}_{qg}^{(0)} \zeta_2 \beta_{0,Q} \gamma_{qq}^{(0)}
-3 \delta m_1^{(0)} \beta_{0,Q} \hat{\gamma}_{qg}^{(0)}
-\frac{1}{2} \hat{\gamma}_{qg}^{(0)} a_{gg,Q}^{(2)}
-\frac{9}{32} \beta_{0,Q} \zeta_2 \hat{\gamma}_{qg}^{(0)} \gamma_{gg}^{(0)}
  \NN\\&&
-\frac{27}{8} \hat{\gamma}_{qg}^{(0)} \zeta_2 \beta_{0,Q}^2
-\frac{9}{16} \hat{\gamma}_{qg}^{(0)} \zeta_2 \beta_{0,Q} \beta_{0}
+4 \beta_{0,Q} a_{Qg}^{(2)}
\Biggr\} \left(L_1+L_2\right)
+8 \overline{a}_{Qg}^{(2)} \beta_{0,Q}
-\frac{1}{32} \hat{\gamma}_{qg}^{(0)} \zeta_2 \hat{\gamma}_{qq}^{\text{PS},(1)}
  \NN\\&&
-\frac{1}{32} \hat{\gamma}_{qg}^{(0)} \zeta_2 \hat{\gamma}_{qq}^{\text{NS},(1)}
+\frac{1}{96} \left(\hat{\gamma}_{qg}^{(0)}\right)^2 \zeta_3 \gamma_{gq}^{(0)}
-\frac{3}{2} \hat{\gamma}_{qg}^{(0)} \beta_{0,Q}^2 \zeta_3
+\frac{1}{8} \hat{\gamma}_{qg}^{(1)} \beta_{0,Q} \zeta_2
-4 \delta m_1^{(1)} \beta_{0,Q} \hat{\gamma}_{qg}^{(0)}
  \NN\\&&
+\frac{1}{8} \hat{\gamma}_{qg}^{(0)} \zeta_2 \beta_{1,Q}
- \hat{\gamma}_{qg}^{(0)} \overline{a}_{gg,Q}^{(2)}
-\frac{1}{12} \hat{\gamma}_{qg}^{(0)} \beta_{0} \beta_{0,Q} \zeta_3
-\frac{1}{24} \beta_{0,Q} \zeta_3 \hat{\gamma}_{qg}^{(0)} \gamma_{gg}^{(0)}
+\frac{1}{2} \hat{\gamma}_{qg}^{(0)} \Bigl(
\tilde{\delta}{m_2}^{1,(0)}
  \NN\\&&
+\tilde{\delta}{m_2}^{2,(0)}
\Bigr)
-\frac{9}{8} \hat{\gamma}_{qg}^{(0)} \zeta_2 \beta_{0,Q} \delta m_1^{(-1)}
+\tilde{a}_{Qg}^{(3)}\left(m_1^2,m_2^2,\mu^2\right)~.
\end{eqnarray}
\section{The main steps of the calculation}
\label{sec:3}

\begin{figure}[H]
  \centering
	\includegraphics[width=0.24\textwidth]{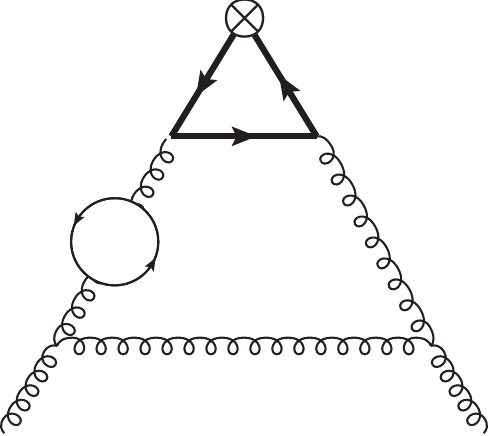}
	\includegraphics[width=0.24\textwidth]{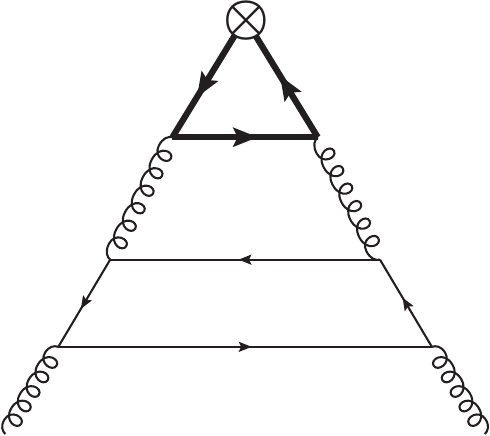}
	\includegraphics[width=0.24\textwidth]{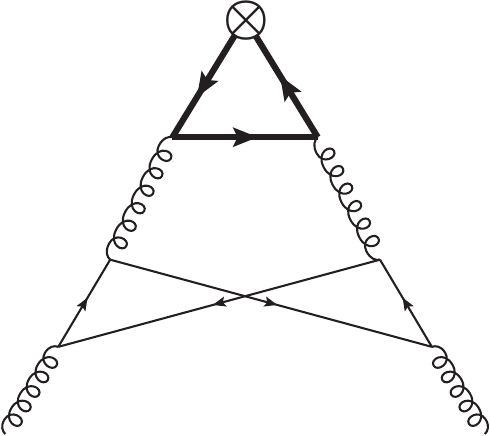}
	\includegraphics[width=0.24\textwidth]{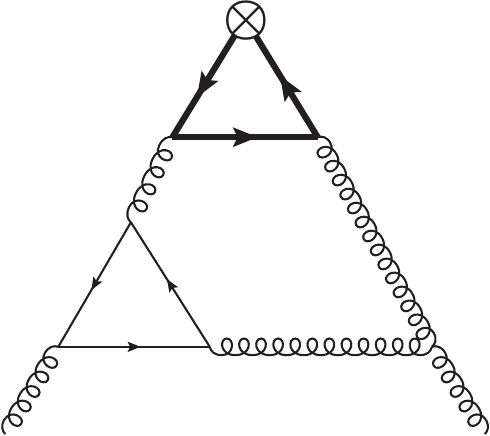}
  \caption{\sf Sample diagrams contributing to the two-mass contributions to the operator matrix
element $\tilde{A}_{Qg}^{(3)}$. The thin fermionic lines correspond to heavy quarks of a mass $m_1$ and the 
thick ones to quarks of a mass $m_2$. The amplitudes are symmetric under the interchange $m_1 \leftrightarrow m_2$.
}
  \label{fig:AQgdiags} 
\end{figure}

\vspace*{1mm}
\noindent
The main technical steps of the calculation of the OMEs in $x$-space are very similar to those described 
in
Ref.~\cite{Ablinger:2023ahe}.
Typical Feynman diagrams contributing to the two-mass corrections of $(\Delta) \tilde{A}_{Qg}^{(3)}$ are shown in 
Figure~\ref{fig:AQgdiags}.
We use the workflow described in Refs.~\cite{YND,Bierenbaum:2009mv,IBP1,IBP2,IBP3,IBP4,Chetyrkin:1981qh,
Laporta:2001dd,Karr:1981,Bron:00,Schneider:01,Schneider:04a,Schneider:05a,Schneider:05b,Schneider:07d,
Schneider:2009rcr,Schneider:10c,Schneider:15a,Schneider:08d,Schneider:08e,Schneider:2017,ABPET1, 
Ablinger:2015tua,Ablinger:2018zwz} and the packages 
{\tt QGRAF} \cite{Nogueira:1991ex}, {\tt Form} \cite{Vermaseren:2000nd,Tentyukov:2007mu}, 
and {\tt color} \cite{vanRitbergen:1998pn} to 
construct the amplitude.
In order to work with (effective) propagators instead of local operator insertions, we 
compute 
the generating function $F(t,\eta)$ \cite{Ablinger:2012qm,Ablinger:2014yaa}
\begin{eqnarray}
F(t,\eta) = \sum_{N=1}^\infty t^N F(N,\eta),~\text{with}~t \in \mathbb{R}.
\label{eq:FT}
\end{eqnarray}
This turns, e.g., the local operator 
\begin{eqnarray}
	\text{OP}_1^{(n)} (\tilde{p}_1) &=&
	( \Delta.\tilde{p}_1 )^n ,
\end{eqnarray}
into the following effective propagator
\begin{align}
	\sum\limits_{N=0}^\infty t^N \text{OP}_1^{(N)} (\tilde{p}_1) &= 
	\sum\limits_{N=0}^\infty t^N ( \Delta.\tilde{p}_1 )^N =
	\frac{1}{1-t \Delta.\tilde{p}_1}.
\end{align}
The Cauchy product 
\begin{align}
	\sum\limits_{i=0}^\infty a_i \sum\limits_{j=0}^\infty b_j &= 
	\sum\limits_{i=0}^\infty \sum\limits_{j=0}^i a_j b_{i-j}
\end{align}
allows to resum more involved structures.
To find the Mellin space result, one has to expand the generating functions around $t=0$ and extract the $N$th coefficient of the series. 
Alternatively, one can use the method described in Ref.~\cite{Behring:2023rlq} to directly 
relate the generating function in $t$ to the solution in $x$-space.

We generate the amplitude with quarks of masses $m_a$ and $m_b$, where the operator always sits on 
a heavy quark line of mass $m_a$. 
This means we have to calculate the amplitude twice. 
Once for the mass assignment $m_a=m_1$, $m_b=m_2$ and once for the assignment $m_a=m_2$, $m_b=m_1$.
The two-mass assignments are related by the symmetry $\eta \leftrightarrow 1/\eta$.

We need three integral families, $B_1$, $B_2$ and $B_3$, to cover all diagrams contributing to $(\Delta) \tilde{A}_{Qg}^{(3)}$.
After resummation their respective inverse propagators are given by
\begin{align}
P_{B_1,1} &= k_1^2 - m_a^2, & 
P_{B_2,1} &= k_1^2 - m_a^2, &
P_{B_3,1} &= k_1^2 - m_a^2, &
\nonumber \\
P_{B_1,2} &= ( k_1 - p )^2 - m_a^2 , & 
P_{B_2,2} &= ( k_1 - p )^2 - m_a^2 , &
P_{B_3,2} &= ( k_1 - p )^2 - m_a^2, &
\nonumber \\
P_{B_1,3} &= k_2^2 - m_b^2 , & 
P_{B_2,3} &= k_2^2 - m_b^2 , &
P_{B_3,3} &= k_2^2 - m_a^2, &
\nonumber \\
P_{B_1,4} &= ( k_2 - p )^2 - m_b^2 , & 
P_{B_2,4} &= ( k_2 - p )^2 - m_b^2 , &
P_{B_3,4} &= ( k_2 - p )^2 - m_a^2, &
\nonumber \\
P_{B_1,5} &= k_3^2 , & 
P_{B_2,5} &= k_3^2 , &
P_{B_3,5} &= k_3^2, &
\nonumber \\
P_{B_1,6} &= ( k_1 - k_3 )^2 - m_a^2 , & 
P_{B_2,6} &= ( k_1 - k_3 )^2 - m_a^2 , &
P_{B_3,6} &= ( k_1 - k_3 )^2 - m_b^2, &
\nonumber \\
P_{B_1,7} &= ( k_2 - k_3 )^2 - m_b^2 , & 
P_{B_2,7} &= ( k_2 - k_3 )^2 - m_b^2 , &
P_{B_3,7} &= ( k_2 - k_3 )^2 - m_b^2 , &
\nonumber \\
P_{B_1,8} &= (k_1 - k_2)^2 , & 
P_{B_2,8} &= (k_1 - k_2)^2 , &
P_{B_3,8} &= (k_1 - k_2)^2, &
\nonumber \\
P_{B_1,9} &= ( k_3 - p )^2 , & 
P_{B_2,9} &= ( k_3 - p )^2 , &
P_{B_3,9} &= ( k_3 - p )^2 , &
\nonumber \\
P_{B_1,10} &= 1 - t \, \Delta . k_1 , & 
P_{B_2,10} &= 1 - t \, \Delta . k_1 , &
P_{B_3,10} &= 1 - t \, \Delta . k_1 , &
\nonumber \\
P_{B_1,11} &= 1 - t \, \Delta . k_3 , & 
P_{B_2,11} &= 1 - t \, \Delta . \left( k_1 - k_3 \right) , &
P_{B_3,11} &= 1 - t \, \Delta . k_3 , &
\nonumber \\
P_{B_1,12} &= 1 - t \, \Delta . k_2 , & 
P_{B_2,12} &= 1 - t \, \Delta . k_2 , &
P_{B_3,12} &= 1 - t \, \Delta . k_2 .
\end{align}
Furthermore, the crossed families, where $p$ goes to $-p$, are needed for the reduction.
At the moment level, these integrals generally have an additional factor of $(-1)^N$ 
compared to the 
non-crossed ones.
At the level of generating functions one has to make the replacement $t \to -t$.
The diagrams with operator insertion on an external gluon need further considerations.
Here terms like
\begin{align}
	\frac{1}{ \left( 1 - t \, \Delta.k_1 \right) \left( 1 - t \, ( \Delta.k_1 - \Delta.p ) \right)}
\end{align}
contribute. Since the propagators $P_{i,10}$ to $P_{i,12}$ do 
not involve the external momentum $p$, these 
terms cannot be attributed to any of the integral families. However, we can use partial fractioning to obtain 
relations like
\begin{align}
\frac{1}{ \left( 1 - t \, \Delta.k_1 \right) \left( 1 - t \, ( \Delta.k_1 - \Delta.p ) \right)}
&=
\frac{1}{t \, \Delta.p} \left( \frac{1}{1 - t \, \Delta.k_1 } \
- \frac{1}{1 - t \, \Delta.( k_1 - 
p )} \right)
.
\end{align}
The two terms can now be handled separately and mapped to one or even several different 
integral families. Operator 
insertions on three and four gluons can be treated similarly and lead to three and four different terms 
respectively. 
We use the code \texttt{Reduze 2} \cite{Studerus:2009ye,vonManteuffel:2012np} for the reduction to 
master integrals and find a set of 150 master integrals,
for which we generate a system of differential equations in the variables $t$ and $m_b$.

One may compute the Mellin moments for $(\Delta) \tilde{a}_{Qg}^{(3)}$ by either setting $N$ to the desired 
integer in the operator insertion or by expanding the resummed amplitude in $t$ to the 
respective order.
The results in $N$- and in $x$-space 
are related by a Mellin transform \cite{RIEMANN,CAHEN,MELLIN1,MELLIN2,TITCHMARSH}
\begin{eqnarray}
\Mvec[f(x)](N) = \int_0^1 dx x^{N-1} f(x).
\end{eqnarray}
When expanding the amplitude in the small mass ratio $\eta$, low Mellin moments can be 
obtained by using the code {\tt Q2E/Exp}. 
The method of arbitrarily high moments 
\cite{Blumlein:2017dxp} allows to obtain a large number of moments expanded in $\eta$ to a certain degree.
We give details on this approach in Section~\ref{sec:4}. 
One can also compute the moments not expanding in $\eta$. 
These moments can be used to compare to the 
to Mellin moments computed numerically from the semi-analytic solutions of $(\Delta)\tilde{a}_{Qg}^{(3)}(x)$
we present in Section~\ref{sec:5}.
The lowest moments expanded and unexpanded in $\eta$ are given in 
Appendix~\ref{sec:A}.
\section{\boldmath The Operator Matrix Elements $\tilde{A}_{Qg}^{(3)}$ in Mellin space}
\label{sec:4}

\vspace*{1mm}
\noindent
The direct calculation of the two-mass contributions in Mellin space 
similar to the single-mass case is not possible at present.
Here recurrences related to the purely rational and $\zeta_3$ terms of the moments 
could not be solved in a closed form yet since analytic solution methods for non-first-order-factorizable 
difference equations are not known at present~\cite{Ablinger:2023ahe}. 
Still, we may analyze the OME $\tilde{A}_{Qg}^{(3)}$ in Mellin space.\footnote{We will only consider the 
unpolarized case here. The results in the polarized case are rather similar.} We use an adapted version of 
the algorithm 
of arbitrary high Mellin moments described in \cite{Blumlein:2017dxp} and calculate a large number of moments 
expanding in $\eta$. Then we use guessing techniques \cite{Blumlein:2009tj,GSAGE,GUESS} to find the recurrences for the 
first few orders of the $\eta$-expansion. It turns out that closed solutions in terms of nested sums
and products can only be found in the region $N > N_0(\eta)$, where $N_0(\eta)$ grows with the depth of 
the $\eta$-expansion. I.e. we do {\it not} obtain a closed form solution for the whole range of $N$,
but still a partial result. Its analytic continuation into $x$-space seems not to be possible, but
all even or odd integer moments can be described.

Let us describe our method of calculating the moments.
To calculate the master integrals, one
first considers their system of coupled differential equations in the variable $t$
which can be written as
\begin{align}
 	\frac{d}{dt} \begin{pmatrix} I_1 (t,\eta) \\ I_2(t,\eta) \\ \vdots \\ I_m (t,\eta) \end{pmatrix} &= 
	M \begin{pmatrix} I_1 (t,\eta) \\ I_2(t,\eta) \\ \vdots \\ I_m (t,\eta) \end{pmatrix}
	+ \begin{pmatrix} r_1 (t,\eta) \\ r_2(t,\eta) \\ \vdots \\ r_m (t,\eta) \end{pmatrix}.
\end{align}
Here  $M$ is an $m \times m$ matrix with entries consisting of rational functions in $\ep$, $\eta$ and $t$.
This system of differential equations can be decoupled to a single higher order differential 
equation using e.g. Z{\"u}rchers's or other algorithms \cite{ORESYS1,ORESYS2} implemented in the package 
\texttt{Oresys} \cite{ORESYS3}.
The master integrals $I_j(t,\eta)$ and the inhomogeneities $r_j(t,\eta)$ can be expanded in a Laurent expansion
in $\ep$, a Taylor series in the resummation parameter $t$ and a logarithmically generalized Laurent series in $\eta$,
yielding the general expressions
\begin{align}
	I_j(t,\eta) &=
	\sum_{n=0}^{\infty} \left[ \sum_{k=o}^{\infty} \left( \sum_{m=0}^{3} \left\{ \sum\limits_{l=o^\prime}^{\infty}
		I_j^{(k,m,l)} \ln^{m}(\eta) 
	\right\} \eta^l \right) \ep^{k} \right] t^n ,
\\
	r_j(t,\eta) &= 
	\sum_{n=0}^{\infty} \left[ \sum_{k=o}^{\infty} \left( \sum_{m=0}^{3} \left\{ \sum\limits_{l=o^\prime}^{\infty}
		r_j^{(k,m,l)} \ln^{m}(\eta) 
	\right\} \eta^l \right) \ep^{k} \right] t^n .
\label{eq:ansatz}
\end{align}
By inserting this ansatz into the decoupled differential equation, it is possible to find 
recurrences for 
the different 
expansion coefficients. Provided enough initial values are known, these recurrences can be 
used to iteratively calculate 
higher and higher moments of the master integrals. Although the ansatz in Eq.~\eqref{eq:ansatz} generalizes the algorithm 
designed for single scale quantities, the generalization is rather immediate.

These sets of Mellin moments provide the basis to guess \cite{Blumlein:2009tj,GUESS,GSAGE} the recurrences for the 
respective functions of the Mellin variable $N$. If these recurrences turn out to be 
first-order factorizable, we 
can find the closed form solutions with \texttt{Sigma} \cite{SIG1,SIG2}
using the algorithms described in 
\cite{Karr:1981,Bron:00,Schneider:01,Schneider:04a,Schneider:05a,Schneider:05b,Schneider:07d,
Schneider:2009rcr,Schneider:10c,Schneider:15a,Schneider:08d,Schneider:08e,Schneider:2017,ABPET1}
under certain conditions. 

Let us now turn to the calculation of the necessary initial values to solve the above system.
In Ref.~\cite{Ablinger:2014uka} a method to calculate initial values based on dimensional shifts was introduced.
Here the master integrals for fixed values of $N$ can be reduced to a small set of
scalar integrals without operator insertion in shifted dimensions.
This small set of scalar integrals can then be calculated using direct integration
techniques like hypergeometric methods \cite{HYPKLEIN,HYPBAILEY,SLATER1,APPEL1,APPEL2,
KAMPE1,EXTON1,EXTON2,SCHLOSSER,Anastasiou:1999ui,Anastasiou:1999cx,SRIKARL,Lauricella:1893,
Saran:1954,Saran:1955,Blumlein:2021hbq,Passarino:2024ugq,Passarino:2025cxi,Passarino:2025pbe}
and Mellin-Barnes integrals 
\cite{POCHHAMMER,KF,BARNES1,MELLIN1}.
The drawback of this procedure is that for every higher moment the scalar integrals
need to be calculated in a higher dimension, i.e.
the shift $N \to N+1$ leads to the dimensional shift $d \to d + 2$.

Another method can be established using the Mellin-Barnes representations of the master integrals.
In the current case, it was possible to find a one-dimensional contour integral 
representation
for all master integrals, which can be calculated analytically in terms of a
linear combination of single infinite sums of the kind
\begin{align}
	\sum\limits_{k=0}^{\infty} \eta^{k + j \pm a \ep} f(k,\ep),
	\label{eq:infsum}
\end{align}
with $j \in \mathbb{Z}$ and $a \in \left[ \tfrac{1}{2} , 1 , \tfrac{3}{2} \right]$.
In the case of more involved topologies, where the operator polynomial has to be split up,
further finite sums over Eq.~\eqref{eq:infsum} have to be applied.
Then the function $f(k,\ep)$ will also depend on the new summation quantifiers.
Fixing the value of $N$ to an integer will lead to a collapse of the finite sums into many terms.
Since we are only interested in the $\eta$-expansion of the initial values, we can cut off 
the
infinite sum in $k$ to the desired order in $\eta$.
Now only the expansion in $\ep$ has to be calculated in order to arrive at the initial
values of the master integrals.
The truncation of the infinite sums and high values of $N$ will lead to a proliferation
of terms.
However, the last step is a simple $\ep$-expansion of ratios of $\Gamma$-functions, which can be
implemented very efficiently and massively parallelized, making this method of calculating
initial values for the master integrals quite efficient.

Let us consider a series of master integrals.
For example, one finds 
\begin{align}
	B_2^{1,1,1,0,0,1,1,0,0,1,1,0}(N) 
	&= \frac{i S_\ep^3}{(4\pi)^6}  e^{\tfrac{3(4-D)}{2} \gamma_E}
	\Gamma(5- \tfrac{3D}{2} )
	\sum_{l=0}^{N} \sum_{i=0}^{l} \binom{l}{i}
	\int\limits_{0}^1 d z_1 
	\int\limits_{0}^1 d z_2 
	\int\limits_{0}^1 d z_3 
	\int\limits_{0}^1 d z_4
\Biggl\{
\nonumber \\ &
	z_1^N \left[ z_2 (1-z_2) \right]^{\tfrac{D}{2}-2}
	z_3^{\tfrac{D}{2}-3+l-i} (1-z_3)^{N-l+i+\tfrac{D}{2}-2}
\nonumber \\ &
	z_4^{2+l-i-\tfrac{D}{2}} (1-z_4)^{N-l+1-\tfrac{D}{2}}
	\left[ \frac{z_4 m_a^2}{z_3(1-z_3)} + \frac{(1-z_4)m_b^2}{z_2(1-z_2)} \right]^{\tfrac{3D}{2}-5} 
\Biggr\}.
\end{align}
This allows to calculate the following first initial values up to ${O}(\eta^5)$. 
For each master integral, we have to consider mass assignment $A$ ($m_a=m_1$, $m_b=m_2$)
and mass assignment $B$ ($m_b=m_1$, $m_a=m_2$), which are represented in the superscript of 
the integral.
\begin{align}
	B_2^{(1,1,1,0,0,1,1,0,0,1,1,0),A}(0)
	&=
\left( \frac{m_1^2}{\mu^2} \right)^{3\ep/2} m_1^2 S_\ep^3
\Biggl\{
-\frac{1}{\ep^3} \frac{8 (2+\eta )}{3 \eta }
+\frac{1}{\ep^2} 
\biggl[
 \frac{8}{9 \eta}
+\frac{4 (5+3 \ln(\eta) )}{3 \eta }
\biggr]
\nonumber \\ &
+ \frac{1}{\ep}
\biggl[
-\frac{2}{3}
-\zeta_2
-\frac{1}{\eta} 
\biggl(
\frac{16}{3}
+2 \zeta_2
+4 \ln(\eta )
+  \ln^2(\eta )
\biggr)
\biggr]
-\frac{16}{3}
+ 2 \ln(\eta )
\nonumber \\ &
-\ln^2(\eta )
+\frac{1}{3} \ln^3(\eta )
+\zeta_2
+\frac{7}{3} \zeta_3
+\frac{1}{\eta}
\biggl(
 \frac{5}{3}
-\frac{10}{3} \zeta_3
+ \ln(\eta )
-\frac{1}{6} \ln^3(\eta )
\nonumber \\ &
+\frac{1}{2} \zeta_2 \bigl(5+3 \ln(\eta )\bigr)
\biggr)
+\eta  
\biggl(
        \frac{7}{4}-\frac{3}{2} \ln(\eta )+\frac{1}{2} \ln^2(\eta )
\biggr)
+\eta ^2 
\biggl(
        \frac{19}{108}
\nonumber \\ &
	-\frac{5}{18} \ln(\eta )+\frac{1}{6} \ln^2(\eta )
\biggr)
+\eta ^3 
\biggl(
        \frac{37}{864}-\frac{7}{72} \ln(\eta )+\frac{1}{12} \ln^2(\eta )
\biggr)
\nonumber \\ &
+\eta ^4 
\biggl(
        \frac{61}{4000}
	-\frac{9}{200} \ln(\eta )+\frac{1}{20} \ln^2(\eta )
\biggr)
\nonumber \\ &
+\eta ^5 
\biggl(
        \frac{91}{13500}-\frac{11}{450} \ln(\eta )+\frac{1}{30} \ln^2(\eta )
\biggr)
\Biggr\}
	+ {O} ( \eta^6 \ln^2(\eta) )
,
\\
	B_2^{(1,1,1,0,0,1,1,0,0,1,1,0),B}(0)
	&=
\left( \frac{m_2^2}{\mu^2} \right)^{3\ep/2} m_2^2 S_\ep^3
\Biggl\{
-\frac{1}{\ep^3} \frac{8 (1+2 \eta )}{3}
+\frac{1}{\ep^2}
\biggl[
\frac{8}{3}
+\frac{4}{3} \eta  \bigl(5-3 \ln(\eta ) \bigr)
\biggr]
\nonumber \\ &
+\frac{1}{\ep}
\biggl[
-\frac{2}{3}
-\zeta_2
-\eta  \biggl(
         \frac{16}{3}
        +2 \zeta_2
        -4 \ln(\eta )
        + \ln(\eta )^2
\biggr)
\biggr]
-\frac{10}{3}
+\zeta_2
+\frac{7}{3} \zeta_3
\nonumber \\ &
+\eta  
\biggl(
        \frac{11}{3}
        +\frac{1}{2} \zeta_2 \bigl(5-3 \ln(\eta ) \bigr)
        -\frac{10}{3} \zeta_3
        -3 \ln(\eta )
        + \ln^2(\eta )
        -\frac{1}{6} \ln^3(\eta )
\biggr)
\nonumber \\ &
+\eta ^2 
\biggl(
        -\frac{7}{4}+\frac{3}{2} \ln(\eta ) -\frac{1}{2} \ln^2(\eta )
\biggr)
+\eta ^3 
\biggl(
        -\frac{19}{108}+\frac{5}{18} \ln(\eta )-\frac{1}{6} \ln^2(\eta )
\biggr)
\nonumber \\ &
+\eta ^4 
\biggl(
        -\frac{37}{864}+\frac{7}{72} \ln(\eta )
	-\frac{1}{12} \ln^2(\eta )
\biggr)
+\eta ^5 
\biggl(
        -\frac{61}{4000}
\nonumber \\ &
	+\frac{9}{200} \ln(\eta )-\frac{1}{20} \ln^2(\eta )
\biggr)
\Biggr\}
	+ {O} ( \eta^6 \ln^2(\eta) )
,
\\
	B_2^{(1,1,1,0,0,1,1,0,0,1,1,0),A}(1)
	&=
\left( \frac{m_1^2}{\mu^2} \right)^{3\ep/2} m_1^2 S_\ep^3
\Biggl\{
-\frac{1}{\ep^3} \frac{4 (3+\eta )}{3 \eta }
+\frac{1}{\ep^2} 
\biggl[
	1+\frac{13+9 \ln(\eta )}{3 \eta }
\biggr]
\nonumber \\ &
+\frac{1}{\ep}
\biggl[
 \frac{3}{4}
-\frac{1}{2} \zeta_2
-\frac{1}{\eta}
\biggl(
 \frac{11}{4}
+\frac{9}{4} \ln(\eta )
+\frac{3}{4} \ln^2(\eta )
+\frac{3}{2} \zeta_2
\biggr)
\biggr]
-\frac{73}{48}
-\frac{1}{2} \ln(\eta )
\nonumber \\ &
-\frac{1}{4} \ln^2(\eta )
+\frac{1}{6} \ln^3(\eta )
+\frac{3}{8} \zeta_2
+\frac{7}{6} \zeta_3
-\frac{1}{\eta}
\biggl(
 \frac{23}{48}
+\frac{5}{2} \zeta_3
+\frac{11}{16} \ln(\eta )
\nonumber \\ &
+\frac{7}{16} \ln^2(\eta )
+\frac{1}{8} \ln^3(\eta )
-\frac{1}{8} \zeta_2 (13+9 \ln(\eta ))
\biggr)
+\eta  
\biggl(
        \frac{415}{432}-\frac{61}{72} \ln(\eta )
\nonumber \\ &
	+\frac{7}{24} \ln^2(\eta )
\biggr)
+\eta ^2 
\biggl(
        \frac{913}{9000}-\frac{49}{300} \ln(\eta )+\frac{1}{10} \ln^2(\eta )
\biggr)
+\eta ^3 
\biggl(
        \frac{29945}{1185408}
\nonumber \\ &
	-\frac{821}{14112} \ln(\eta ) +\frac{17}{336} \ln^2(\eta )
\biggr)
+\eta ^4 
\biggl(
        \frac{53141}{5832000}-\frac{881}{32400} \ln(\eta ) +\frac{11}{360} \ln^2(\eta )
\biggr)
\nonumber \\ &
+\eta ^5 
\biggl(
        \frac{97289}{23958000}-\frac{1079}{72600} \ln(\eta ) +\frac{9}{440} \ln^2(\eta )
\biggr)
\Biggr\}
	+ {O} ( \eta^6 \ln^2(\eta) )
,
\\
	B_2^{(1,1,1,0,0,1,1,0,0,1,1,0),B}(1)
	&=
\left( \frac{m_2^2}{\mu^2} \right)^{3\ep/2} m_2^2 S_\ep^3
\Biggl\{
-\frac{1}{\ep^3} \frac{4 (1+3 \eta )}{3}
+\frac{1}{\ep^2} 
\biggl[
1+\eta  \bigl(\frac{13}{3}-3 \ln(\eta )\bigr)
\biggr]
\nonumber \\ &
+\frac{1}{\ep}
\biggl[
 \frac{3}{4}
-\frac{1}{2} \zeta_2
-\eta  \big(
         \frac{11}{4}
        -\frac{9}{4} \ln(\eta )
        +\frac{3}{4} \ln^2(\eta )
        +\frac{3}{2} \zeta_2
\big)
\biggr]
-\frac{193}{48}
+\frac{3}{8} \zeta_2
\nonumber \\ &
+\frac{7}{6} \zeta_3
+\eta  
\biggl(
        \frac{217}{48}
        -\frac{37}{16} \ln(\eta )
        +\frac{9}{16} \ln^2(\eta )
        -\frac{1}{8} \ln^3(\eta )
        +\frac{1}{8} \zeta_2 \bigl( 13-9 \ln(\eta ) \bigr)
\nonumber \\ &
        -\frac{5}{2} \zeta_3
\biggr)
+\eta ^2 
\biggl(
        -\frac{265}{216}+\frac{37}{36} \ln(\eta ) -\frac{1}{3} \ln^2(\eta )
\biggr)
+\eta ^3 
\biggl(
        -\frac{6397}{54000}
\nonumber \\ &
	+\frac{331}{1800} \ln(\eta ) -\frac{13}{120} \ln^2(\eta )
\biggr)
+\eta ^4 
\biggl(
        -\frac{5585}{197568}+\frac{149}{2352} \ln(\eta ) -\frac{3}{56} \ln^2(\eta )
\biggr)
\nonumber \\ &
+\eta ^5 
\biggl(
        -\frac{116063}{11664000}+\frac{1883}{64800} \ln(\eta ) -\frac{23}{720} \ln^2(\eta )
\biggr)
\Biggr\}
	+ {O} ( \eta^6 \ln^2(\eta) )
.
\end{align}

We can also choose not to expand in $\eta$ in order to obtain moments exact in the mass ratio.
To obtain initial values in this case, we do not truncate the infinite sum as before, but 
evaluate it 
with \texttt{HarmonicSums} and \texttt{Sigma}. 
The resulting infinite sums can be rewritten in terms of harmonic polylogarithms using the 
capabilities of  \texttt{HarmonicSums}. 

However, a more effective approach is to perform the expansion in $t \to 0$ and $p^2=0$, 
which leads to two-mass tadpoles, and employ an IBP reduction. We reveal four master 
integrals in the limit $t \to 0$ which are easily computed from their Mellin-Barnes 
representation.

When we do not expand in the mass ratio $\eta$, higher moments in $N$ develop larger and 
larger 
rational function coefficients in $\eta$, leading to very large final expressions for the moments. 
We have calculated the moments up to $N=60$ to cross-check the $x$-space results presented in the 
next section. 
We note that the expansion in $\eta$ for a fixed moment converges very quickly.

The initial values allow to compute a large number of moments. We generated 2000 Mellin moments by expanding 
the master integrals to 
$O(\eta^5)$. This number is sufficient to determine the corresponding 
recurrences for the color-$\zeta$-$\eta$-log 
expressions up to $O(\eta^3)$, because of additional poles in $\eta$ in the amplitudes. 
For the cases
$\textcolor{blue}{C_A T_F^2}$ and $\textcolor{blue}{C_A T_F^2} \eta$, we 
needed 3000 moments.

All recurrences can be solved by {\tt Sigma} in product-sum expressions for $N > N_0,~~N_0 \in 
\mathbb{N}$ in terms of generalized harmonic sums \cite{Ablinger:2013cf,Moch:2001zr}. The boundary 
value $N_0$ grows with the power of the expansion parameter $\eta$. 
I.e. no closed analytic representation 
can be given for all even moments in $N$ in the unpolarized case. Therefore, the present representation 
in Mellin $N$ space does not allow the inversion to $x$-space.
Table~\ref{tab1} summarizes the characteristics of the different contributions. The corresponding 
recurrences are valid for $N \geq N_0$. In most of the cases, non-removable poles limit 
the validity of the
recurrence below $N_0$, but not in all cases.

\section{\boldmath The semi-analytic approach in $x$--space}
\label{sec:5}

\vspace*{1mm}
\noindent
In the previous section, we have seen that the expansion around $\eta=0$ introduces 
severe problems for the 
transition of the results from $N$- to $x$-space.
One way to find a viable $x$-space solution is to fix $\eta$ to a physical value and use the semi-analytic approach already used in 
the single mass case to calculate a numerical approximation. 
While this is viable, phenomenological applications need to scan over different values of the masses or want to 
use different renormalization schemes so that the differential equation needs to be solved for several different 
values of the quark masses in order to obtain a grid. 
Another technical issue are poles in the differential equation which move very close to $t=1$ for physical 
quark masses. This makes the precise numerical solution of the system a non-trivial task.

In order to find a more flexible representation of the OMEs we will not expand them around 
$\eta \to 0$, but around the unphysical limit $\eta \to 1$ for which we introduce the 
expansion variable 
\begin{eqnarray}
\delta = 1 - \frac{m_c}{m_b}.
\end{eqnarray}
This approach was already successfully applied to the calculation of inclusive semi-leptonic $b$ 
quark decays in 
Ref.~\cite{Fael:2020tow}.
We can also have a look at the analytically known moments to examine the convergence in this limit.
Using the results for fixed moments (see Appendix~\ref{sec:A}) and setting
\begin{eqnarray}
m_c = 1.59~\text{GeV},~~~~~m_b = 4.78~\text{\rm GeV}, 
\end{eqnarray}
cf.~\cite{Alekhin:2012vu,ParticleDataGroup:2022pth},
and, $\mu = 10~\text{GeV}$ 
\begin{table}[H]
\centering
{\small
\renewcommand*{\arraystretch}{1.2}
\begin{tabular}{|l|r|r|r|r|r|}
\hline
Unpolarized case     &              &            &              &       &\\
Coefficient          & $\#$ moments & rec. order &  rec. degree & $N_0$ & pole position  \\
\hline
$\textcolor{blue}{C_F T_F^2}$             &        1550  &        17  &          320 &     2 &     \\
$\textcolor{blue}{C_F T_F^2} \ln(\eta)$   &         168  &         4  &           43 &     2 &     \\
$\textcolor{blue}{C_F T_F^2} \ln^2(\eta)$ &          56  &         2  &           19 &     2 &     \\
$\textcolor{blue}{C_F T_F^2} \ln^3(\eta)$ &          36  &         2  &            9 &     2 &     \\
$\textcolor{blue}{C_F T_F^2} \zeta_2$     &         294  &         7  &           69 &     2 &     \\
$\textcolor{blue}{C_F T_F^2} \zeta_3$     &         126  &         4  &           31 &     2 &     \\
$\textcolor{blue}{C_F T_F^2} \zeta_2 \ln(\eta)$  &   36  &         2  &            9 &     2 &     \\
$\textcolor{blue}{C_F T_F^2} \eta$        &         384  &         8  &           95 &     4 & 2   \\
$\textcolor{blue}{C_F T_F^2} \eta \ln(\eta)$     &   72  &         2  &           23 &     4 &     \\
$\textcolor{blue}{C_F T_F^2} \eta \ln^2(\eta)$   &   32  &         2  &           8  &     2 &     \\
$\textcolor{blue}{C_F T_F^2} \eta^2$      &         600  &        13  &         192  &     4 & 2   \\
$\textcolor{blue}{C_F T_F^2} \eta^2 \ln(\eta)$   &  336  &         8  &          84  &     4 & 2   \\
$\textcolor{blue}{C_F T_F^2} \eta^2 \ln^2(\eta)$ &   42  &         2  &          13  &     4 &     \\
$\textcolor{blue}{C_F T_F^2} \eta^3$      &         561  &         9  &         151  &     6 & 2,4 \\
$\textcolor{blue}{C_F T_F^2} \eta^3 \ln(\eta)$   &  221  &         5  &          71  &     6 & 2,4 \\
$\textcolor{blue}{C_F T_F^2} \eta^3 \ln^2(\eta)$ &   54  &         2  &          17  &     6 &     \\
\hline
$\textcolor{blue}{C_A T_F^2}$             &         2622 &         26 &          513 &     2 &      \\
$\textcolor{blue}{C_A T_F^2} \ln(\eta)$   &          748 &         12 &          161 &     2 &      \\
$\textcolor{blue}{C_A T_F^2} \ln^2(\eta)$ &          308 &          8 &           70 &     2 &      \\
$\textcolor{blue}{C_A T_F^2} \ln^3(\eta)$ &          108 &          4 &           27 &     2 &      \\
$\textcolor{blue}{C_A T_F^2} \zeta_2$     &          360 &         10 &           76 &     2 &      \\
$\textcolor{blue}{C_A T_F^2} \zeta_3$     &          108 &          4 &           26 &     2 &      \\
$\textcolor{blue}{C_A T_F^2} \zeta_2 \ln(\eta)$  &   108 &          4 &           27 &     2 &      \\
$\textcolor{blue}{C_A T_F^2} \eta$        &         2080 &         23 &          400 &     4 & 2    \\
$\textcolor{blue}{C_A T_F^2} \eta \ln(\eta)$     &   840 &         14 &          160 &     4 &      \\
$\textcolor{blue}{C_A T_F^2} \eta \ln^2(\eta)$   &   280 &          8 &           55 &     2 &      \\
$\textcolor{blue}{C_A T_F^2} \eta^2$      &         1548 &         20 &          311 &     4 & 2    \\
$\textcolor{blue}{C_A T_F^2} \eta^2 \ln(\eta)$   &   500 &         10 &          113 &     4 & 2    \\
$\textcolor{blue}{C_A T_F^2} \eta^2 \ln^2(\eta)$ &   117 &          4 &           31 &     4 &      \\
$\textcolor{blue}{C_A T_F^2} \eta^3$      &         1457 &         17 &          324 &     6 & 2,4  \\
$\textcolor{blue}{C_A T_F^2} \eta^3 \ln(\eta)$   &   525 &          9 &          128 &     6 & 2,4  \\
$\textcolor{blue}{C_A T_F^2} \eta^3 \ln^2(\eta)$ &   117 &          4 &          33  &     6 &   \\
\hline
\end{tabular}
}
\caption[]{\sf Characteristics of the recurrences and the parameter $N_0 \leq N,~N$ even,
for which the respective recurrence is valid in the unpolarized case The poles of the solution of the 
recurrence
are also given.
\label{tab1}}
\renewcommand*{\arraystretch}{1}
\end{table}

\noindent
we obtain for the 10th moment 
\begin{align}
        \tilde{A}_{Qg}^{(3)}(N=10) &= -74.215863973188462672...
        \nonumber\\ 
        &\approx
		  4.69095...
		-52.87648... \chi
		-13.51460... \chi^2
		- 5.77578... \chi^3
		- 2.91093... \chi^4
		\nonumber \\ &
		- 1.58651... \chi^5
		- 0.90329... \chi^6
		- 0.52844... \chi^7
		- 0.31488... \chi^8
		- 0.19014... \chi^9
		\nonumber \\ &
		- 0.11599... \chi^{10}
		- 0.07134... \chi^{11}
		- 0.04417... \chi^{12}
		- 0.02750... \chi^{13}
		\nonumber \\ &
                - 0.01720... \chi^{14}
		- 0.01081... \chi^{15}
		- 0.00681... \chi^{16}
		- 0.00431... \chi^{17}
                \nonumber \\ &
                - 0.00273... \chi^{18}
		- 0.00174... \chi^{19}
		+ \mathcal{O}(\delta^{20})
		\nonumber \\ 
		&= -74.212774664755273018...,
        \end{align}
where we set 
\begin{equation}
\delta = \left(1-\frac{m_c}{m_b}\right) \chi 
\end{equation}
in order to 
see the individual terms in the expansion.

We see that keeping the first 20 terms in the expansion allows for an accuracy of $0.004\%$ for the 
individual moment. 
Using Aitken extrapolation \cite{AITKEN}
we can push the accuracy even further.
Using the partial sums 
\begin{align}
        \tilde{A}^{(3)}_{Qg}(N=10) &= \sum_{j=0}^{\infty} 
\tilde{A}^{(3),\delta,k}_{Qg}(N=10) \delta^k ~, 
        \\ 
        \tilde{A}^{(3),k}_{Qg}(N=10) &= \sum_{j=0}^{k} 
\tilde{A}^{(3),\delta,k}_{Qg}(N=10) \delta^k,
        \label{eq:partial}
\end{align}
we use one step in the series acceleration to obtain 
\begin{align*}
        \tilde{A}^{(3),\text{acc}}_{Qg}(N=10) &= \frac{\left[\tilde{A}^{(3),k}_{Qg}(N=10)\right]^2 - \left[\tilde{A}^{(3),k-1}_{Qg}(N=10)\right]^2 }{
                \tilde{A}^{(3),k}_{Qg}(N=10) - 2\tilde{A}^{(3),k-1}_{Qg}(N=10) + \tilde{A}^{(3),k-2}_{Qg}(N=10)
        }~,
\end{align*}
which converges to the same limit as $k \to \infty$, but usually shows a better 
convergence.
In our case we find 
\begin{align}
        \tilde{A}^{(3),\text{acc}}_{Qg}(N=10) &= -74.215816598482...
        ~,
\end{align}
which shows a relative deviation of only $0.00006 \%$ from the exact result. 
In principle, this procedure can be repeated, however, one step of the sum acceleration
already leads to sufficiently precise predictions.

To obtain the OMEs in an expansion around the equal mass limit, we proceed as follows.
\begin{itemize}
        \item We use the differential equation in $m_b$ for the two-mass master integrals to obtain 
        an expansion to 21 terms 
        in the equal mass limit. To do this, we plug in an ansatz of the form 
        \begin{align}
                \vec{I}(t,m_a,m_b,\varepsilon) &= \sum\limits_{i=0}^{\infty} \delta^i 
\vec{I}_i(t,m_a,m_a,\varepsilon) ~,
        \end{align}
        where $\vec{I}$ is the vector of master integrals,
        into the differential equation, and solve for the coefficients 
$\vec{I}_0(t,m_a,m_a,\varepsilon)$ as boundary 
        conditions.
        \item The boundary conditions $\vec{I}_0(t,m_a,m_a,\varepsilon)$ are the 
integrals we already encountered in the 
        single mass case (cf.~Ref.~\cite{Ablinger:2023ahe,Ablinger:2024xtt}), where they 
        have been calculated using semi-analytic methods.
        \item We plug all expansions back into our amplitude expression and expand consistently in $\delta$.
        This allows us to compute $\tilde{A}_{Qg}^{(3)}$ to order $\delta^{19}$ and, due to higher poles in the 
        amplitude, $\Delta \tilde{A}_{Qg}^{(3)}$ to order $\delta^{18}$. 
        If more precise results are needed in the future, we can extend the expansion in 
$\delta$ easily.
\end{itemize}
\begin{figure}[H]
	\centering
	\includegraphics[width=0.75 \textwidth]{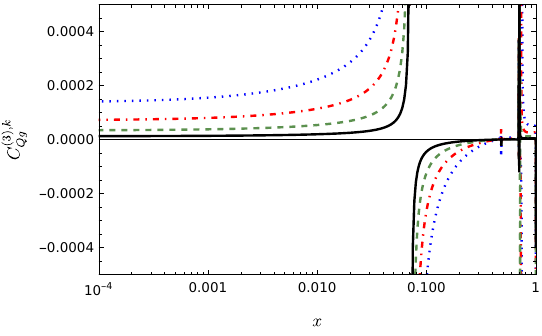}
	\caption{\sf The relative difference between different expansions depths of 
$\tilde{A}_{Qg}^{(3)}$ in $\delta$
        to our best approximation which uses 20 expansion terms for 
        $Q^2=50~\GeV^2$. 
        The curves show the relative difference to the expansion up to $\delta^{15}$, $k=15$ 
(blue dotted line),
        $\delta^{16}$, $k=16$ (red dash-dotted line),
        $\delta^{17}$, $k=17$ (green dashed line),
        $\delta^{18}$, $k=18$ (black full line).
        }
\label{fig:AQgCONV}
\end{figure}

\begin{table}[H]
        \centering
	{\small
	\renewcommand*{\arraystretch}{1.9}
	\begin{tabular}{|c|r|r|c|r|r|}
	        \hline
		N       & $\tilde{A}^{(3),\text{exact}}_{Qg}(N)$ & $r^{(3)}_{Qg}(N)[\%]$ & N & $\Delta \tilde{A}^{(3),\text{exact}}_{Qg}(N)$ & $\Delta r^{(3)}_{Qg}(N)[\%]$ \\
		\hline
		2       & --4236.5 & 0.00004 & 1  &       0  & (0.00004)  \\
		6       & --3091.6 & 0.00005 & 5  &--2988.0 & 0.00008  \\
		12      & --1970.7 & 0.00005 & 11 &--2026.4 & 0.00008  \\
		\hline
		\end{tabular}
		}
	\renewcommand*{\arraystretch}{1}
        \caption{\sf
                Relative deviation of exact moments and the ones obtained by numerically 
                integrating over the $x$-space approximation as defined in Eq.~\eqref{eq:momcomp}.
                Since $\Delta \tilde{A}^{(3),\text{exact}}_{Qg}(N=1)$ vanishes, we give 
the absolute 
                difference for this value.
        }
        \label{tab:momcomp}
\end{table}

The approximation shows good perturbative convergence, which is illustrated in 
Figure~\ref{fig:AQgCONV}, 
where we plot the quantity 
\begin{align}
        C_{Qg}^{(3),k}(x,Q^2) &=  1 - 
\frac{\tilde{A}^{(3),k}_{Qg}(x,Q^2)}{\tilde{A}^{(3),19}_{Qg}(x,Q^2)}.
\end{align}
Here, we borrow the notation of Eq.~\eqref{eq:partial} also for the partial sums in $x$-space.
Since the denominator vanishes at $x \sim 0.07$ the relative difference diverges there. 
However, the value of all approximations and the absolute differences are small in this region.
Overall, one sees that higher expansion orders converge homogeneously. 
At $x \sim 0.75$, we see a slower convergence of the series. However, the relative 
difference between the 
two last expansion terms is smaller than $0.001$, sufficient for any practical purpose. 
The convergence in the polarized case is similar.
\begin{figure}[H]
	\centering
	\includegraphics[width=0.75 \textwidth]{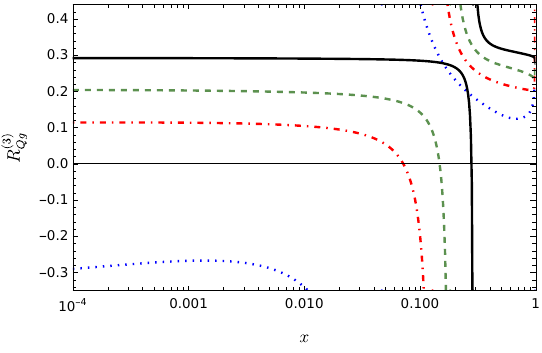}
	\caption{\sf The ratio $R_{Qg}^{(3)}(x,Q^2)$ of the 2-mass contributions to $\tilde{A}_{Qg}^{(3)}$ to the complete contribution of 
\textcolor{blue}{$O(T_F^2)$} and \textcolor{blue}{$O(T_F^3)$} to ${A}_{Qg}^{(3)}$ 
as defined in Eq.~\eqref{eq:ratio} in the unpolarized case.
$Q^2=30~\GeV^2$ dotted line (blue), $Q^2=50~\GeV^2$ dash-dotted line (red),  
$Q^2=100~\GeV^2$ dashed line (green), $Q^2=1000~\GeV^2$ full line (black) .}
\label{fig:3}
\end{figure}

\noindent
Note that for the evaluation of the OMEs, we use one step of 
the sum acceleration at each 
point of $x$ separately. 
From the difference between the approximations using our highest available 
expansion depth and one term less 
we can even estimate the residual error of our approximation. 
However, due to the high expansion orders already available, the residual error is rather small and will not be 
considered further.
As a check, we can compare the relative difference between fixed moments which we have 
computed analytically and the moments we obtain by numerically integrating over our 
approximation
\begin{align}
       (\Delta) r^{(3)}_{Qg} (N,Q^2) &= \frac{(\Delta)\tilde{A}^{(3),\text{exact}}_{Qg}(N,Q^2) - (\Delta)\tilde{A}^{(3),\text{approx}}_{Qg}(N,Q^2) }{(\Delta)\tilde{A}_{Qg}^{(3),\text{exact}}(N,Q^2)}
       ~.
       \label{eq:momcomp}
\end{align}

The results for some moments are shown in Table~\ref{tab:momcomp}. There is excellent 
agreement between the 
exact moments of the OME and the ones obtained by numerically integrating over the 
$x$-space approximation obtained 
after expanding in $\delta$.
Note that the achieved relative error is of similar size as the one obtained from the expansion of 
the moment directly. This shows that the errors due to the approximate solution of the $x$-space
through generalized series expansions are negligible.

\begin{figure}[H]
	\centering
	\includegraphics[width=0.75 \textwidth]{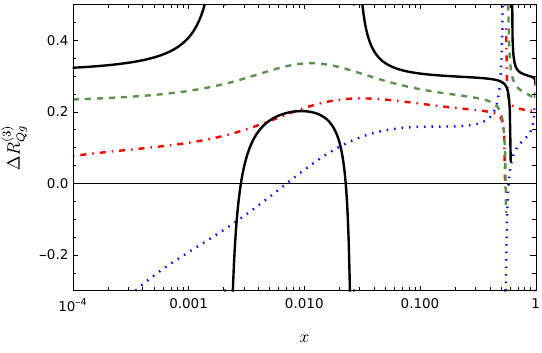}
	\caption{\sf The ratio $\Delta R_{Qg}^{(3)}(x,Q^2)$ of the 
2-mass contributions to $\Delta \tilde{A}_{Qg}^{(3)}$ to the complete contribution of 
\textcolor{blue}{$O(T_F^2)$} and \textcolor{blue}{$O(T_F^3)$} to $\Delta {A}_{Qg}^{(3)}$ 
as defined in Eq.~\eqref{eq:ratio} in the polarized case in the Larin scheme.
$Q^2=30~\GeV^2$ dotted line (blue), $Q^2=50~\GeV^2$ dash-dotted line (red),  
$Q^2=100~\GeV^2$ dashed line (green), $Q^2=1000~\GeV^2$ full line (black) .}
\label{fig:4}
\end{figure}

In order to illustrate the phenomenological results, we show the ratio of the two-mass 
contributions over 
the contributions of the color factors $T_F^2$ and $T_F^3$ to the full OME
\begin{align}
        (\Delta) R_{Qg}^{(3)}(x,Q^2) &= \frac{(\Delta) \tilde{A}_{Qg}^{(3)}(x,Q^2)}{(\Delta){A}_{Qg}^{(3), T_F^2+T_F^3}(x,Q^2)}~,
        \label{eq:ratio}
\end{align}
in the unpolarized case in Figure~\ref{fig:3} and in the polarized case in Figure~\ref{fig:4}, where we work 
in the Larin scheme.
We see that the two-mass contributions make up 
$-30\% $ to $30\%$ in the small $x$ region moving from $Q^2$ from $30~\GeV^2$ to $1000~\GeV^2$.
At large $x$, the ratio varies between $ 20\%$ and $ 50\%$. In the polarized case the picture is similar
and one observes variations between $-20\%$ to $ 40\%$.

The polarized OME $\Delta \tilde{A}_{Qg}$ contributes to the 2-mass variable flavor number scheme 
\cite{Ablinger:2017err} and the polarized heavy-flavor Wilson coefficients for $Q^2 \gg 
m_Q^2$.
By referring to the polarized massless Wilson coefficients \cite{Blumlein:2022gpp}, the anomalous dimensions 
\cite{Moch:2014sna,Ablinger:2022wbb}, and the parton distributions \cite{Blumlein:2024euz}, one can 
represent the structure function $g_1(x,Q^2)$ as an observable. 
\section{Conclusions}
\label{sec:6}

\vspace*{1mm}
\noindent
We have calculated the two-mass corrections to the constant part of the three-loop massive OMEs 
$\tilde{A}_{Qg}^{(3)}$ and $\Delta \tilde{A}_{Qg}^{(3)}$ using a semi-analytic method in $x$-space. 
Initially, we have considered the structure of this contribution in Mellin $N$-space by expanding in the 
variable $\eta$ to the first few powers. It turned out that a closed $N$-space solution covering
all physical values of $N = 2k, k \in \mathbb{N}$ in the unpolarized case and of  $N = 2k+1$ in 
the polarized case cannot be found at present. Like known in the $x$-space solution, one may conjecture that
the $N$-space solution contains letters which are higher transcendental functions in the whole range of $N$. 
Sum-product solutions are only possible for $N > N_0$, 
where the value of 
$N_0$ is not always caused by poles in the corresponding expressions. In 
general, it is expected that in the case of the present amplitude, non-first-order-factorizing 
contributions are present,
which will not thoroughly lead to sum-product solutions, since these obey 
first-order-factorizing recurrences.
Nevertheless, by employing first-order
differential equations of the master integrals in the generating-function-variable $t$, and by 
fixing
the value of $\eta$, we can solve the two-parameter problem. Like in the single-mass case 
Ref.~\cite{Ablinger:2024xtt} for $x \in ]0,1]$, highly precise local series expansions can be 
derived. The ratios $(\Delta) \tilde{A}_{Qg}^{(3)}/(\Delta) {A}_{Qg}^{(3), \rm T_F^2}$ 
illustrate the importance of this contribution. They amount to about $-30\%$ to $50\%$ of all 
$T_F^2 C_i$ contributions, with $C_i = C_F, C_A$ or $T_F$.
A numerical program to evaluate the two-mass three-loop contributions to the unpolarized and polarized 
massive OMEs 
will be published in a later paper.

\appendix
\section{\boldmath Mellin moments for $\tilde{a}_{Qg}^{(3)}$ and  $\Delta \tilde{a}_{Qg}^{(3)}$ at fixed values 
of $N$}
\label{sec:A}

\vspace*{1mm}
\noindent 
The Mellin moments given in Refs.~\cite{Ablinger:2011pb,Ablinger:2011gdv} concerned only 
the contributions due to irreducible Feynman diagrams in the unpolarized case up to terms of 
$O(\eta^3)$. Here we present the complete results including terms of $O(\eta^5)$. We use the 
abbreviations 
\begin{eqnarray} 
L_1 = \ln\left(\frac{m_b^2}{\mu^2}\right),~~~ L_2 = 
\ln\left(\frac{m_c^2}{\mu^2}\right),~~~ L_\eta = \ln\left(\frac{m_c^2}{m_b^2}\right),~~~\xi = 
\frac{m_c}{m_b}. 
\end{eqnarray} 
One obtains 
\begin{eqnarray} 
\tilde{a}_{Qg}^{(3)} (N=2) &\propto& 
\textcolor{blue}{T_F^2} \Biggl\{
        \textcolor{blue}{C_A} \Biggl[
                \frac{59314}{2187}
                +\frac{1964 L_1}{243}
                -\frac{4009 L_1^2}{216}
                +\frac{ 1276 L_1^3}{81}
                +\frac{1964 L_2}{243}
                -\frac{1351 L_1 L_2}{108}
\nonumber\\ &&           
     +\frac{440 L_1^2 L_2}{27}
                -\frac{4009 L_2^2}{216}
                +\frac{616 L_1 L_2^2}{27}
                +\frac{1100 L_2^3}{81}
                -\frac{4880 L_\eta}{243}
                +\frac{3307 L_\eta^2}{648}
\nonumber\\ &&               
                -\frac{8 \eta ^3}{843908625}
\big(
                        -7010842-6475770 L_\eta+31652775 
L_\eta^2\big)
\nonumber\\ && 
 -\frac{4 \eta ^2}{496125}
 \big(
                        -391259+157710 L_\eta+22050 
L_\eta^2\big)
                +\eta  \Biggl(
                        \frac{256304}{10125}
                        +\frac{161 L_1}{108}
                        -\frac{161 L_1^2}{432}
\nonumber\\ &&
    -\frac{161 L_2}{108}
                        +\frac{161 L_1 L_2}{216}
                        -\frac{161 L_2^2}{432}
                        -\frac{8237 L_\eta}{900}
                        +\frac{421 L_\eta^2}{2160}
                \Biggr)
\nonumber\\ &&
                +\Biggl(
                        -
                        \frac{1340}{81}
                        +\frac{308 L_1}{9}
                        +\frac{308 L_2}{9}
                \Biggr) \zeta_2
                -\frac{176}{81} \zeta_3
        \Biggr]
\nonumber\\ &&
        +\textcolor{blue}{C_F} \Biggl[
                \frac{25556}{729}
                +\frac{13636 L_1}{243}
                +\frac{16957 L_1^2}{324}
                -\frac{992 L_1^3}{81}
                +\frac{13636 L_2}{243}
                +\frac{10691 L_1 L_2}{162}
\nonumber\\ &&               
 -\frac{64 L_1^2 L_2}{27}
                +\frac{16957 L_2^2}{324}
                -\frac{320 L_1 L_2^2}{27}
                -\frac{736 L_2^3}{81}
                +\frac{3184 L_\eta}{243}
                -\frac{925 L_\eta^2}{324}
\nonumber\\ &&
                +\frac{32 \eta ^3}{843908625}
 \big(
                        25837687-58011030 L_\eta+74021850 
L_\eta^2\big)
\nonumber\\ &&
                +\frac{32 \eta ^2}{10418625}
\big(
                        5309152-3373755 L_\eta+1642725 
L_\eta^2\big)
                +\eta  \Biggl(
                        -\frac{758944}{30375}
                        +\frac{409 L_1}{54}
\nonumber\\ && 
                        -\frac{409 L_1^2}{216}
                        -\frac{409 L_2
                        }{54}
                        +
                        \frac{409 L_1 L_2}{108}
                        -\frac{409 L_2^2}{216}
                        -\frac{15277 L_\eta}{4050}
                        +\frac{5629 L_\eta^2}{1080}
                \Biggr)
\nonumber\\ &&
                +\Biggl(
                        \frac{416}{9}
                        -\frac{160 L_1}{9}
                        -\frac{160 L_2}{9}
                \Biggr) \zeta_2
                +\frac{1408}{81} \zeta_3
        \Biggr]
\Biggr\}
\nonumber\\ &&
+\textcolor{blue}{T_F^3} \Biggl[
        -\frac{32 L_1^3}{3}
        -\frac{64 L_1^2 L_2}{3}
        -\frac{64 L_1 L_2^2}{3}
        -\frac{32 L_2^3}{3}
        -32 (L_1
+        L_2)
        \zeta_2
        -\frac{128}{9} \zeta_3
\Biggr],
\\
\tilde{a}_{Qg}^{(3)} (N=4) &\propto& 
\textcolor{blue}{T_F^2} \Biggl\{
        \textcolor{blue}{C_A} \Biggl[
                \frac{4366284317}{36450000}
                +\frac{64863121 L_1}{972000}
                +\frac{965364173 L_1^2}{82944000}
                +\frac{37642 L_1^3}{2025}
                +\frac{64817671 L_2}{972000}
\nonumber\\ && 
                +\frac{1259765299 L_1 L_2}{41472000}
                +\frac{2596 L_1^2 L_2}{135}
                +\frac{965364173 L_2^2}{82944000}
                +\frac{18172 L_1 L_2^2}{675}
                +\frac{1298 L_2^3}{81}
\nonumber\\ && 
                -\frac{120332519 L_\eta}{4860000}
                +\frac{518749747 L_\eta^2}{82944000}
                +\eta ^3 \Biggl(
                        \frac{250077164867}{11232423798750}
                        +\frac{101 L_1}{2160}
                        -\frac{101 L_1^2}{1440}
\nonumber\\ && 
                        -\frac{101 L_2}{2160}
                        +\frac{101 L_1 L_2}{720}
                        -\frac{101 L_2^2}{1440}
                        +\frac{2461291549 L_\eta}{25933446000}
                        -\frac{4904369 L_\eta^2}{14968800}
                \Biggl)
\nonumber\\ &&                 
+\eta ^2 \Biggl(
                        \frac{1255194149}{468838125}
                        +\frac{80203 L_1}{2764800}
                        -\frac{468043 L_1^2}{11059200}
                        -\frac{80203 L_2}{2764800}
                        +\frac{468043 L_1 L_2}{5529600}
\nonumber\\ &&                
         -\frac{468043 L_2^2}{11059200}
                        -\frac{6519186689 L_\eta}{6096384000}
                        -\frac{17662451 L_\eta
                        ^2}{77414400}
                \Biggl)
\nonumber\\ &&                
                +\eta  \Biggl(
                        \frac{496855133}{14883750}
                        +\frac{559627 L_1}{460800}
                        -\frac{144839 L_1^2}{460800}
                        -\frac{559627 L_2}{460800}
\nonumber\\ &&                
         +\frac{144839 L_1 L_2}{230400}
                        -\frac{144839 L_2^2}{460800}
                        -\frac{1746174071 L_\eta}{145152000}
                        +\frac{796501 L_\eta^2}{1382400}
                \Biggl)
                +\Biggl(
                        \frac{120721}{6750}
\nonumber\\ &&                
         +\frac{9086 L_1}{225}
                        +\frac{9086 L_2}{225}
                \Biggl) \zeta_2
                -\frac{5192 \zeta_3}{2025}
        \Biggl]
\nonumber\\ && 
        +\textcolor{blue}{C_F} \Biggl[
                -\frac{12930316237}{2187000000}
                +
                \frac{417741347 L_1}{24300000}
                +\frac{352861499 L_1^2}{17280000}
                -\frac{360041 L_1^3}{40500}
\nonumber\\ &&                
 +\frac{424203347 L_2}{24300000}
                +\frac{14360107 L_1 L_2}{576000}
                -\frac{33209 L_1^2 L_2}{6750}
                +\frac{352861499 L_2^2}{17280000}
\nonumber\\ &&                 
-\frac{70411 L_1 L_2^2}{6750}
                -\frac{285637 L_2^3}{40500}
                +\frac{25621379 L_\eta}{3037500}
                -\frac{98948977 L_\eta^2}{51840000}
\nonumber\\ &&                
                +\eta ^3 \Biggl(
                        \frac{23024568781}{44929695195}
                        -
                        \frac{359 L_1}{1350}
                        +\frac{359 L_1^2}{900}
                        +\frac{359 L_2}{1350}
\nonumber\\ &&                
         -\frac{359 L_1 L_2}{450}
                        +\frac{359 L_2^2}{900}
                        -\frac{742502683 L_\eta}{648336150}
                        +\frac{2772871 L_\eta^2}{1871100}
                \Biggl)
                +\eta ^2 \Biggl(
                        \frac{582667691}{75014100}
\nonumber\\ &&                
         -\frac{441313 L_1}{1728000}
                        +\frac{1819873 L_1^2}{6912000}
                        +\frac{441313 L_2}{1728000}
                        -\frac{1819873 L_1 L_2}{3456000}
                        +\frac{1819873 L_2^2}{6912000}
\nonumber\\ &&                
         -\frac{3876440473 L_\eta}{762048000}
                        +\frac{14002001 L_\eta^2}{5376000}
                \Biggl)
                +\eta  \Biggl(
                        -\frac{59657237}{4134375}
                        +\frac{2875369 L_1}{864000}
                        -\frac{55159 L_1^2}{72000}
\nonumber\\ &&                
                        -\frac{2875369 L_2}{864000}
         +\frac{55159 L_1 L_2}{36000}
                        -\frac{55159 L_2^2}{72000}
                        -\frac{40838437 L_\eta}{30240000}
                        +\frac{65917 L_\eta^2}{24000}
                \Biggl)
\nonumber\\ &&                
 +\Biggl(
                        \frac{6503111}{405000}
                        -\frac{70411 L_1}{4500}
                        -\frac{70411 L_2}{4500}
                \Biggl) \zeta_2
                +\frac{78397 \zeta_3}{10125}
        \Biggl]
\Biggl\}
\nonumber\\ && 
+\textcolor{blue}{T_F^3} \Biggl[
        -\frac{88 L_1
        ^3}{15}
        -\frac{176 L_1^2 L_2}{15}
        -\frac{176 L_1 L_2^2}{15}
        -\frac{88 L_2^3}{15}
        +\Biggl(
                -\frac{88 L_1}{5}
                -\frac{88 L_2}{5}
        \Biggl) \zeta_2
\nonumber\\ &&
        -\frac{352}{45} \zeta_3
\Biggl],
\\
\tilde{a}_{Qg}^{(3)} (N=6) &\propto& 
 \textcolor{blue}{T_F^2} \Biggl\{
        \textcolor{blue}{C_A} \Biggl[
                \frac{8699108665601}{73513818000}
                +\frac{499483057039 L_1}{7468070400}
                +\frac{2934742201991 L_1^2}{182078668800}
                +\frac{128557 L_1^3}{7938}
\nonumber\\ && 
                +\frac{499300102897 L_2}{7468070400}
                +\frac{3078895880569 L_1 L_2}{91039334400}
                +\frac{22165 L_1^2 L_2}{1323}
                +\frac{2934742201991 L_2
                ^2}{182078668800}
\nonumber\\ &&                
 +
                \frac{4433 L_1 L_2^2}{189}
                +\frac{110825 L_2^3}{7938}
                -\frac{154918449457 L_\eta}{7468070400}
                +\frac{949106952569 L_\eta^2}{182078668800}
\nonumber\\ &&      
                +\eta ^3 \Biggl(
                        -\frac{84840004938801319}{1381947564807810000}
                        +\frac{217742347 L_1}{4335206400}
                        -\frac{282650969 L_1^2}{5780275200}
                        -\frac{217742347 L_2}{4335206400}
\nonumber\\ &&           
         +\frac{282650969 L_1 
L_2}{2890137600}
                        -\frac{282650969 L_2^2}{5780275200}
                        +\frac{50096069455446961 
L_\eta}{251324552134656000}
\nonumber\\ &&      
                        -\frac{9178336044211 
L_\eta^2}{22317642547200}
                \Biggr)
          +\eta ^2 \Biggl(
                        \frac{755537213056}{624023544375}
                        +\frac{62549797 L_1}{4335206400}
    -\frac{398602763 L_1^2}{17340825600}
\nonumber\\ &&                     
                        -\frac{62549797 L_2}{4335206400}
                        +\frac{398602763 L_1 
L_2}{8670412800}
                        -\frac{398602763 L_2^2}{17340825600}
                        -\frac{6551048225287 
L_\eta}{23605198848000}
\nonumber\\ && 
                        -\frac{86215219591 L_\eta^2}{190749081600}
                \Biggr)
                +\eta  \Biggl(
                        \frac{832369820129}{29172150000}
                        +\frac{448073153 L_1}{867041280}
                        -\frac{2325070613 L_1^2}{17340825600}
\nonumber\\ && 
                        -\frac{448073153 L_2}{867041280}
                        +\frac{2325070613 L_1 L_2
                        }{8670412800}
                        -
                        \frac{2325070613 L_2^2}{17340825600}
                        -\frac{21862063585279 
L_\eta}{2275983360000}
\nonumber\\ &&
                        +\frac{19009228649 L_\eta^2}{86704128000}
                \Biggr)
                +\Biggl(
                        \frac{6117389}{277830}
                        +\frac{4433 L_1}{126}
                        +\frac{4433 L_2}{126}
                \Biggr) \zeta_2
\nonumber\\ &&                
 -\frac{8866 \zeta_3}{3969}
        \Biggr]
\nonumber\\ &&
        +\textcolor{blue}{C_F} \Biggl[
                -\frac{9883289655671}{428830605000}
                +\frac{275930321 L_1}{131712000}
                +\frac{25268241945419 L_1^2}{2549101363200}
                -\frac{1106501 L_1^3}{138915}
\nonumber\\ &&               
 +\frac{903763853 L_2}{395136000}
                +\frac{1959440025187 L_1 L_2}{182078668800}
                -\frac{43142 L_1^2 L_2}{9261}
                +\frac{25268241945419 L_2^2}{2549101363200}
\nonumber \\ &&                
 -\frac{439406 L_1 L_2^2}{46305}
                -\frac{25223 L_2^3}{3969}
                +\frac{4266955293521 L_\eta}{522764928000}
                -\frac{4852991775563 L_\eta^2}{2549101363200}
\nonumber\\ &&                   
 +\eta ^3 \Biggl(
                        \frac{990283034941336}{2467763508585375}
                        -\frac{12005277251 L_1}{40461926400}
                        +\frac{18378891457 L_1^2}{53949235200}
\nonumber\\ &&                
                        +\frac{12005277251 L_2}{40461926400}
         -\frac{18378891457 L_1 
L_2}{26974617600}
                        +\frac{18378891457 L_2^2}{53949235200}
\nonumber\\ &&                
         -\frac{174872727202100857 L_\eta
                        }{201059641707724800}
                        +
                        \frac{145121184651853 
L_\eta^2}{124978798264320}
                \Biggr)
                +\eta ^2 \Biggl(
                        \frac{524351089261}{97070329125}
\nonumber\\ &&               
         -\frac{99760601 L_1}{421478400}
                        +\frac{50312021201 L_1^2}{242771558400}
                        +\frac{99760601 L_2}{421478400}
                        -\frac{50312021201 L_1 
L_2}{121385779200}
\nonumber\\ && 
                        +\frac{50312021201 L_2^2}{242771558400}
                        -\frac{5027545889867 L_\eta}{1376969932800}
                        +\frac{212963348429 L_\eta^2}{106819485696}
                \Biggr)
\nonumber\\ &&                
 +\eta  \Biggl(
                        -\frac{32427817736}{2552563125}
                        +\frac{273911981947 L_1}{121385779200}
                        -\frac{251102611049 L_1^2}{485543116800}
\nonumber\\ &&                
                        -\frac{273911981947 L_2}{121385779200}
         +\frac{251102611049 L_1 
L_2}{242771558400}
                        -\frac{251102611049 L_2^2}{485543116800}
\nonumber\\ && 
                        -\frac{4936024379021 
L_\eta}{12745506816000}
                        +\frac{1040084747881 
L_\eta^2}{485543116800}
                \Biggr)
                +\Biggl(
                        \frac{29196929}{4862025}
                        -\frac{219703 L_1}{15435}
\nonumber\\ &&                
         -\frac{219703 L_2}{15435}
                \Biggr) \zeta_2
                +\frac{130108 \zeta_3}{19845}
        \Biggr]
\Biggr\}
+\textcolor{blue}{T_F^3} \Biggl[
        -\frac{88 L_1^3}{21}
        -\frac{176 L_1^2 L_2}{21}
        -\frac{176 L_1 L_2^2}{21}
\nonumber\\ &&  
      -\frac{88 L_2^3}{21}
        +\Biggl(
                -\frac{88 L_1}{7}
                -\frac{88 L_2}{7}
        \Biggr) \zeta_2
        -\frac{352}{63} \zeta_3
\Biggr].
\end{eqnarray}
The corresponding unexpanded quantities are given by
\begin{eqnarray}
\lefteqn{\tilde{a}_{Qg}^{(3)} (N=2) =} 
\nonumber\\ &&
 \textcolor{blue}{T_F^3} P_{29}
+\textcolor{blue}{T_F^2} \Biggl\{
        \textcolor{blue}{C_A} \Biggl[
                -
                \frac{28043}{2187}
                +\frac{1276 L_1^3}{81}
                +\frac{440 L_1^2 L_2}{27}
                +\frac{616 L_1 L_2^2}{27}
                +\frac{1100 L_2^3}{81}
                -\frac{161}{54 \xi ^2}
\nonumber\\ && 
                -\frac{161 \xi ^2}{54}
                +\frac{L_1^2 \big(
                        -161-8018 \xi ^2-161 \xi ^4\big)}{432 \xi ^2}
                +\frac{L_2^2 \big(
                        -161-8018 \xi ^2-161 \xi ^4\big)}{432 \xi ^2}
\nonumber\\ &&                
 +\frac{7 L_1 L_2 \big(
                        23-386 \xi ^2+23 \xi ^4\big)}{216 \xi ^2}
                +\frac{H_{1,0,0}(\xi ) P_1}{108 \xi ^3}
                +\frac{H_{-1,0,0}(\xi ) P_2}{108 \xi ^3}
\nonumber\\ &&                
 +L_2 \Bigg(
                        \frac{1449+7856 \xi ^2-1449 \xi ^4}{972 \xi ^2}
                        +\frac{308 \zeta_2}{9}
                \Bigg)
                +L_1 \Bigg(
                        \frac{-1449+7856 \xi ^2+1449 \xi ^4}{972 \xi ^2}
                        +\frac{308 \zeta_2}{9}
                \Bigg)
\nonumber\\ &&                
 -\frac{1340}{81} \zeta_2
                +\frac{248}{27} \zeta_3
                -\frac{920 \zeta_3}{81}
        \Biggr]
        +\textcolor{blue}{C_F} \Biggl[
                \frac{45478}{729}
                -\frac{992 L_1^3}{81}
                -\frac{64 L_1^2 L_2}{27}
                -\frac{320 L_1 L_2^2}{27}
\nonumber\\ &&                
 -\frac{736 L_2^3}{81}
                -\frac{409}{27 \xi ^2}
                -\frac{409 \xi ^2}{27}
                +\frac{L_1^2 \big(
                        -1227+33914 \xi ^2-1227 \xi ^4\big)}{648 \xi ^2}
\nonumber\\ &&                
 +
                \frac{L_2^2 \big(
                        -1227+33914 \xi ^2-1227 \xi ^4\big)}{648 \xi ^2}
                +\frac{L_1 L_2 \big(
                        1227+21382 \xi ^2+1227 \xi ^4\big)}{324 \xi ^2}
\nonumber\\ &&                
 +\frac{H_{-1,0,0}(\xi ) P_3}{54 \xi ^3}
                +\frac{H_{1,0,0}(\xi ) P_4}{54 \xi ^3}
                +L_2 \Bigg(
                        \frac{3681+27272 \xi ^2-3681 \xi ^4}{486 \xi ^2}
                        -\frac{160 \zeta_2}{9}
                \Bigg)
\nonumber\\ && 
                +L_1 \Bigg(
                        \frac{-3681+27272 \xi ^2+3681 \xi ^4}{486 \xi ^2}
                        -\frac{160 \zeta_2}{9}
                \Bigg)
                +\frac{416}{9} \zeta_2
                +\frac{1408}{81} \zeta_3
        \Biggr]
\Biggr\},
\\ 
\lefteqn{\tilde{a}_{Qg}^{(3)} (N=4) =}
\nonumber\\ &&
\frac{11 \textcolor{blue}{T_F^3} P_{29}}{20}
+\textcolor{blue}{T_F^2} \Biggl\{
        \textcolor{blue}{C_A} \Biggl[
                \frac{32864913953}{466560000}
                +\frac{37642 L_1^3}{2025}
                +\frac{2596 L_1^2 L_2}{135}
                +\frac{18172 L_1 L_2^2}{675}
                +\frac{1298 L_2^3}{81}
\nonumber\\ && 
                +\frac{5453}{153600 \xi ^4}
                -\frac{303011}{129600 \xi ^2}
                -\frac{303011 \xi ^2}{129600}
                +\frac{5453 \xi ^4}{153600}
                +\frac{H_{1,0,0}(\xi ) P_7}{2764800 \xi ^5}
                +\frac{H_{-1,0,0}(\xi ) P_8}{2764800 \xi ^5}
\nonumber\\ &&                 
+\frac{L_1 L_2 P_{13}}{82944000 (1-\xi^2)^2 \xi ^4}
                +
                \frac{L_1^2 P_{16}}{165888000 (1-\xi^2 )^2 \xi ^4 }
                +\frac{L_2^2 P_{16}}{165888000 (1-\xi^2 )^2 \xi ^4}
\nonumber\\ &&                 
+L_1 \Bigg(
                        \frac{P_6}{124416000 (\xi-1 ) \xi ^4 (1+\xi )}
                        +\frac{9086 \zeta_2}{225}
                \Bigg)
                +L_2 \Bigg(
                        \frac{P_{11}}{124416000 (\xi-1 ) \xi ^4 (1+\xi )}
                        +\frac{9086 \zeta_2}{225}
                \Bigg)
\nonumber\\ &&               
 +\frac{120721 \zeta_2}{6750}
                +\frac{6182}{675} \zeta_3
                -\frac{23738 \zeta_3}{2025}
        \Biggr]
        +\textcolor{blue}{C_F} \Biggl[
                \frac{10051995341}{874800000}
                -\frac{360041 L_1^3}{40500}
-\frac{33209 L_1^2 L_2}{6750}
\nonumber\\ &&                 
                -\frac{70411 L_1 L_2^2}{6750}
                -\frac{285637 L_2^3}{40500}
                -\frac{2023}{96000 \xi ^4}
                -\frac{2328341}{324000 \xi ^2}
                -\frac{2328341 \xi ^2}{324000}
                -\frac{2023 \xi ^4}{96000}
 +\frac{H_{-1,0,0}(\xi ) P_9}{1728000 \xi ^5}
\nonumber\\ &&               
                +\frac{H_{1,0,0}(\xi ) P_{10}}{1728000 \xi ^5}
                +\frac{L_1^2 P_{14}}{34560000 (1-\xi^2 )^2 \xi ^4}
                +\frac{L_2^2 P_{14}}{34560000 (1-\xi^2 )^2 \xi ^4}
                +\frac{L_1 L_2 P_{15}}{3456000 (1-\xi^2 )^2 \xi ^4}
\nonumber\\ &&               
 +L_2 \Bigg(
                        \frac{P_5}{388800000 (\xi^2-1 ) \xi ^4 }
                        -\frac{70411 \zeta_2}{4500}
                \Bigg)
                +L_1 \Bigg(
                        \frac{P_{12}}{388800000 (\xi^2-1 ) \xi ^4}
                        -\frac{70411 \zeta_2}{4500}
                \Bigg)
\nonumber\\ && 
                +\frac{6503111 \zeta_2}{405000}
+\frac{78397}{10125} \zeta_3
        \Biggr]
\Biggr\},
\\
\lefteqn{\tilde{a}_{Qg}^{(3)} (N=6) =}
\nonumber\\ && 
\frac{11 \textcolor{blue}{T_F^3} P_{29}}{28}
+\textcolor{blue}{T_F^2} \Biggl\{
        \textcolor{blue}{C_A} \Biggl[
                \frac{128557 L_1^3}{7938}
                +\frac{22165 L_1^2 L_2}{1323}
                +\frac{4433 L_1 L_2^2}{189}
                +\frac{110825 L_2^3}{7938}
\nonumber\\ &&                
 +\frac{23764414511324327}{301112598528000 (1-\xi^2)^2}
                -\frac{18297}{3211264 (1-\xi^2 )^2 \xi ^6}
                +\frac{503795}{9633792 (1-\xi^2 )^2 \xi ^4}
\nonumber\\ &&                
                -\frac{17392516271}{16257024000 (1-\xi^2)^2 \xi ^2}
 -\frac{4691336790174373 \xi ^2}{30111259852800 (1-\xi^2)^2}
                +\frac{23764414511324327 \xi ^4}{301112598528000 (1-\xi^2)^2}
\nonumber\\ &&                
 -\frac{17392516271 \xi ^6}{16257024000 (1-\xi^2)^2}
                +\frac{503795 \xi ^8}{9633792 (1-\xi^2)^2}
                -\frac{18297 \xi ^{10}
                }{3211264 (1-\xi^2 )^2}
                +
                \frac{H_{1,0,0}(\xi ) P_{19}}{4335206400 \xi ^7}
\nonumber\\ &&                
 +\frac{H_{-1,0,0}(\xi ) P_{20}}{4335206400 \xi ^7}
                +\frac{L_1^2 P_{26}}{364157337600 (1-\xi^2 )^4 \xi ^6 }
                +\frac{L_2^2 P_{26}}{364157337600 (1-\xi^2 )^4 \xi ^6 }
\nonumber\\ &&                
                +\frac{L_1 L_2 P_{28}}{182078668800 (1-\xi^2 )^4 \xi ^6 }
 +L_2 \Biggl(
                        \frac{P_{21}}{1911826022400 (\xi^2-1 )^3 \xi ^6 }
                        +\frac{4433 \zeta_2}{126}
                \Biggr)
\nonumber\\ && 
                +L_1 \Bigg(
                        \frac{P_{24}}{1911826022400 (\xi^2-1 )^3 \xi ^6}
                        +\frac{4433 \zeta_2}{126}
                \Bigg)
                +\frac{6117389 \zeta_2}{277830}
                +\frac{10252 \zeta_3}{1323}
                -\frac{39622 \zeta_3}{3969 (1-\xi^2 )^2}
\nonumber\\ &&           
                +\frac{79244 \xi ^2 \zeta_3}{3969 (1-\xi^2)^2}
      -\frac{39622 \xi ^4 \zeta_3}{3969 (1-\xi^2 )^2}
        \Biggr] 
\nonumber\\ && 
        +\textcolor{blue}{C_F} \Biggl[
                -\frac{1106501 L_1^3}{138915}
                -\frac{43142 L_1^2 L_2}{9261}
                -\frac{439406 L_1 L_2^2}{46305}
                -\frac{25223 L_2^3}{3969}
\nonumber\\ && 
                +\frac{6110987626095533}{1756490158080000 (1-\xi^2 )^2}
                -\frac{33885}{89915392 (1-\xi^2 )^2 \xi ^6}
                -
                \frac{189583}{56197120 (1-\xi^2 )^2 \xi ^4 }
\nonumber\\ &&                
 -\frac{74541151381}{15173222400 (1-\xi^2 )^2 \xi ^2}
                +\frac{20279035268810711 \xi ^2}{7025960632320000 (1-\xi^2 )^2}
                +\frac{6110987626095533 \xi ^4}{1756490158080000 (1-\xi^2 )^2}
\nonumber\\ &&                
 -\frac{74541151381 \xi ^6}{15173222400 (1-\xi^2 )^2}
                -\frac{189583 \xi ^8}{56197120 (1-\xi^2 )^2 }
                -\frac{33885 \xi ^{10}}{89915392 (1-\xi^2 )^2 }
                +\frac{H_{-1,0,0}(\xi ) P_{17}}{24277155840 \xi ^7}
\nonumber\\ &&             
    +\frac{H_{1,0,0}(\xi ) P_{18}}{24277155840 \xi ^7}
                +\frac{L_1^2 P_{25}}{10196405452800 (1-\xi^2 )^4 \xi ^6}
                +\frac{L_2^2 P_{25}}{10196405452800 (1-\xi^2 )^4 \xi ^6}
\nonumber\\ &&                
 +\frac{L_1 L_2 P_{27}}{728314675200 (1-\xi^2 )^4 \xi ^6}
                +L_2 \Bigg(
                        \frac{P_{22}}{606928896000 (\xi^2-1 )^3 \xi ^6 }
                        -\frac{219703 \zeta_2}{15435}
                \Bigg)
\nonumber\\ && 
                +L_1 \Bigg(
                        \frac{P_{23}}{606928896000 (\xi^2-1 )^3 \xi ^6}
                        -\frac{219703 \zeta_2}{15435}
                \Bigg)
                +\frac{29196929 \zeta_2}{4862025}
                -\frac{31196 \zeta_3}{6615}
                +\frac{223696 \zeta_3}{19845 (1-\xi^2 )^2}
\nonumber\\ &&               
 -\frac{447392 \xi ^2 \zeta_3}{19845 (1-\xi^2 )^2}
                +\frac{223696 \xi ^4 \zeta_3}{19845 (1-\xi^2 )^2}
        \Biggr]
\Biggr\}.
\end{eqnarray}
The polynomials $P_i$ are
\begin{eqnarray}
P_1&=&161 \xi ^6+2151 \xi ^4-5632 \xi ^3+2151 \xi ^2+161, \\
P_2&=&161 \xi ^6+2151 \xi ^4+5632 \xi ^3+2151 \xi ^2+161, \\
P_3&=&409 \xi ^6-753 \xi ^4-4096 \xi ^3-753 \xi ^2+409, \\
P_4&=&409 \xi ^6-753 \xi ^4+4096 \xi ^3-753 \xi ^2+409, \\
P_5&=&-4096575 \xi ^{10}-1393211475 \xi ^8+8081169602 \xi ^6-5389945502 \xi ^4-1393211475 \xi^2
\nonumber\\ &&
-4096575, \\
P_6&=&-2208465 \xi ^{10}+147490155 \xi ^8+8151380198 \xi ^6-8447761178 \xi ^4+147490155 \xi ^2
\nonumber\\ &&
-2208465, \\
P_7&=&-49077 \xi ^{10}+3233935 \xi ^8+68098470 \xi ^6-170131456 \xi ^5+68098470 \xi ^4
+3233935 \xi ^2
\nonumber\\ &&
-49077, \\
P_8&=&-49077 \xi ^{10}+3233935 \xi ^8+68098470 \xi ^6+170131456 \xi ^5+68098470 \xi ^4
+3233935 \xi ^2
\nonumber\\ &&
-49077,\\
P_9&=&18207 \xi ^{10}+6208235 \xi ^8-15266250 \xi ^6-76189696 \xi ^5-15266250 \xi ^4+6208235
\xi ^2
\nonumber\\ &&
+18207, \\
P_{10}&=&18207 \xi ^{10}+6208235 \xi ^8-15266250 \xi ^6+76189696 \xi ^5-15266250 \xi ^4
+6208235 \xi ^2 \nonumber\\ &&
+18207, \\
P_{11}&=&2208465 \xi ^{10}-147490155 \xi ^8+8447761178 \xi ^6-8151380198 \xi ^4-147490155 
\xi ^2
\nonumber\\ &&
+2208465, \\
P_{12}&=&4096575 \xi ^{10}+1393211475 \xi ^8+5389945502 \xi ^6-8081169602 \xi ^4+1393211475 
\xi ^2 \nonumber\\ &&
+4096575, \\
P_{13}&=&-736155 \xi ^{12}+49735950 \xi ^{10}+2422267163 \xi ^8-4938655516 \xi ^6+2422267163 
\xi ^4
\nonumber\\ && 
+49735950 \xi ^2
-736155, \\
P_{14}&=&-91035 \xi ^{12}-30889450 \xi ^{10}+767775003 \xi ^8-1468993836 \xi ^6
+767775003 \xi ^4
\nonumber\\ &&
-30889450 \xi ^2
-91035, \\
P_{15}&=&18207 \xi ^{12}+6177890 \xi ^{10}+73750241 \xi ^8-160811716 \xi ^6+73750241 \xi ^4
+6177890 \xi ^2
\nonumber\\ &&
+18207, \\
P_{16}&=&736155 \xi ^{12}-49735950 \xi ^{10}+2027991781 \xi ^8-3961862372 \xi ^6
+2027991781 \xi ^4 \nonumber\\ &&
-49735950 \xi ^2+736155, \\
P_{17}&=&4574475 \xi ^{14}+49929453 \xi ^{12}+59726658551 \xi ^{10}-204795402735 \xi ^8
-938249027584 \xi ^7
\nonumber\\ &&
-204795402735 \xi ^6+59726658551 \xi ^4
+49929453 \xi ^2+4574475, \\
P_{18}&=&4574475 \xi ^{14}+49929453 \xi ^{12}+59726658551 \xi ^{10}-204795402735 \xi ^8
+938249027584 \xi ^7
\nonumber\\ &&
-204795402735 \xi ^6+59726658551 \xi ^4
+49929453 \xi ^2+4574475, \\
P_{19}&=&12350475 \xi ^{14}-89110350 \xi ^{12}+2132547424 \xi ^{10}+89696346915 \xi ^8
-232416870400 \xi ^7
\nonumber\\ &&
+89696346915 \xi ^6+2132547424 \xi ^4-89110350 \xi ^2+12350475, \\
P_{20}&=&12350475 \xi ^{14}-89110350 \xi ^{12}+2132547424 \xi ^{10}+89696346915 \xi ^8
+232416870400 \xi ^7
\nonumber\\ &&
+89696346915 \xi ^6+2132547424 \xi ^4-89110350 \xi ^2+12350475, \\
P_{21}&=&-5446559475 \xi ^{18}+55032169500 \xi ^{16}-1068722020863 \xi ^{14}
+130798101242323 \xi ^{12}
\nonumber\\ &&
-385462593602301 \xi ^{10}+381602873228547 \xi ^8-124890387701293 \xi ^6
\nonumber\\ &&
-1068722020863 \xi ^4+55032169500 \xi ^2-5446559475, \\
P_{22}&=&-114361875 \xi ^{18}-917857575 \xi ^{16}-1489869987200 \xi ^{14}
+5747747644968 \xi ^{12}
\nonumber\\ &&
-6923569098094 \xi ^{10}+1055435494034 \xi ^8+3088079447592 \xi ^6-1489869987200 \xi ^4
\nonumber\\ &&
-917857575 \xi ^2-114361875, \\
P_{23}&=&114361875 \xi ^{18}+917857575 \xi ^{16}+1489869987200 \xi ^{14}-3088079447592 \xi ^{12}
\nonumber\\ &&
-1055435494034 \xi ^{10}
+6923569098094 \xi ^8-5747747644968 \xi ^6+1489869987200 \xi ^4
\nonumber\\ &&
+917857575 \xi ^2
+114361875, \\
P_{24}&=&5446559475 \xi ^{18}-55032169500 \xi ^{16}+1068722020863 \xi ^{14}+124890387701293 \xi ^{12}
\nonumber\\ &&
-381602873228547 \xi ^{10}+385462593602301 \xi ^8-130798101242323 \xi ^6+1068722020863 \xi ^4
\nonumber\\ &&
-55032169500 \xi ^2+5446559475, \\
P_{25}&=&-480319875 \xi ^{20}-3481419690 \xi ^{18}-6254413865175 \xi ^{16}
+126135043836776 \xi ^{14}
\nonumber\\ &&
-447182751937958 \xi ^{12}
+654730699180164 \xi ^{10}-447182751937958 \xi ^8
\nonumber\\ &&
+126135043836776 \xi ^6
-6254413865175 \xi ^4
-3481419690 \xi ^2-480319875, \\
P_{26}&=&-259359975 \xi ^{20}+2822303925 \xi ^{18}-52907211399 \xi ^{16}+6058077409828 \xi ^{14}
\nonumber\\ &&
-23799432487570 \xi ^{12}
+35582472943182 \xi ^{10}-23799432487570 \xi ^8+6058077409828 \xi ^6
\nonumber\\ &&
-52907211399 \xi ^4
+2822303925 \xi ^2-259359975, \\
P_{27}&=&68617125 \xi ^{20}+497345670 \xi ^{18}+893487695025 \xi ^{16}+4257463521448 \xi ^{14}
\nonumber\\ &&
-25223771715670 \xi ^{12}+40127575963044 \xi ^{10}-25223771715670 \xi ^8+4257463521448 \xi ^6
\nonumber\\ &&
+893487695025 \xi ^4+497345670 \xi ^2+68617125, \\
P_{28}&=&259359975 \xi ^{20}-2822303925 \xi ^{18}+52907211399 \xi ^{16}+5969198755292 \xi ^{14}
\nonumber\\ && 
-24309672172910 \xi ^{12}+36581184047538 \xi ^{10}-24309672172910 \xi ^8+5969198755292 \xi ^6
\nonumber\\ &&
+52907211399 \xi ^4-2822303925 \xi ^2+259359975, \\
P_{29}&=&-\frac{32 L_1^3}{3}-\frac{64 L_1^2 L_2}{3}-\frac{64 L_1 L_2^2}{3}-32 L_1 \zeta_2
-\frac{32 L_2^3}{3}-32 L_2 \zeta_2 -\frac{128 \zeta_3}{9}.
\end{eqnarray}


\noindent
The first expanded moments in the polarized case are
\begin{eqnarray}
\lefteqn{\Delta \tilde{a}_{Qg}^{(3)} (N=3) =} 
\nonumber\\ 
&&
\textcolor{blue}{T_F^2} \Biggl\{
        \textcolor{blue}{C_A} \Biggl[
                \frac{1377865}{17496}
                +\frac{311183 L_1}{7776}
                +\frac{73615 L_1^2}{13824}
                +\frac{1160 L_1^3}{81}
                +\frac{310913 L_2}{7776}
                +\frac{127345 L_1 L_2}{6912}
\nonumber\\ && 
                +\frac{400 L_1^2 L_2}{27}
                +\frac{73615 L_2^2}{13824}
                +\frac{560 L_1 L_2^2}{27}
                +\frac{1000 L_2^3}{81}
                -\frac{149369 L_\eta}{7776}
                +\frac{202835 L_\eta^2}{41472}
\nonumber\\ &&                
 +\eta ^3 \Bigg(
                        \frac{83626153}{843908625}
                        +\frac{5 L_1}{144}
                        -\frac{5 L_1^2}{96}
                        -\frac{5 L_2}{144}
                        +\frac{5 L_1 L_2}{48}
                        -\frac{5 L_2^2}{96}
                        -\frac{1276873 L_\eta}{42865200}
                        -\frac{45377 L_\eta^2}{272160}
                \Bigg)
\nonumber\\ &&               
 +\eta ^2 \Bigg(
                        \frac{35770738}{10418625}
                        +\frac{5 L_1}{144}
                        -\frac{5 L_1^2}{144}
                        -\frac{5 L_2}{144}
                        +\frac{5 L_1 L_2}{72}
                        -\frac{5 L_2^2}{144}
                        -\frac{2446699 L_\eta}{1587600}
                        +\frac{461 L_\eta^2}{15120}
                \Bigg)
\nonumber\\ &&                
                +\eta  \Bigg(
                        \frac{742837}{30375}
                        +\frac{11269 L_1}{6912}
                        -\frac{11509 L_1^2}{27648}
                        -\frac{11269 L_2}{6912}
                        +\frac{11509 L_1 L_2}{13824}
\nonumber\\ &&                
         -\frac{11509 L_2^2}{27648}
                        -\frac{4851337 L_\eta}{518400}
                        +\frac{68809 L_\eta^2}{138240}
                \Bigg)
                +\Bigg(
                        \frac{785}{81}
                        +\frac{280 L_1}{9}
                        +\frac{280 L_2}{9}
                \Bigg) \zeta_2
                -\frac{160}{81} \zeta_3
        \Bigg]
\nonumber\\ &&
        +\textcolor{blue}{C_F} \Bigg[
                \frac{463987}{23328}
                +\frac{124775 L_1}{3888}
                +\frac{216931 L_1^2}{10368}
                -\frac{1055 L_1^3}{162}
                +\frac{125315 L_2}{3888}
                +\frac{142349 L_1 L_2}{5184}
\nonumber\\ &&                
 -\frac{95 L_1^2 L_2}{27}
                +\frac{216931 L_2^2}{10368}
                -\frac{205 L_1 L_2^2}{27}
                -\frac{835 L_2^3}{162}
                +\frac{8233 L_\eta}{1944}
                -\frac{9523 L_\eta^2}{10368}
\nonumber\\ &&                
 +\eta ^3 \Bigg(
                        \frac{335165512}{843908625}
                        -\frac{5 L_1}{36}
                        +\frac{5 L_1^2}{24}
                        +\frac{5 L_2}{36}
                        -\frac{5 L_1 L_2}{12}
                        +\frac{5 L_2^2}{24}
                        -\frac{10119223 L_\eta}{10716300}
                        +\frac{85153 L_\eta^2}{68040}
                \Bigg)
\nonumber\\ && 
 +\eta ^2 \Bigg(
                        \frac{13141981}{2315250}
                        -\frac{5 L_1}{36}
                        +\frac{5 L_1^2}{36}
                        +\frac{5 L_2}{36}
                        -\frac{5 L_1 L_2}{18}
                        +\frac{5 L_2^2}{36}
                        -\frac{57787 L_\eta}{14700}
                        +\frac{2717 L_\eta^2}{1260}
                \Bigg)
\nonumber\\ &&                
 +\eta  \Bigg(
                        -\frac{21016}{3375}
                        +\frac{4123 L_1}{1728}
                        -\frac{3883 L_1^2}{6912}
                        -\frac{4123 L_2}{1728}
                        +\frac{3883 L_1 L_2}{3456}
                        -\frac{3883 L_2^2}{6912}
                        -\frac{139613 L_\eta}{43200}
\nonumber\\ &&                
        +\frac{93143 L_\eta^2}{34560}
                \Bigg)
                +\Bigg(
                        \frac{1919}{108}
                        -\frac{205 L_1}{18}
                        -\frac{205 L_2}{18}
                \Bigg) \zeta_2
                +\frac{470}{81} \zeta_3
        \Bigg]
\Bigg\}
\nonumber\\ && 
+\textcolor{blue}{T_F^3} \Bigg[
        -
        \frac{16 L_1^3}{3}
        -\frac{32 L_1^2 L_2}{3}
        -\frac{32 L_1 L_2^2}{3}
        -\frac{16 L_2^3}{3}
        +(-16 L_1
        -16 L_2
        ) \zeta_2
        -\frac{64}{9} \zeta_3
\Bigg],
\\
\lefteqn{\Delta \tilde{a}_{Qg}^{(3)} (N=5) =}
\nonumber\\
&&
 \textcolor{blue}{T_F^2} \Biggl\{
        \textcolor{blue}{C_A} \Biggl[
                \frac{2875905577}{27337500}
                +\frac{450503027 L_1}{7776000}
                +\frac{9047504867 L_1^2}{663552000}
                +\frac{30856 L_1^3}{2025}
                +\frac{450408077 L_2}{7776000}
\nonumber\\ && 
                +\frac{9835907101 L_1 L_2}{331776000}
                +\frac{2128 L_1^2 L_2}{135}
                +\frac{9047504867 L_2^2}{663552000}
                +\frac{14896 L_1 L_2^2}{675}
                +\frac{1064 L_2^3}{81}
                -\frac{757990081 L_\eta}{38880000}
\nonumber\\ &&                
 +\frac{3256879133 L_\eta^2}{663552000}
                +\eta ^2 \Bigg(
                        \frac{676581544}{468838125}
                        +\frac{153173 L_1}{22118400}
                        -\frac{66139 L_1^2}{5898240}
                        -\frac{153173 L_2}{22118400}
                        +\frac{66139 L_1 L_2}{2949120}
\nonumber\\ &&                
         -\frac{66139 L_2^2}{5898240}
                        -\frac{20502832607 L_\eta}{48771072000}
                        -\frac{237897901 L_\eta^2}{619315200}
                \Bigg)
                +\eta ^3 \Bigg(
                        -\frac{91188920849}{5616211899375}
                        +
                        \frac{1811 L_1}{86400}
\nonumber\\ &&                
         -\frac{1307 L_1^2}{57600}
                        -\frac{1811 L_2}{86400}
                        +\frac{1307 L_1 L_2}{28800}
                        -\frac{1307 L_2^2}{57600}
                        +\frac{33142148327 L_\eta}{207467568000}
                        -\frac{44268859 L_\eta^2}{119750400}
                \Bigg)
\nonumber\\ &&                    
                +\eta  \Bigg(
                        \frac{983886373}{37209375}
                        +\frac{7044007 L_1}{11059200}
                        -\frac{1788229 L_1^2}{11059200}
                        -\frac{7044007 L_2}{11059200}
         +\frac{1788229 L_1 L_2}{5529600}
\nonumber\\ &&                
                        -
                        \frac{1788229 L_2^2}{11059200}
                        -\frac{52251929221 L_\eta}{5806080000}
                        +\frac{11128409 L_\eta^2}{55296000}
                \Bigg)
                +\Bigg(
                        \frac{192092}{10125}
                        +\frac{7448 L_1}{225}
                        +\frac{7448 L_2}{225}
                \Bigg) \zeta_2
\nonumber\\ && 
                -\frac{4256 \zeta_3}{2025}
        \Bigg]
\nonumber\\ && 
        +\textcolor{blue}{C_F} \Bigg[
                -\frac{221066081}{22781250}
                +\frac{87833107 L_1}{8100000}
                +\frac{1227439897 L_1^2}{103680000}
                -\frac{74032 L_1^3}{10125}
                +\frac{88285357 L_2}{8100000}
\nonumber\\ &&                
 +\frac{146865403 L_1 L_2}{10368000}
                -\frac{14336 L_1^2 L_2}{3375}
                +\frac{1227439897 L_2^2}{103680000}
                -\frac{29344 L_1 L_2^2}{3375}
                -\frac{59024 L_2^3}{10125}
\nonumber\\ && 
                +\frac{159295001 L_\eta}{24300000}
 -\frac{157462297 L_\eta^2}{103680000}        
                +\eta ^3 \Bigg(
                        \frac{402156040976}{1123242379875}
                        -\frac{109 L_1}{1200}
                        +\frac{81 L_1^2}{800}
                        +\frac{109 L_2}{1200}
\nonumber\\ &&                
         -\frac{81 L_1 L_2}{400}
                        +\frac{81 L_2^2}{800}
                        -\frac{4378699309 L_\eta}{5186689200}
                        +\frac{17063353 L_\eta^2}{14968800}
                \Bigg)
                +\eta ^2 \Bigg(
                        \frac{151435948}{31255875}
                        -\frac{82681 L_1}{1152000}
\nonumber\\ &&                
         +\frac{844843 L_1^2}{13824000}
                        +\frac{82681 L_2}{1152000}
                        -\frac{844843 L_1 L_2}{6912000}
                        +\frac{844843 L_2^2}{13824000}
                        -\frac{1728724721 L_\eta}{508032000}
                        +\frac{187253459 L_\eta^2}{96768000}
                \Bigg)
\nonumber\\ &&                
 +\eta  \Bigg(
                        -\frac{672368}{70875}
                        +\frac{1816457 L_1}{864000}
                        -\frac{1179509 L_1^2}{2304000}
                        -\frac{1816457 L_2}{864000}
                        +\frac{1179509 L_1 L_2}{1152000}
                        -
                        \frac{1179509 L_2^2}{2304000}
\nonumber\\ &&                 
         -\frac{1235063 L_\eta}{864000}
                        +\frac{1758503 L_\eta^2}{768000}
                \Bigg)
                +\Bigg(
                        \frac{422596}{50625}
                        -\frac{14672 L_1}{1125}
                        -\frac{14672 L_2}{1125}
                \Bigg) \zeta_2
                +\frac{61376 \zeta_3}{10125}
        \Biggr]
\Biggr\}
\nonumber\\ && 
+\textcolor{blue}{T_F^3} \Biggl[
        -\frac{64 L_1^3}{15}
        -\frac{128 L_1^2 L_2}{15}
        -\frac{128 L_1 L_2^2}{15}
        -\frac{64 L_2^3}{15}
        +\Bigg(
                -\frac{64 L_1}{5}
                -\frac{64 L_2}{5}
        \Bigg) \zeta_2
        -\frac{256}{45} \zeta_3
\Biggr],
\\
\lefteqn{\Delta \tilde{a}_{Qg}^{(3)} (N=7) =} 
\nonumber\\ 
&&
\textcolor{blue}{T_F^2} \Biggl\{
        \textcolor{blue}{C_A} \Biggl[
                \frac{23334644017021}{217818720000}
                +\frac{64027714809077 L_1}{1062125568000}
                +\frac{125636708578021 L_1^2}{7768689868800}
                +\frac{10382 L_1^3}{735}
\nonumber\\ &&              
  +\frac{64003435143947 L_2}{1062125568000}
                +\frac{124025409719579 L_1 
L_2}{3884344934400}
                +\frac{716 L_1^2 L_2}{49}
                +\frac{125636708578021 L_2^2}{7768689868800}
\nonumber\\ &&                
 +\frac{716 L_1 L_2^2}{35}
                +\frac{1790 L_2^3}{147}
                -\frac{56215293776161 L_\eta}{3186376704000}
                +
                \frac{34360837098779 L_\eta^2}{7768689868800}
\nonumber\\ &&                
 +\eta ^2 \Bigg(
                        \frac{14352439212829}{23296878990000}
                        +\frac{1592793023 L_1}{123312537600}
                        -\frac{613388477 L_1^2}{29595009024}
                        -\frac{1592793023 L_2}{123312537600}
\nonumber\\ &&               
         +\frac{613388477 L_1 
L_2}{14797504512}
                        -\frac{613388477 L_2^2}{29595009024}
                        +\frac{334835304263447 
L_\eta}{14100172111872000}
                        -\frac{4130848867577 
L_\eta^2}{8138627481600}
                \Bigg)
\nonumber\\ && 
                +\eta ^3 \Bigg(
                        -\frac{14624127854214079}{230324594134635000}
                        +\frac{7986605831 L_1}{246625075200}
                        -\frac{2535517577 L_1^2}{65766686720}
                        -\frac{7986605831 L_2}{246625075200}
\nonumber\\ &&                        
+\frac{2535517577 L_1 
L_2}{32883343360}
                        -\frac{2535517577 L_2^2}{65766686720}
                        +\frac{9420751076138605999 
L_\eta}{42892723564314624000}
\nonumber\\ && 
       -\frac{1585041603186541 L_\eta^2}{3808877661388800}
                \Bigg)
                +\eta  \Bigg(
                        \frac{166013524319}{6806835000}
                        +\frac{188096578493 L_1}{739875225600}
\nonumber\\ &&                
         -\frac{202758388859 L_1^2}{2959500902400}
                        -\frac{188096578493 L_2}{739875225600}
                        +\frac{202758388859 L_1 
L_2}{1479750451200}
                        -\frac{202758388859 L_2^2}{2959500902400}
\nonumber\\ &&                
         -\frac{4292717431702621 
L_\eta}{543808290816000}
                        +\frac{800363669341 
L_\eta^2}{20716506316800}
                \Bigg)
                +\Bigg(
                        \frac{529103}{24696}
                        +\frac{1074 L_1}{35}
                        +\frac{1074 L_2}{35}
                \Bigg) \zeta_2
                -\frac{1432}{735} \zeta_3
        \Bigg]
\nonumber\\ && 
        +\textcolor{blue}{C_F} \Biggl[
                -\frac{292368887666357}{12197848320000}
                -\frac{1944703337 L_1}{2458624000}
                +\frac{167874076860551 L_1^2}{27190414540800}
                -\frac{115915 L_1^3}{16464}
\nonumber\\ &&                
 -\frac{15058164463 L_2}{22127616000}
                +\frac{11776150078591 L_1 
L_2}{1942172467200}
                -\frac{57107 L_1^2 L_2}{13720}
                +\frac{167874076860551 L_2^2}{27190414540800}
\nonumber\\ &&            
     -\frac{115429 L_1 L_2^2}{13720}
                -\frac{66133 L_2^3}{11760}
                +\frac{2005296301549 L_\eta}{278807961600}
                -\frac{9248094576667 L_\eta^2}{5438082908160}
\nonumber\\ &&                
 +\eta ^3 \Bigg(
                        \frac{787942400862074}{2617324933348125}
                        -\frac{18467618651 L_1}{129478164480}
                        +\frac{156484147949 L_1^2}{863187763200}
                        +\frac{18467618651 L_2}{129478164480}
\nonumber\\ &&                     
    -\frac{156484147949 L_1 
L_2}{431593881600}
                        +\frac{156484147949 L_2
                        ^2}{863187763200}
                        -
                        \frac{348761937329434169 
L_\eta}{487417313230848000}
\nonumber\\ && 
                        +\frac{296565060769037 
L_\eta^2}{302978904883200}
                \Bigg)
                +\eta ^2 \Bigg(
                        \frac{2256072641521}{575231580000}
                        -\frac{16002091283 L_1}{129478164480}
                        +\frac{297330438653 L_1^2}{2589563289600}
\nonumber\\ &&                 
        +\frac{16002091283 L_2}{129478164480}
                        -\frac{297330438653 L_1 
L_2}{1294781644800}
                        +\frac{297330438653 L_2^2}{2589563289600}
                        -\frac{156623887360933 
L_\eta}{55953063936000}
\nonumber\\ && 
                        +\frac{46763955788321 
L_\eta^2}{28485196185600}
                \Bigg)
                +\eta  \Bigg(
                        -\frac{119203663}{11576250}
                        +\frac{1062066777019 L_1}{647390822400}
                        -\frac{991362614047 L_1^2}{2589563289600}
\nonumber\\ &&                
         -\frac{1062066777019 L_2}{647390822400}
                        +\frac{991362614047 L_1 
L_2}{1294781644800}
                        -\frac{991362614047 L_2^2}{2589563289600}
                        -\frac{392516586611 L_\eta}{1078984704000}
\nonumber\\ &&                
         +\frac{4858443793183 
L_\eta^2}{2589563289600}
                \Bigg)
                +\Bigg(
                        \frac{31215727}{11524800}
                        -\frac{346287 L_1}{27440}
                        -\frac{346287 L_2}{27440}
                \Bigg) \zeta_2
                +\frac{16837 \zeta_3}{2940}
        \Biggr]
\Biggr\}
\nonumber\\ && 
+\textcolor{blue}{T_F^3} \Biggl[
        -\frac{24 L_1^3}{7}
        -\frac{48 L_1^2 L_2}{7}
        -\frac{48 L_1 L_2^2}{7}
        -\frac{24 L_2^3}{7}
        +\Bigg(
                -\frac{72 L_1}{7}
                -\frac{72 L_2}{7}
        \Bigg) \zeta_2
        -\frac{32}{7} \zeta_3
\Biggr].
\end{eqnarray}

The first unexpanded moments read
\begin{eqnarray}
\lefteqn{\Delta \tilde{a}_{Qg}^{(3)} (N=3) =}
\nonumber\\ &&
 \textcolor{blue}{T_F^3} Q_{37}
+\textcolor{blue}{T_F^2} \Bigg\{
        \textcolor{blue}{C_A} \Bigg[
                \frac{5678723}{139968}
                +\frac{1160 L_1^3}{81}
                +\frac{400 L_1^2 L_2}{27}
                +\frac{560 L_1 L_2^2}{27}
                +\frac{1000 L_2^3}{81}
                -\frac{11029}{3456 \xi ^2}
\nonumber\\ && 
                -\frac{11029 \xi ^2}{3456}
                +\frac{L_2 Q_1}{62208 (\xi^2-1 ) \xi ^2 }
                +\frac{H_{1,0,0}(\xi ) Q_5}{6912 \xi ^3}
                +\frac{H_{-1,0,0}(\xi ) Q_6}{6912 \xi ^3}
                +\frac{L_1 Q_8}{62208 (\xi^2-1 ) \xi ^2}
\nonumber\\ &&               
 +\frac{L_1^2 Q_{10}}{27648 (\xi^2-1 )^2 \xi ^2}
                +\frac{L_2^2 Q_{10}}{27648 (\xi^2-1 )^2 \xi ^2 }
                +\frac{L_1 L_2 Q_{11}}{13824 (\xi^2-1 )^2 \xi ^2 }
                +\Bigg(
                        \frac{785}{81}
                        +\frac{280 L_1}{9}
\nonumber\\ &&                    
     +\frac{280 L_2}{9}
                \Bigg) \zeta_2
                -\frac{160}{81} \zeta_3
        \Bigg]
        +\textcolor{blue}{C_F} \Bigg[
                \frac{223435}{7776}
                -\frac{1055 L_1^3}{162}
                -\frac{95 L_1^2 L_2}{27}
                -\frac{205 L_1 L_2^2}{27}
                -\frac{835 L_2^3}{162}
\nonumber\\ &&                
 -\frac{4363}{864 \xi ^2}
                -\frac{4363 \xi ^2}{864}
                +\frac{L_2 Q_2}{15552 (\xi^2-1 ) \xi ^2}
                +\frac{H_{-1,0,0}
                (\xi ) Q_3}{1728 \xi ^3}
                +
                \frac{H_{1,0,0}(\xi ) Q_4}{1728 \xi ^3}
                +\frac{L_1 Q_7}{15552 (\xi^2-1 ) \xi ^2}
\nonumber\\ &&                
 +\frac{L_1^2 Q_9}{20736 (\xi^2-1 )^2 \xi ^2 }
                +\frac{L_2^2 Q_9}{20736 (\xi^2-1 )^2 \xi ^2 }
                +\frac{L_1 L_2 Q_{12}}{10368 (\xi^2-1 )^2 \xi ^2 }
                +\Bigg(
                        \frac{1919}{108}
                        -\frac{205 L_1}{18}
\nonumber\\ &&                
         -\frac{205 L_2}{18}
                \Bigg) \zeta_2
                +\frac{470}{81} \zeta_3
        \Bigg]
\Bigg\}
\\
\lefteqn{\Delta \tilde{a}_{Qg}^{(3)} (N=5) =}
\nonumber\\ &&
\frac{4 \textcolor{blue}{T_F^3} Q_{37}}{5}
+\textcolor{blue}{T_F^2} \Bigg\{
        \textcolor{blue}{C_A} \Bigg[
                \frac{30856 L_1^3}{2025}
                +\frac{2128 L_1^2 L_2}{135}
                +\frac{14896 L_1 L_2^2}{675}
                +\frac{1064 L_2^3}{81}
\nonumber\\ && 
                +\frac{1541144309981}{22394880000 (\xi^2-1 )^2}
                +\frac{22547}{1228800 (\xi^2-1 )^2 \xi ^4 }
                -\frac{21325961}{16588800 (\xi^2-1 )^2 \xi ^2 }
\nonumber\\ &&                
 -\frac{756388034053 \xi ^2}{5598720000 (\xi^2-1 )^2 }
                +\frac{1541144309981 \xi ^4}{22394880000 (\xi^2-1 )^2 }
                -\frac{21325961 \xi ^6}{16588800 (\xi^2-1 )^2 }
\nonumber\\ &&           
     +\frac{22547 \xi ^8}{1228800 (\xi^2-1 )^2 }
                +
                \frac{H_{1,0,0}(\xi ) Q_{13}}{22118400 \xi ^5}
                +\frac{H_{-1,0,0}(\xi ) Q_{14}}{22118400 \xi ^5}
                +\frac{L_1 Q_{17}}{995328000 (\xi^2-1 )^3 \xi ^4 }
\nonumber\\ &&                
 +\frac{L_2 Q_{22}}{995328000 (\xi^2-1 )^3 \xi ^4 }
                +\frac{L_1 L_2 Q_{25}}{663552000 (\xi^2-1 )^4 \xi ^4 }
                +\frac{L_1^2 Q_{28}}{1327104000 (\xi^2-1 )^4 \xi ^4 }
\nonumber\\ &&                
 +\frac{L_2^2 Q_{28}}{1327104000 (\xi^2-1 )^4 \xi ^4 }
                +\Bigg(
                        \frac{192092}{10125}
                        +\frac{7448 L_1}{225}
                        +\frac{7448 L_2}{225}
                \Bigg) \zeta_2
                +\Bigg(
                        \frac{4976}{675}
\nonumber\\ && 
         -\frac{19184}{2025 (\xi^2-1 )^2 }
                        +\frac{38368 \xi ^2}{2025 (\xi^2 -1)^2 }
                        -\frac{19184 \xi ^4}{2025 (\xi^2-1 )^2 }
                \Bigg) \zeta_3
        \Bigg]
\nonumber\\ && 
        +\textcolor{blue}{C_F} \Bigg[
                -\frac{74032 L_1^3}{10125}
                -\frac{14336 L_1^2 L_2}{3375}
                -\frac{29344 L_1 L_2^2}{3375}
                -\frac{59024 L_2^3}{10125}
                +\frac{4811953349}{388800000 (\xi^2-1 )^2 }
\nonumber\\ &&                
 +\frac{49}{64000 (\xi^2-1 )^2 \xi ^4 }
                -\frac{1868743}{432000 (\xi^2-1 )^2 \xi ^2 }
                -
                \frac{782406581 \xi ^2}{48600000 (\xi^2-1 )^2 }
\nonumber\\ &&                
 +\frac{4811953349 \xi ^4}{388800000 (\xi^2-1 )^2 }
                -\frac{1868743 \xi ^6}{432000 (\xi^2-1 )^2 }
                +\frac{49 \xi ^8}{64000 (\xi^2-1 )^2 }
                +\frac{H_{-1,0,0}(\xi ) Q_{15}}{3456000 \xi ^5}
\nonumber\\ &&                
 +\frac{H_{1,0,0}(\xi ) Q_{16}}{3456000 \xi ^5}
                +\frac{L_1 Q_{20}}{259200000 (\xi^2-1 )^3 \xi ^4 }
                +\frac{L_2 Q_{21}}{259200000 (\xi^2-1 )^3 \xi ^4 }
\nonumber\\ &&            
     +\frac{L_1 L_2 Q_{26}}{20736000 (\xi^2-1 )^4 \xi ^4 }
                +\frac{L_1^2 Q_{27}}{207360000 (\xi^2-1 )^4 \xi ^4 }
                +\frac{L_2^2 Q_{27}}{207360000 (\xi^2-1 )^4 \xi ^4 }
\nonumber\\ &&                
 +\Bigg(
                        \frac{422596}{50625}
                        -\frac{14672 L_1}{1125}
                        -\frac{14672 L_2}{1125}
                \Bigg) \zeta_2
                +\Bigg(
                        -\frac{2912}{675}
                        +\frac{105056}{10125 (\xi^2-1 )^2 }
\nonumber\\ &&                
         -\frac{210112 \xi ^2}{10125 (\xi^2-1 )^2 }
                        +\frac{105056 \xi ^4}{10125 (\xi^2-1 )^2 }
                \Bigg) \zeta_3
        \Bigg]
\Bigg\}
\\
\lefteqn{\Delta \tilde{a}_{Qg}^{(3)} (N=7) =}
\nonumber\\ &&
\frac{9 \textcolor{blue}{T_F^3} Q_{37}}{14}
+\textcolor{blue}{T_F^2} \Bigg\{
        \textcolor{blue}{C_A}
         \Bigg[
                \frac{10382 L_1^3}{735}
                +\frac{716 L_1^2 L_2}{49}
                +\frac{716 L_1 L_2^2}{35}
                +\frac{1790 L_2^3}{147}
\nonumber\\ && 
 +\frac{395611716643088909}{5353112862720000 (\xi^2-1 )^4 }
                -\frac{4958901}{1644167168 (\xi^2-1 )^4 \xi ^6 }
                +\frac{40076219}{1027604480 (\xi^2-1 )^4 \xi ^4 }
\nonumber\\ &&                
 -\frac{3267395061527}{5549064192000 (\xi^2-1 )^4 \xi ^2 }
                -\frac{12456159186762222973 \xi ^2}{42824902901760000 (\xi^2-1 )^4 }
\nonumber\\ &&                
 +\frac{776244888381666553 \xi ^4}{1784370954240000 (\xi^2-1 )^4}
                -\frac{12456159186762222973 \xi ^6}{42824902901760000 (\xi^2-1 )^4 }
\nonumber\\ &&               
 +\frac{395611716643088909 \xi ^8}{5353112862720000 (\xi^2-1 )^4 }
                -\frac{3267395061527 \xi ^{10}}{5549064192000 (\xi^2-1 )^4 }
  +\frac{40076219 \xi ^{12}}{1027604480 (\xi^2-1 )^4 }
\nonumber\\ && 
                -\frac{4958901 \xi ^{14}}{1644167168 (\xi^2-1 )^4 }
                +\frac{H_{1,0,0}(\xi ) Q_{23}}{147975045120 \xi ^7}
                +\frac{H_{-1,0,0}(\xi ) Q_{24}}{147975045120 \xi ^7}
\nonumber\\ &&                
 +\frac{L_2 Q_{29}
                }{543808290816000 (\xi^2-1 )^5 \xi ^6 }
                +
                \frac{L_1 Q_{32}}{543808290816000 (\xi^2-1 )^5 \xi ^6 }
\nonumber\\ &&                
 +\frac{L_1^2 Q_{33}}{62149518950400 (\xi^2-1 )^6 \xi ^6 }
                +\frac{L_2^2 Q_{33}}{62149518950400 (\xi^2-1 )^6 \xi ^6 }
\nonumber\\ &&                
 +\frac{L_1 L_2 Q_{36}}{31074759475200 (\xi^2-1 )^6 \xi ^6 }
                +\Bigg(
                        \frac{529103}{24696}
                        +\frac{1074 L_1}{35}
                        +\frac{1074 L_2}{35}
                \Bigg) \zeta_2
\nonumber\\ &&        
 +\Bigg(
                        \frac{1642}{245}
                        -\frac{6358}{735 (\xi^2-1 )^4 }
                        +\frac{25432 \xi ^2}{735 (\xi^2-1 )^4 }
                        -\frac{12716 \xi ^4}{245 (\xi^2-1 )^4 }
                        +\frac{25432 \xi ^6}{735 (\xi^2-1 )^4 }
\nonumber\\ &&                
         -\frac{6358 \xi ^8}{735 (\xi^2-1 )^4 }
                \Bigg) \zeta_3
        \Bigg]
        +\textcolor{blue}{C_F} \Bigg[
                -\frac{115915 L_1^3}{16464}
                -\frac{57107 L_1^2 L_2}{13720}
                -\frac{115429 L_1 L_2^2}{13720}
\nonumber\\ &&                
 -\frac{66133 L_2^3}{11760}
                +\frac{88119076890915193}{18735895019520000 (\xi^2-1 )^4 }
                +\frac{30375}{1438646272 (\xi^2-1 )^4 \xi ^6 }
\nonumber\\ &&                
 -\frac{1750311}{1438646272 (\xi^2-1 )^4 \xi ^4 }
                -
                \frac{226840183453}{64739082240 (\xi^2-1 )^4 \xi ^2 }
                +\frac{477564512027081881 \xi ^2}{37471790039040000 (\xi^2-1 )^4 }
\nonumber\\ &&             
 -\frac{348302141626761803 \xi ^4}{12490596679680000 (\xi^2-1 )^4 }
                +\frac{477564512027081881 \xi ^6}{37471790039040000 (\xi^2-1 )^4 }
\nonumber\\ &&                
 +\frac{88119076890915193 \xi ^8}{18735895019520000 (\xi^2-1 )^4 }
                -\frac{226840183453 \xi ^{10}}{64739082240 (\xi^2-1 )^4 }
                -\frac{1750311 \xi ^{12}}{1438646272 (\xi^2-1 )^4 }
\nonumber\\ &&                
 +\frac{30375 \xi ^{14}}{1438646272 (\xi^2-1 )^4 }
                +\frac{H_{-1,0,0}(\xi ) Q_{18}}{129478164480 \xi ^7}
                +\frac{H_{1,0,0}(\xi ) Q_{19}}{129478164480 \xi ^7}
\nonumber\\ &&               
  +\frac{L_1 Q_{30}}{22658678784000 (\xi^2-1 )^5 \xi ^6 }
                +\frac{L_2 Q_{31}}{22658678784000 (\xi^2-1 )^5 \xi ^6 }
\nonumber\\ &&                
 +\frac{L_1 L_2 Q_{34}}{3884344934400 (\xi^2-1 )^6 \xi ^6 }
                +\frac{L_1^2 Q_{35}}{54380829081600 (\xi^2-1 )^6 \xi ^6 }
\nonumber\\ &&               
  +\frac{L_2^2 Q_{35}}{54380829081600 (\xi^2-1 )^6 \xi ^6 }
                +\Bigg(
                        \frac{31215727}{11524800}
                        -\frac{346287 L_1}{27440}
                        -\frac{346287 L_2}{27440}
                \Bigg) \zeta_2
\nonumber\\ &&                
 +\Bigg(
                        -\frac{1027}{245}
                        +\frac{29161}{2940 (\xi^2-1 )^4 }
                        -\frac{29161 \xi ^2}{735 (\xi^2 -1 )^4 }
                        +\frac{29161 \xi ^4}{490 (\xi^2-1 )^4 }
                        -\frac{29161 \xi ^6}{735 (\xi^2-1 )^4 }
\nonumber\\ &&                
         +\frac{29161 \xi ^8}{2940 (\xi^2-1 )^4 }
                \Bigg) \zeta_3
        \Bigg]
\Bigg\},
\end{eqnarray}
with the polynomials $Q_i$
\begin{eqnarray}
Q_1&=&-99261 \xi ^6+2588725 \xi ^4-2388043 \xi ^2-99261, \\
Q_2&=&-39267 \xi ^6+538367 \xi ^4-461993 \xi ^2-39267, \\
Q_3&=&4363 \xi ^6-7803 \xi ^4-56320 \xi ^3-7803 \xi ^2+4363, \\
Q_4&=&4363 \xi ^6-7803 \xi ^4+56320 \xi ^3-7803 \xi ^2+4363, \\
Q_5&=&11029 \xi ^6+131547 \xi ^4-327680 \xi ^3+131547 \xi ^2+11029, \\
Q_6&=&11029 \xi ^6+131547 \xi ^4+327680 \xi ^3+131547 \xi ^2+11029, \\
Q_7&=&39267 \xi ^6+461993 \xi ^4-538367 \xi ^2+39267, \\
Q_8&=&99261 \xi ^6+2388043 \xi ^4-2588725 \xi ^2+99261, \\
Q_9&=&-13089 \xi ^8+460040 \xi ^6-892462 \xi ^4+460040 \xi ^2-13089, \\
Q_{10}&=&-11029 \xi ^8+169288 \xi ^6-316998 \xi ^4+169288 \xi ^2-11029, \\
Q_{11}&=&11029 \xi ^8+232632 \xi ^6-486842 \xi ^4+232632 \xi ^2+11029, \\
Q_{12}&=&13089 \xi ^8+258520 \xi ^6-544658 \xi ^4+258520 \xi ^2+13089, \\
Q_{13}&=&-202923 \xi ^{10}+13818977 \xi ^8+429684810 \xi ^6-1115684864 \xi ^5
+429684810 \xi ^4
\nonumber\\ &&
+13818977 \xi ^2-202923, \\
Q_{14}&=&-202923 \xi ^{10}+13818977 \xi ^8+429684810 \xi ^6+1115684864 \xi ^5
+429684810 \xi ^4
\nonumber\\ &&
+13818977 \xi ^2-202923, \\
Q_{15}&=&-1323 \xi ^{10}+7472375 \xi ^8-23485500 \xi ^6-122945536 \xi ^5-23485500 \xi ^4
\nonumber\\ &&
+7472375 \xi ^2-1323, \\
Q_{16}&=&-1323 \xi ^{10}+7472375 \xi ^8-23485500 \xi ^6+122945536 \xi ^5-23485500 
\xi ^4
\nonumber\\ &&
+7472375 \xi ^2-1323, \\
Q_{17}&=&-9131535 \xi ^{14}+648233955 \xi ^{12}+55762321201 \xi ^{10}
-171719012613 \xi ^8
\nonumber\\ &&
+174230851323 \xi ^6
-59554300111 \xi ^4
+648233955 \xi ^2-9131535, \\
Q_{18}&=&-1366875 \xi ^{14}+73347120 \xi ^{12}+227138865058 \xi ^{10}
-956498341239 \xi ^8
\nonumber\\ &&
-4403163561984 \xi ^7-956498341239 \xi ^6
+227138865058 \xi ^4+73347120 \xi ^2
\nonumber\\ &&
-1366875, \\
Q_{19}&=&-1366875 \xi ^{14}+73347120 \xi ^{12}+227138865058 \xi ^{10}-956498341239 \xi ^8+4403163561984 \xi ^7
\nonumber\\ &&
-956498341239 \xi ^6+227138865058 \xi ^4+73347120 \xi ^2-1366875, \\
Q_{20}&=&-99225 \xi ^{14}+560714775 \xi ^{12}+1143582449 \xi ^{10}-7338980847 \xi ^8
+9568391697 \xi ^6
\nonumber\\ &&
-4492208399 \xi ^4+560714775 \xi ^2-99225, \\
Q_{21}&=&99225 \xi ^{14}-560714775 \xi ^{12}+4492208399 \xi ^{10}-9568391697 \xi ^8+7338980847 \xi ^6
\nonumber\\ &&
-1143582449 \xi ^4-560714775 \xi ^2+99225, \\
Q_{22}&=&9131535 \xi ^{14}-648233955 \xi ^{12}+59554300111 \xi ^{10}-174230851323 \xi ^8
+171719012613 \xi ^6
\nonumber\\ && 
-55762321201 \xi ^4-648233955 \xi ^2+9131535, \\
Q_{23}&=&223150545 \xi ^{14}-2001150423 \xi ^{12}+34327153343 \xi ^{10}+2606894507847 \xi ^8
\nonumber\\ &&
-6919192313856 \xi ^7+2606894507847 \xi ^6+34327153343 \xi ^4-2001150423 \xi ^2
\nonumber\\ &&
+223150545, \\
Q_{24}&=&223150545 \xi ^{14}-2001150423 \xi ^{12}+34327153343 \xi ^{10}+2606894507847 \xi ^8
\nonumber\\ && 
+6919192313856 \xi ^7
+2606894507847 \xi ^6+34327153343 \xi ^4-2001150423 \xi ^2
\nonumber\\ &&
+223150545, \\
Q_{25}&=&-3043845 \xi ^{16}+218445420 \xi ^{14}+18828470972 \xi ^{12}-77222873708 \xi ^{10}
\nonumber\\ &&
+116359292562 \xi ^8-77222873708 \xi ^6+18828470972 \xi ^4+218445420 \xi ^2-3043845, \\
Q_{26}&=&-3969 \xi ^{16}+22431678 \xi ^{14}+204043784 \xi ^{12}-1019181374 \xi ^{10}
+1585258482 \xi ^8
\nonumber\\ &&
-1019181374 \xi ^6+204043784 \xi ^4+22431678 \xi ^2-3969, \\
Q_{27}&=&19845 \xi ^{16}-112158390 \xi ^{14}+2903314904 \xi ^{12}-10598228426 \xi ^{10}
+15614910534 \xi ^8
\nonumber\\ &&
-10598228426 \xi ^6+2903314904 \xi ^4-112158390 \xi ^2+19845, \\
Q_{28}&=&3043845 \xi ^{16}-218445420 \xi ^{14}+18938352964 \xi ^{12}-73844422036 \xi 
^{10}
\nonumber\\ &&
+110241651054 \xi ^8-73844422036 \xi ^6+18938352964 \xi ^4-218445420 \xi ^2+3043845, \\
Q_{29}&=&-820078252875 \xi ^{22}+11363499263025 \xi ^{20}-169884277764915 \xi ^{18}
\nonumber\\ &&
+33489861620977249 \xi ^{16}-165103109949724790 \xi ^{14}+328428484284419170 \xi ^{12}
\nonumber\\ &&
-327091003475063710 \xi ^{10}+162656633930016650 \xi ^8-32062087154971039 \xi ^6
\nonumber\\ &&
-169884277764915 \xi ^4+11363499263025 \xi ^2-820078252875, \\
Q_{30}&=&-239203125 \xi ^{22}+14005183500 \xi ^{20}+39684280140400 \xi ^{18}-214042666755237 \xi ^{16}
\nonumber\\ &&
+437303357270770 \xi ^{14}-362976447216020 \xi ^{12}-29556983576980 \xi ^{10}
\nonumber\\ &&
+270593625451250 \xi ^8
-180700720391333 \xi ^6+39684280140400 \xi ^4+14005183500 \xi ^2
\nonumber\\ &&
-239203125, \\
Q_{31}&=&239203125 \xi ^{22}-14005183500 \xi ^{20}-39684280140400 \xi ^{18}
+180700720391333 \xi ^{16}
\nonumber\\ &&
-270593625451250 \xi ^{14}+29556983576980 \xi ^{12}
+362976447216020 \xi ^{10}
\nonumber\\ &&
-437303357270770 \xi ^8
+214042666755237 \xi ^6-39684280140400 \xi ^4
-14005183500 \xi ^2
\nonumber\\ &&
+239203125, \\
Q_{32}&=&820078252875 \xi ^{22}-11363499263025 \xi ^{20}+169884277764915 \xi ^{18}
\nonumber\\ &&
+32062087154971039 \xi ^{16}-162656633930016650 \xi ^{14}+327091003475063710 \xi ^{12}
\nonumber\\ &&
-328428484284419170 \xi ^{10}+165103109949724790 \xi ^8-33489861620977249 \xi ^6
\nonumber\\ &&
+169884277764915 \xi ^4-11363499263025 \xi ^2+820078252875, \\
Q_{33}&=&-23430807225 \xi ^{24}+342895368690 \xi ^{22}-5104322258070 \xi ^{20}
+1029830924634698 \xi ^{18}
\nonumber\\ &&
-6092302565491047 \xi ^{16}+15174619653233604 \xi ^{14}-20214731883198100 \xi ^{12}
\nonumber\\ &&
+15174619653233604 \xi ^{10}
-6092302565491047 \xi ^8+1029830924634698 \xi ^6
\nonumber\\ &&
-5104322258070 \xi ^4
+342895368690 \xi^2
-23430807225, \\
Q_{34}&=&-20503125 \xi ^{24}+1216391175 \xi ^{22}+3400577829420 \xi ^{20}+3124437140987 \xi ^{18}
\nonumber\\ &&
-87250203933651 \xi ^{16}+266415595377150 \xi ^{14}-371385991522312 \xi ^{12}
\nonumber\\ &&
+266415595377150 \xi ^{10}
-87250203933651 \xi ^8+3124437140987 \xi ^6+3400577829420 \xi ^4
\nonumber\\ &&
+1216391175 \xi ^2-20503125, \\
Q_{35}&=&143521875 \xi ^{24}-8514738225 \xi ^{22}-23804044805940 \xi ^{20}+478743194834467 \xi ^{18}
\nonumber\\ &&
-2392934101392699 \xi ^{16}+5644304654680590 \xi ^{14}-7412583155771336 \xi ^{12}
\nonumber\\ &&
+5644304654680590 \xi ^{10}-2392934101392699 \xi ^8+478743194834467 \xi ^6
\nonumber\\ &&
-23804044805940 \xi ^4
-8514738225 \xi ^2+143521875, \\
Q_{36}&=&23430807225 \xi ^{24}-342895368690 \xi ^{22}+5104322258070 \xi ^{20}+967466021746102 \xi ^{18}
\nonumber\\ &&
-5891479112793753 \xi ^{16}+14784834542478396 \xi ^{14}-19731207044417900 \xi ^{12}
\nonumber\\ &&
+14784834542478396 \xi ^{10}
-5891479112793753 \xi ^8+967466021746102 \xi ^6
\nonumber\\ &&
+5104322258070 \xi ^4-342895368690 \xi ^2+23430807225, \\
Q_{37}&=&-\frac{16 L_1^3}{3}-\frac{32 L_1^2 L_2}{3}-\frac{32 L_1 L_2^2}{3}- 16 ( L_1 +L_2 ) 
\zeta_2 -\frac{16 L_2^3}{3}-\frac{64}{9} \zeta_3.
\end{eqnarray}


\noindent
The first moment $\Delta \tilde{a}_{Qg}^{(3)}(N=1)$ vanishes.

\vspace*{5mm}
\noindent
{\bf Acknowledgments}.\\
We would like to thank P.~Marquard for discussions.
This work was supported by the European Research Council (ERC)
under the European Union's Horizon 2020 research and innovation programme
grant agreement 101019620 (ERC Advanced Grant TOPUP), the UZH Postdoc Grant,
grant no.~[FK-24-115], and Austrian Science Fund (FWF) Grant-DOI 10.55776/P20347.
KS is support by the European Union under the
HORIZON program in Marie Sk\l{}odowska-Curie project 
No. 101204018 
\hspace*{-2cm} \parbox{180pt}{\centering\includegraphics[height=5mm]{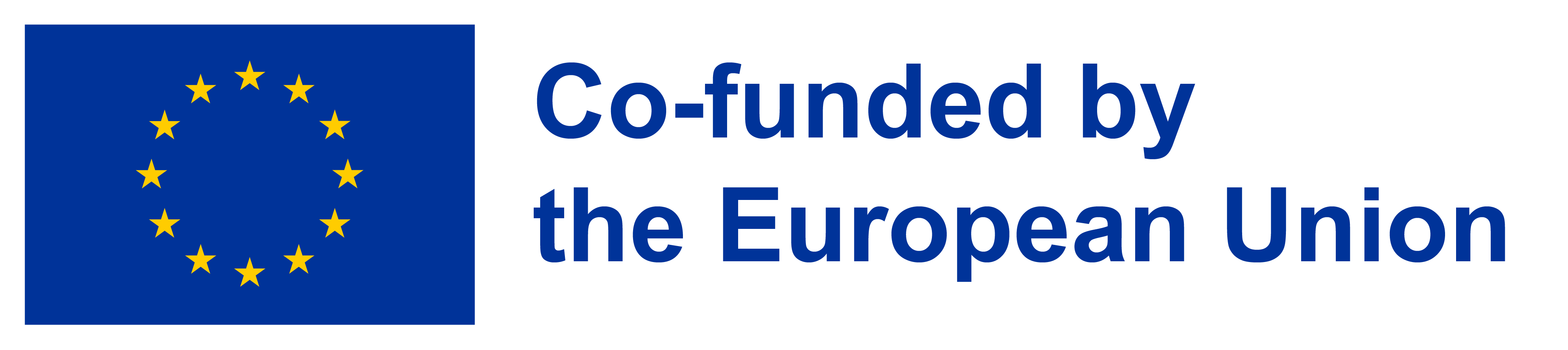}.} 


\end{document}